\documentclass[aps,pra,twocolumn,superscriptaddress]{revtex4-1}
\usepackage{amsmath}
\usepackage{graphicx}
\usepackage{amsfonts}
\usepackage{color}
\usepackage{upgreek}
\usepackage{mathtools}
\usepackage{hyperref}
\usepackage{comment}
\hypersetup{
    colorlinks = true,
    linkcolor =blue,
	citecolor=blue, 
	urlcolor=blue 
}

\makeatletter
\def\captionof#1#2{{\def\@captype{#1}#2}}
\makeatother
\usepackage{array}

\begin{document}

\title{Lyapunov equation in open quantum systems and non-Hermitian physics}

\author{Archak Purkayastha}
\email{archak.p@phys.au.dk}
\affiliation{School of Physics, Trinity College Dublin, College Green, Dublin 2, Ireland}
\affiliation{Centre for complex quantum systems, Aarhus University, Nordre Ringgade 1, 8000 Aarhus C, Denmark}

\date{\today}

\begin{abstract}
The continuous-time differential Lyapunov equation is widely used in linear optimal control theory, a branch of mathematics and engineering. In quantum physics, it is known to appear in Markovian descriptions of linear (quadratic Hamiltonian, linear equations of motion) open quantum systems, typically from quantum master equations. Despite this,  the Lyapunov equation is seldom considered a fundamental formalism for linear open quantum systems. In this work we aim to change that. We establish the Lyapunov equation as a fundamental and efficient formalism for linear open quantum systems that can go beyond the limitations of various standard quantum master equation descriptions, while remaining of much less complexity than general exact formalisms. This also provides valuable insights for non-Hermitian quantum physics. In particular, we derive the Lyapunov equation for the most general number conserving linear system in a lattice of arbitrary dimension and geometry, connected to an arbitrary number of baths at different temperatures and chemical potentials. Three slightly different forms of the Lyapunov equation are derived via an equation of motion approach, by making increasing levels of controlled approximations, without reference to any quantum master equation.  Then we discuss their relation with quantum master equations, positivity, accuracy and additivity issues, the possibility of describing dark states, general perturbative solutions in terms of single-particle eigenvectors and eigenvalues of the system, and quantum regression formulas. Our derivation gives a clear understanding of the origin of the non-Hermitian Hamiltonian describing the dynamics and separates it from the effects of quantum and thermal fluctuations. Many of these results would have been hard to obtain via standard  quantum master equation approaches.  
\end{abstract}

\maketitle

\section{Introduction}

\noindent
{\it General background ---} As we go towards the age of quantum technology, it has become crucial to formulate the theory to describe microscopic (i.e, containing finite number of degrees of freedom) quantum systems coupled to multiple macroscopic (i.e, containing infinite number of degrees of freedom) thermal environments (baths) which can all be at different temperatures and chemical potentials. This is relevant across quantum optics  \cite{LeHur2016}, thermodynamics \cite{quantum_thermodynamics}, chemistry \cite{quantum_chemistry}, biology\cite{quantum_biology} and engineering \cite{quantum_engineering}.  However, this is an extremely challenging problem in general because the dimension of Hilbert space scales exponentially with the number of degrees of freedom. 

In absence of coupling to any macroscopic bath, the situation can be considerably simplified if the Hamiltonian is quadratic in fermionic or bosonic creation and annihilation operators. In that case, many properties of the system can be obtained by calculating the so-called single-particle density matrix or the correlation matrix, whose elements are the equal time two-point correlation functions of the system \cite{Mahan_book}. The dynamics of the correlation matrix is governed by the so-called single-particle Hamiltonian of the system, whose dimension scales linearly with the number of degrees of freedom in the system. This therefore leads to an exponential simplification of the problem. 

However, in presence of coupling to multiple macroscopic thermal baths, describing even dynamics of systems governed by quadratic Hamiltonians in generality becomes quite complicated because of the baths having infinite degrees of freedom. In this case, standard technique in all open system formalisms, like non-equilibrium Green's functions (NEGF) \cite{Jauho_book, Wang_2014}, Feynman-Vernon influence functional approach \cite{Kamenev_book}, quantum Langevin equations \cite{Ford_Kac_Mazur_1965,Kac_1981,Cortes_1985,Ford_1988,Dhar_2003,
Dhruba_2004,Dhar_Sen_2006,Dhar_Roy_2006} and quantum master equations (QME) \cite{Breuer_book, Rivas_book}, is to derive effective equations for dynamics of the system after analytically integrating out the baths. For quadratic Hamiltonians, this can be done in generality, and leads to non-Markovian equations of motion \cite{Dhar_Sen_2006,Dhar_Roy_2006,WMZ_2008,Jin_2010,WMZ_2012,WMZ_2012_comment,Martinez_2013,WMZ_2013,
Ribeiro_2015}. Due to this non-Markovianity, for a system with large (but finite) number of degrees of freedom, obtaining the full dynamics again becomes quite challenging. If the long-time non-equilibrium steady state (NESS) is unique, and only the NESS properties are desired, they can be relatively easily found via a Fourier transform. But obtaining full dynamics requires a Laplace transform that needs to be finally inverted.  Inverting a Laplace transform can become difficult depending on the number of degrees of freedom and the spectrum of the system. As a result, to obtain the full dynamics, in many cases it becomes useful to make further approximations to have a much simpler effective Markovian description.
\\

\noindent
{\it Weak-coupling Markovian descriptions and their limitations ---} The standard approach to obtain a Markovian description is to assume weak coupling between the system and the baths and implement the so-called Born-Markov-secular approximations to obtain a QME in the so-called Lindblad  form \cite{Breuer_book}. But several drawbacks of Born-Markov-secular approximations have been pointed out, in particular, for describing a system coupled to multiple thermal baths \cite{Walls1970,Novotny_2002,Wichterich_2007,Novotny_2010,
Rivas_2010,barranco_2014,Levy2014,
archak_2016,
Trushechkin_2016,
Eastham_2016,Hofer_2017,Gonzalez_2017,Mitchison_2018,
Cattaneo_2019,Hartmann_2020_1,Benatti_2020,konopik_2020local,
Scali_2021,Floreanini_2021,trushechkin2021,Archak_2021}. It has been shown that not implementing the secular approximation, thereby obtaining a QME in so-called Redfield form \cite{redfield1965}, can be more accurate (for example, \cite{archak_2016,Hartmann_2020_1,Archak_2021,Davidovic_2022}). On the other hand, it is known that the Redfield equation violates complete positivity \cite{Davidovic_2022, Archak_2020, Hartmann_2020_1,Eastham_2016,anderloni_2007,Gaspard_Nagaoka_1999,Kohen_1997,
Gnutzmann_1996,Suarez_1992}. This means it can lead to having unphysical negative eigenvalues of the system density matrix. But such negative eigenvalues will be quite close to zero, and can be shown to be below accuracy level of the Redfield equation \cite{fleming_cummings_accuracy, Hartmann_2020_1, Archak_2021, Davidovic_2022}. Several more-refined Lindblad equations have been proposed to circumvent these drawbacks \cite{ule,Kleinherbers_2020,Davidovic_2020,mozgunov2020,mccauley2020,
kirvsanskas2018}. Nevertheless, it has been shown that all of these approaches have inherent limitations even to the leading order in the system-bath coupling strength when describing a system coupled to multiple thermal baths \cite{Archak_2021}.
\\

\noindent
{\it The Lyapunov equation from weak-coupling Markovian descriptions ---}
Regardless of the several limitations, because of their simplicity and the valuable insight they provide, weak coupling QME descriptions of thermal baths remain of wide use. Interestingly, for systems governed by time-independent quadratic Hamiltonians, the weak system-bath coupling QMEs referred to above give an amazing connection to seemingly unrelated fields in mathematics and engineering. It turns out that the equation of motion for the correlation matrix as obtained from such QMEs has the form of a continuous-time Lyapunov equation (see for example \cite{Prosen_2008, Prosen_2010, Prosen_2012, Koga_2012, Nicacio_2012, Landi_2013, Nicacio_2016, Landi_2017, Landi_2019, Mcdonald_2021, Bernal_Garcia_2021,Roccati_2021}). The Lyapunov equation is an extremely well-studied equation in mathematics and engineering, where appears extensively in the field of linear optimal control theory \cite{Control_theory_book1,Control_theory_book2,Yuan_2021_lyapunov_control}. It is surprising that the same equation appears in a completely different context, that of linear (i.e, governed by quadratic Hamiltonians) open quantum systems. Further, efficient numerical methods for solving Lyapunov equations already exist in all high-level scientific programming languages, like python, Mathematica, Matlab etc. 
\\

\noindent
{\it Summary of our main results ---}
Although this above fact is already known and used, Lyapunov equations are not usually discussed as a fundamental formalism for general linear open quantum systems. This is presumably due to the various limitations of the underlying QME descriptions. In this manuscript, we attempt to change that.  We establish the continuous-time differential Lyapunov equation as a rigorously derived efficient description of linear open quantum systems, that is more fundamental than many of the existing QME descriptions and can go beyond their limitations.

\begin{figure}
\includegraphics[width=0.9\columnwidth]{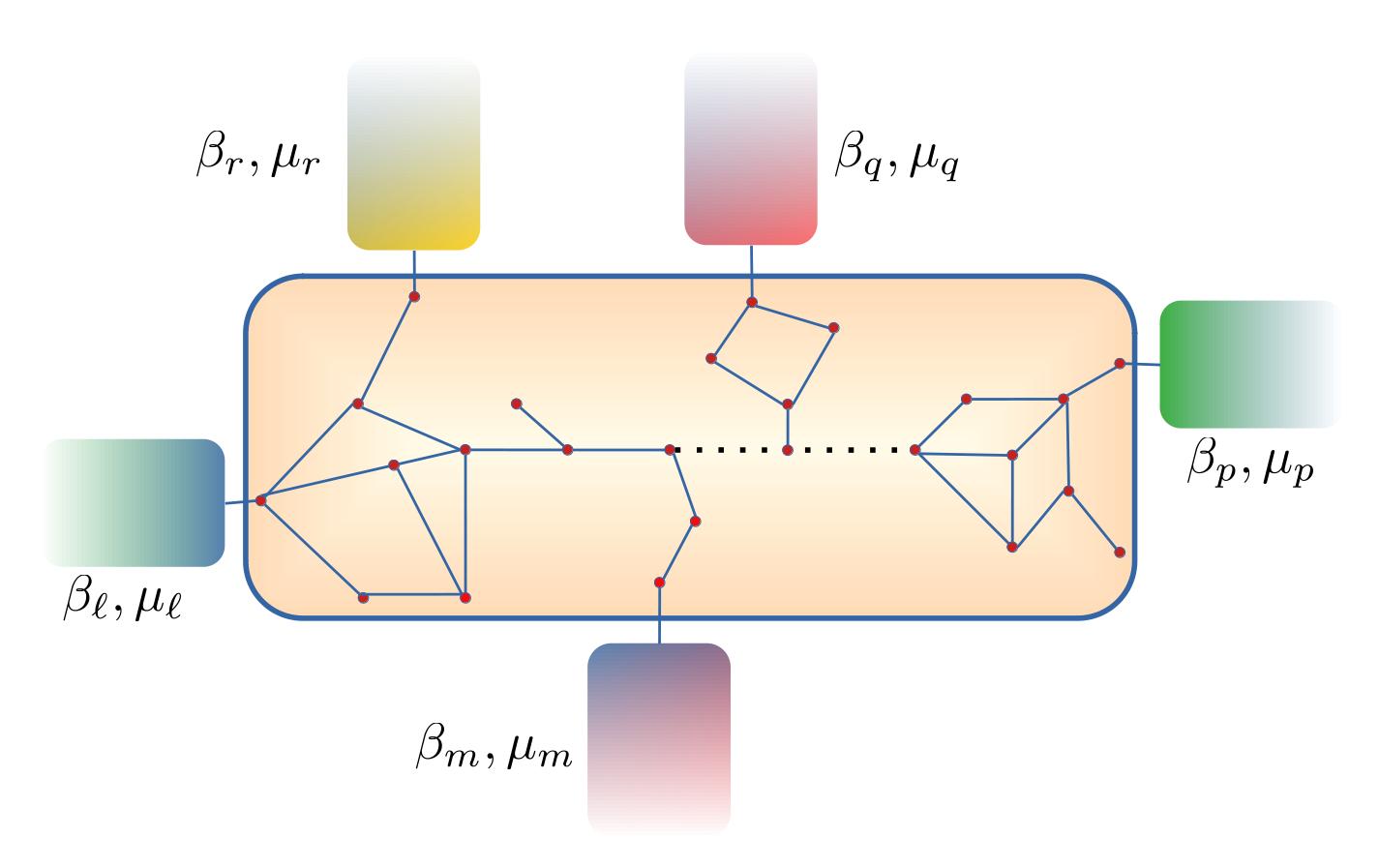}
\caption{ A system on a lattice of arbitrary geometry and dimension coupled to at an arbitrary number of sites to thermal baths which can all be at different temperatures (inverse temperatures, say, $\beta_\ell,\beta_m, \beta_p, \beta_q, \beta_r$)  and chemical potentials (say $\mu_\ell,\mu_m, \mu_p, \mu_q, \mu_r$) . \label{fig:schematic}}
\end{figure}

To this end, we consider a system governed by a number conserving quadratic time-independent Hamiltonian, bosonic or fermionic, in a lattice of arbitrary dimension and geometry bilinearly coupled at an arbitrary number of sites to thermal baths which can all be at different temperatures and chemical potentials (see Fig.~\ref{fig:schematic}). The baths are assumed to be modelled by an infinite number of bosonic or fermionic modes.  By systematically carrying out various controlled approximations on bath spectral functions and strength of system-bath couplings, we follow an equation of motion approach to obtain the continuous-time differential Lyapunov equation without referring to any QME. In particular, we derive three slightly different Lyapunov equations at three different increasing levels of approximations. At the first level of approximation, the Lyapunov equation does not have any positivity problem at all times, at the second level of approximation, the Lyapunov equation does not have a positivity problem at the NESS, while at the third level of approximation, the Lyapunov equation has the same positivity problem as the Redfield equation. This shows that the Lyapunov equation is more fundamental for describing such set-ups than the standard QMEs. In fact, the Lyapunov equations at first two levels suggest corresponding QMEs, which would be difficult to obtain otherwise. For fermionic systems, at the first two level of approximations, even the weak system-bath-coupling approximation is not required, but rather a so-called wide-band-limit approximation on the bath spectral functions suffices. The controlled microscopic derivation allows us to specify the validity regime of the Lyapunov equations and the accuracy of the solutions.  A plethora of semi-analytical results follow, some of which reduces the complexity of the problem of simulating dynamics of this non-equilibrium open quantum system to the same level as simulating dynamics of the isolated system. We also give generalized regression formulas for two-time correlations, which can be easily obtained via our operator equation of motion approach. These formulas do not correspond to those that would be obtained by naively applying the standard quantum regression formula \cite{Breuer_book} at the level of the associated QMEs. 
\\

\noindent
{\it The Lyapunov equation and non-Hermitian quantum physics ---}
The continuous-time differential Lyapunov equation is a specific form of a linear differential equation with a homogeneous and an inhomogeneous part. In our microscopic derivation, it becomes completely clear that homogeneous part is associated with time-evolution via a non-Hermitian Hamiltonian, while the inhomogeneous part is associated with quantum and thermal fluctuations due to sources of loss. The Lyapunov equation therefore gives a natural way of identifying the non-Hermitian Hamiltonian governing the dynamics and study the effect of quantum and thermal fluctuations. Non-Hermitian physics is an extremely rapidly growing field at present (see for example,  \cite{Bergholtz_2021,Ashida_2020, El_Ganainy2018,El_Ganainy2019,Feng2017}). But majority of works phenomenologically assume a non-Hermitian Hamiltonian, which is very often expected to govern a classical system, and quantum and thermal fluctuations are ignored. Only recently, there has been growing interest in going beyond such descriptions, and realizing dynamics governed by specific non-Hermitian Hamiltonians in quantum systems (for example, \cite{Nori_2018, Archak_2020_PT, Huber_2020, Huber_2020_PRA, Avila_2020, Mcdonald_2021, Gomez_Leon_2021,Arkhipov_2020,Arkhipov_2021,Roccati_2021}). On the other hand, classifying all kinds of non-Hermitian Hamiltonians and their relations to topology has remained an extremely active direction of research (for example, \cite{Wang2021,Bergholtz_2021,Kawabata_2019,Foa_Torres_2019,Gong_2018,Harari_2019,
Miguel_2019}). The Lyapunov equation then gives the unified way to account for quantum and thermal fluctuations in all such non-Hermitian systems, as long as there is no source of gain, i.e, for all passive non-Hermitian systems. (The theory presented here does not describe sources of gain. See however, \cite{Archak_2020_PT}.) If, for some physical reason, the baths can be considered empty initially, then there are no quantum and thermal fluctuations, perfectly realizing passive non-Hermitian systems. Interestingly, we find that, at all the three levels of approximation for the Lyapunov equation, the non-Hermitian Hamiltonian is the same. The three levels of approximations only change the inhomogeneous part of the Lyapunov equation. Thus, they specify at what level of accuracy the quantum and thermal fluctuations are considered.
\\

\noindent
{\it Plan of the paper ---}
In Sec.~\ref{sec:Lyapunov_equation} we formally introduce the Lyapunov equation. In Sec.~\ref{sec:set_up} we give the set-up and the exact  equations of motion. In Sec.~\ref{sec:deriving_Lyapunov_equations} we derive the non-Hermitian Hamiltonian and the Lyapunov equations. In Sec.~\ref{sec:deriving_associated_QMEs} we obtain the associated QMEs. In Sec.~\ref{sec:positivity_accuracy_additivity} we discuss positivity, accuracy and additivity issues. In Sec.~\ref{sec:dark_states_perturbative_solution_thermalization}, we discuss the possibility of dark states, give perturbative solutions and discuss thermalization. In Sec.~\ref{sec:regression_formulas} we derive the generalized regression formulas. In Sec.~\ref{sec:examples}, we discuss two insightful examples. In Sec.~\ref{sec:summary_and_outlook}, we summarize and  discuss the implications of our results. A summary of our main results is given in Table.~\ref{table1}. Finally, there is an Appendix, where details of some proofs are given.

\section{The Lyapunov equation}
\label{sec:Lyapunov_equation}

The continuous-time differential Lyapunov equation is an equation of the form
\begin{align}
\frac{d\mathbf{C}}{dt}= - \left(\mathcal{G} \mathbf{C}+ \mathbf{C}\mathcal{G}^\dagger \right) + \epsilon^2\mathbf{Q},
\end{align}
where $\mathbf{C}$, $\mathbf{Q}$ and $\mathcal{G}$ are $N \times N$ matrices, for some given $N$, and $\mathcal{G}^\dagger$ is the conjugate transpose of $\mathcal{G}$. The $\epsilon^2$ is used for notational convenience to be  consistent with that used later in the paper. Usually, as will also be in our case, $\mathbf{C}$ is required to be Hermitian and positive semi-definite at all times. The necessary and sufficient condition for this $\mathbf{Q}$ being Hermitian and positive semi-definite. The formal solution of the Lyapunov equation is 
\begin{align}
\label{C_solution}
\mathbf{C}(t) = e^{-\mathcal{G}t} \mathbf{C}(0) e^{-\mathcal{G}^\dagger t} +  \epsilon^2\int_0^t dt^\prime e^{-\mathcal{G}t^\prime} \mathbf{Q} e^{-\mathcal{G}^\dagger t^\prime}.
\end{align}
If the real parts of eigenvalues of $\mathcal{G}$ are positive, then are is a unique steady solution in the long time limit, given by
\begin{align}
\mathbf{C}(\infty) =   \epsilon^2\int_0^\infty dt^\prime e^{-\mathcal{G}t^\prime} \mathbf{Q} e^{-\mathcal{G}^\dagger t^\prime}.
\end{align}
Since this is the steady state, this is the solution of 
\begin{align}
\label{algebraic_Lyapunov}
\mathcal{G} \mathbf{C}(\infty)+ \mathbf{C}(\infty)\mathcal{G}^\dagger = \epsilon^2\mathbf{Q},
\end{align}
which is called the algebraic Lyapunov equation. This greatly simplifies the problem of finding $\mathbf{C}(\infty)$ numerically. It is complete set of $N^2$ linear equations. Usually, time taken to solve such a system of equations scales as $N^6$. However, for the algebraic Lyapunov equation, standard efficient algorithms (standard packages in Mathematica, python, Matlab etc.) are available for which the time taken to solve scales as $N^3$ \cite{Control_theory_book1,Behr_2019}. Further, recently, even more efficient Krylov subspace methods for solving Lyapunov equations have been investigated \cite{Behr_2019}.  In the following, we derive equations of the above form for dynamics of a very general linear open quantum system.

\section{The set-up and exact equations of motion}
\label{sec:set_up}

We consider the most general number-conserving linear system (quadratic Hamiltonian)  in a lattice of $N$ sites in arbitrary dimension and geometry, 
\begin{align}
\label{general_non_int}
\hat{\mathcal{H}}_S= \sum_{\ell,m=1}^N \mathbf{H}_{\ell m} \hat{c}_\ell^\dagger \hat{c}_m. 
\end{align} 
where $\hat{c}_{\ell}$ is fermionic or bosonic annihilation operator, and $\mathbf{H}$ is a Hermitian matrix, often called the single-particle Hamiltonian. We are interested in the case where an arbitrary number of sites can be attached to baths (see Fig.~\ref{fig:schematic}). However, in deriving the theory, for notational simplicity and generality, we consider that each site of the system is coupled to a bath, each of which is described by an infinite number of non-interacting modes,
\begin{align}
\label{system_bath_coupling}
&\hat{\mathcal{H}}_{SB}=\epsilon\sum_{\ell=1}^{N}\hat{\mathcal{H}}_{SB_\ell},~\hat{\mathcal{H}}_{SB_\ell}= \epsilon\sum_{r=1}^\infty \left(\kappa_{rl}\hat{c}_\ell^\dagger \hat{B}_{r\ell} + \kappa_{rl}^*\hat{B}_{r\ell}^\dagger \hat{c}_\ell\right) \nonumber \\
& \hat{\mathcal{H}}_{B}=\sum_{\ell=1}^{N}\hat{\mathcal{H}}_{B_\ell},~ \hat{\mathcal{H}}_{B_\ell}=\sum_{r=1}^\infty \Omega_{rl}\hat{B}_{r\ell}^\dagger \hat{B}_{r\ell}.
\end{align}
Here $\hat{B}_{r\ell}$ is the fermionic or bosonic annihilation operator of the $r$th mode of the bath attached to the $\ell$th site of the system. The factor $\epsilon$ is a dimensionless parameter controlling the strength of system-bath coupling. The initial state of the whole set-up is taken to be
\begin{align}
\label{initial_state}
\hat{\rho}_{tot}(0) = \hat{\rho}(0) \prod_{\ell=1}^N \frac{e^{-\beta_{\ell}\left(\hat{\mathcal{H}}_{B_\ell} - \mu_{\ell} \hat{N}_{B_\ell}\right)}}{Z_{B_\ell}},
\end{align}
where $\hat{N}_{B_\ell}=\sum_{r=1}^\infty \hat{B}_{r\ell}^\dagger \hat{B}_{r\ell}$ is the operator for total number of particles in the bath attached at the $\ell$th site and $Z_{B_\ell}$ is the partition function. So, the system is initially in an arbitrary state while the baths are all in the thermal states with their individual temperatures and chemical potentials. The spectral functions of the baths are defined by
\begin{align}
\mathfrak{J}_{\ell}(\omega) = 2\pi \sum_{r=1}^\infty | \kappa_{r \ell} |^2 \delta (\omega - \Omega_{r \ell}).
\end{align}
For future reference, we also introduce the notation $f^H(\omega)$ for the Hilbert transform of any arbitrary function $f(\omega)$,
\begin{align}
\label{def_Hilbert_transform}
f^H(\omega)= \frac{1}{\pi}\mathcal{P}\int_{-\infty}^{\infty} d\omega^\prime \frac{f(\omega^\prime)}{\omega -\omega^\prime},
\end{align} 
where $\mathcal{P}$ denotes the principal value.

The dynamics of the system can be described by the an exact quantum Langevin equation \cite{Dhar_2003,Dhar_Sen_2006,Dhar_Roy_2006}. This is derived in two steps. First, the formal solution for annihilation operators of the baths is written down,
\begin{align}
\label{bath_formal_solution}
\hat{B}_{r\ell}(t) = e^{-i\Omega_{r\ell}t}\hat{B}_{r\ell}(0)-i\epsilon\kappa_{r\ell}\int_0^t dt^\prime e^{-\Omega_{r\ell}(t-t^\prime)} \hat{c}_\ell(t^\prime).
\end{align}
Then this solution is used in the equation of motion for the annihilation operators of the system
\begin{align}
\frac{d \hat{c}_\ell}{dt} &= -i \sum_{\ell,m=1}^N \mathbf{H}_{\ell m}\hat{c}_m(t)  - i\epsilon\sum_{r=1}^\infty \kappa_{r\ell}^*\hat{B}_{r\ell}(t),
\end{align}
to obtain the so-called quantum Langevin equation,
\begin{align}
\label{exact_QLE}
\frac{d \hat{c}_\ell}{dt} &= -i \sum_{\ell,m=1}^N \mathbf{H}_{\ell m}\hat{c}_m(t)  - i \epsilon\hat{\xi}_\ell(t) \nonumber \\
&- \epsilon^2 \int_0^t dt^\prime \int \frac{d\omega}{2\pi} \mathfrak{J}_{\ell}(\omega) e^{-i\omega(t-t^\prime)} \hat{c}_\ell(t^\prime),
\end{align}
where the integration is over all bath frequencies and $\hat{\xi}_\ell(t)=\sum_{r=1}^\infty \kappa_{r\ell} e^{i\Omega_{r\ell}t}\hat{B}_{r\ell}(0)$ is the noise operator. The noise correlation functions are given by
\begin{align}
\label{noise_corrs}
& \langle \hat{\xi}_\ell (t) \rangle =0,~\langle \hat{\xi}_\ell^\dagger (t) \hat{\xi}_m (t^\prime) \rangle = \int \frac{d\omega}{2\pi}  \mathbf{F}_{\ell m}(\omega) e^{i\omega (t-t^\prime)}, 
\end{align}
with
\begin{align}
\label{def_F}
& \mathbf{F}_{\ell m}(\omega) = \mathfrak{J}_\ell(\omega) \mathfrak{n}_{B_\ell}(\omega) \delta_{\ell m}.
\end{align}
where 
\begin{align}
\mathfrak{n}_{B_\ell}(\omega)=[e^{\beta_{\ell}(\omega-\mu_{\ell})} \pm 1]^{-1}
\end{align}
is the fermi or bose distribution function, $\delta_{\ell m}$ is Kronecker delta function and $\langle ... \rangle= {\rm Tr}\left( ... \hat{\rho}_{tot}(0)\right)$. Note that the quantum Langevin equation,  Eq.(\ref{exact_QLE}), is completely exact for our set-up. It does not require on any further approximation. There is no Markovian approximation, no weak system-bath coupling approximation, also no approximation on bath spectral functions, except that they are continuous, and no approximation on the temperatures and the chemical potentials of the baths. It also holds for both fermionic and bosonic set-ups. This generality and fundamental nature of such quantum Langevin equations, although known for a long time \cite{Ford_Kac_Mazur_1965,Cortes_1985,Ford_1988,Dhar_2003,Dhar_Sen_2006,
Dhar_Roy_2006}, is often not emphasized in standard textbooks of quantum optics (for example, \cite{Milburn_book1, Milburn_book2}), where quantum Langevin equations are instead derived from weak system-bath coupling Lindblad equations. 

Another point to mention is that the effect of baths are additive at the level of the exact quantum Langevin equation. This additivity, which stems from the fact that the system-bath couplings are bilinear,  allows for easy generalization.  We have assumed that each site is coupled to single bath. It is clear that if some sites are not connected to a bath, it can be easily incorporated just by setting the system-bath coupling of those sites to zero in the quantum Langevin equation. This leads to dropping the last two terms Eq.(\ref{exact_QLE}) for the corresponding sites. Further, it can also easily be generalized to the case one or more sites are coupled to more than one bath. This can be done just by adding the corresponding terms at the corresponding sites in the quantum Langevin equation (an example of this will be given later in Sec.~\ref{example:resonant_level}). Keeping this in mind, for notational simplicity, in derivation of our results, we stick to the set-up of having each site coupled to a single bath.

The Eq.(\ref{exact_QLE}) can be exactly solved for NESS via a Fourier transform (assuming the bandwidth of the bath includes all the system modes, a necessary condition for unique NESS). It yields exactly the same expressions as obtained via a NEGF approach \cite{Dhar_Roy_2006,Dhar_Sen_2006}. However, the dynamics of approach to NESS requires Laplace transform. This can be difficult, because, inverse Laplace transform is hard. Depending on the number of lattice sites and the spectral properties of $\mathbf{H}$, this can essentially be intractable. 

On the other hand, in absence of the baths, the dynamics of the system is numerically tractable up to very large number of sites in the lattice. The $N \times N$ matrix $\mathbf{H}$ can be diagonalized
\begin{align}
\label{diag_Hs}
\Phi^\dagger \mathbf{H} \Phi = \mathbf{D}, 
\end{align} 
where $\mathbf{D}={\rm diag}\{\omega_\alpha\}$ is a diagonal matrix containing the eigenvalues of the matrix $\mathbf{H}$, which are the single-particle eigenvalues, $\Phi$ is a unitary matrix whose columns are the corresponding single-particle eigenvectors  and $\Phi^\dagger$ represents the conjugate transpose. The dynamics of the isolated system is given in terms of these is given as 
\begin{align}
\label{isolated_evolution}
\hat{c}_\ell(t) = \sum_{\alpha,m=1}^N \Phi_{\ell \alpha} \Phi_{m \alpha}^* e^{-\omega_\alpha t} \hat{c}_m(0).
\end{align}
 Thus, given the single-particle eigenvectors and eigenvalues, the isolated system dynamics can be obtained exactly. This is not true for the exact open system dynamics. In the following, using various levels of Markov approximations, we simplify the open system time evolution greatly, deriving the Lyapunov equation. 

\section{Deriving the non-Hermitian Hamiltonian and the Lyapunov equations}
\label{sec:deriving_Lyapunov_equations}

The Eq.(\ref{exact_QLE}) is our starting point for further Born-Markov-like  approximations. The non-Markovianity of Eq.(\ref{exact_QLE}) is encoded in two places, (i) the last term of Eq.(\ref{exact_QLE}), (ii) in the fact that the noise correlation is not a delta function.  
 Correspondingly, we will do two levels of Markov approximations, involving two time scales, $\tau_{B_1}$ and $\tau_{B_2}$. The corresponding Markov approximations are valid for times larger than these time scales.   As a collorary, we will also identify the non-Hermitian Hamiltonian governing the time evolution of open quantum system under such approximations.

\subsection{First Markov approximation}
\subsubsection{Equation of motion and the non-Hermitian Hamiltonian}
We call the approximation that deals with the last term in Eq.(\ref{exact_QLE}) the first Markov approximation. The main goal is to make the last term depend only on $\hat{c}_\ell(t)$, and not on its past history. There can be two different approximations to obtain this. The first is the so-called wide-band limit, with 
\begin{align}
\label{wide-band-limit}
\mathfrak{J}_\ell(\omega)=\Gamma_\ell.
\end{align}
This directly gives 
\begin{align}
\int_0^t dt^\prime \int \frac{d\omega}{2\pi} \mathfrak{J}_{\ell}(\omega) e^{-i\omega(t-t^\prime)} \hat{c}_\ell(t^\prime)= \frac{\Gamma_\ell}{2}\hat{c}_\ell,
\end{align}
where the factor of $1/2$ appears because the time integration is from $0$ to $t$. This approximation does not depend on strength of system-bath coupling, $\Gamma_\ell$ can be arbitrary. While this approximation is often used for fermionic systems, it is artificial for bosonic systems, where $\mathfrak{J}_{\ell}(\omega)$ must go to zero at $\omega=0$. So, in the following, we will focus on the other way of doing the first Markov approximation, which explicitly depends on weak system-bath coupling.  Later, we will discuss which of the approximations hold  in the wide-band limit for fermionic systems without assuming weak system-bath coupling, and which do not. 

We will write Eq.(\ref{exact_QLE}) correct to $O(\epsilon^2)$. The last term of the equation is already or $O(\epsilon^2)$. So, following Eq.(\ref{isolated_evolution}) we use the result,
\begin{align}
\hat{c}_\ell(t^\prime) = \sum_{\alpha,m=1}^N \Phi_{\ell \alpha} \Phi_{m \alpha}^* e^{\omega_\alpha (t-t^\prime)} \hat{c}_m(t) + O(\epsilon), 
\end{align} 
and neglect the $O(\epsilon)$ term. This gives
\begin{align}
&\epsilon^2\int_0^t dt^\prime \int \frac{d\omega}{2\pi} \mathfrak{J}_{\ell}(\omega) e^{-i\omega(t-t^\prime)} \hat{c}_\ell(t^\prime)  \\
\simeq &\epsilon^2 
\sum_{\alpha,m=1}^N \Phi_{\ell \alpha} \Phi_{m \alpha}^*  \hat{c}_m(t)\int_0^t dt^\prime \int \frac{d\omega}{2\pi} \mathfrak{J}_{\ell}(\omega) e^{-i(\omega-\omega_\alpha)t^\prime} \nonumber.
\end{align}
The above is essentially a Born approximation. Next we make a Markov approximation. Let $\tau_{B_1}$ be defined via the condition,  
\begin{align}
\label{def_tau_B1}
\left| \int \frac{d\omega}{2\pi} \mathfrak{J}_{\ell}(\omega) e^{-i\omega t} \right|<\textrm{some tolerance, say $O(\epsilon)$, } \forall~t>\tau_{B_1}
\end{align}
for all $\ell$.
If $t\gg \tau_{B_1}$, we can essentially take the upper limit of the time integral to infinity since this only causes changes in higher order terms. This, with a little algebra, leads to
\begin{align}
\label{QLE_first_Markov}
\frac{d c_{vec}}{dt} = -i \mathbf{H}_{\rm NH}~c_{vec}(t) - i \epsilon\xi_{vec}(t), 
\end{align} 
where $c_{vec}(t)$ ($\xi_{vec}(t)$) is a column vector with the $\ell$th element being $\hat{c}_\ell(t)$ ($\hat{\xi}_\ell(t)$), and
\begin{align}
\label{def_non_Hermitian_Hamiltonian}
\mathbf{H}_{\rm NH}=\mathbf{H}-i\epsilon^2 \mathbf{v},
\end{align}
 is the non-Hermitian Hamiltonian governing the open system dynamics. The elements of the matrix $\mathbf{v}$ are given by,
\begin{align}
\label{def_v}
\mathbf{v}_{\ell m} = \frac{1}{2}\sum_{\alpha=1}^N \Phi_{\ell \alpha} \Phi_{m \alpha}^* \Big( \mathfrak{J}_\ell (\omega_\alpha) + i \mathfrak{J}_\ell^H (\omega_\alpha) \Big), 
\end{align}
with $\mathfrak{J}_\ell^H (\omega)$ being the Hilbert transform of $\mathfrak{J}_\ell(\omega)$ (see Eq.(\ref{def_Hilbert_transform})). Thus, the first Markov approximation allows identification of the (single-particle) non-Hermitian Hamiltonian that governs the dynamics of the open quantum system. All kinds of non-Hermitian Hamiltonians with losses can be derived by changing the relative strengths of system-bath couplings and the choosing the spectral functions. However, baths of this kind are unable to describe sources of gain. So, this gives a microscopic way of generating all kinds of passive non-Hermitian systems. In accordance with fluctuation-dissipation theorem, the sources of loss also give sources of noise, embodied in the term $\xi_{vec}(t)$. Thus, the effects of quantum and thermal fluctuations on all the classes of passive non-Hermitian systems, as well as possible transitions between them \cite{Bergholtz_2021,Kawabata_2019,Foa_Torres_2019,Gong_2018}, can be studied in this set-up. It can also be seen that if temperatures and chemical potentials of the baths are such that they can be considered empty, i.e, $\mathfrak{n}_{B_\ell}(\omega)=0$, then there will be no noise, thereby perfectly realizing quantum passive non-Hermitian systems.

Note that, the matrix $\mathbf{v}$ in Eq.(\ref{def_v}) is not diagonal in the site basis in general. However, for fermionic wide-band baths, Eq.(\ref{wide-band-limit}), $\mathbf{v}$ is diagonal in the site basis,
\begin{align}
\label{wide_band_v}
\mathbf{v}_{\ell m} = \frac{\Gamma_\ell}{2}\delta_{\ell m}~~~\textrm{for wide band limit}.
\end{align}
As mentioned, this case does not require weak system-bath coupling approximation (i.e, holds with $\epsilon=1$). 

Having derived the non-Hermitian Hamiltonian, we now derive the Lyapunov equations.

\subsubsection{The correlation matrix and the Lyapunov-like form}
The formal solution of Eq.(\ref{QLE_first_Markov}) can be written as
\begin{align}
\label{formal_soln_first_markov}
c_{vec}(t) = e^{-i\mathbf{H}_{\rm NH} t} c_{vec}(0) - i\epsilon \int_0^t dt^\prime e^{-i\mathbf{H}_{\rm NH} (t-t^\prime)} \xi_{vec}(t^\prime).
\end{align}
We will be interested in the the correlation matrix $\mathbf{C}$ whose elements are
\begin{align}
\mathbf{C}_{\ell m}(t) = \langle \hat{c}_\ell^\dagger (t) \hat{c}_m(t) \rangle.
\end{align}
This matrix is Hermitian and positive semi-definite by construction.
The expression for $\mathbf{C}(t)$ can be obtained from Eq.(\ref{formal_soln_first_markov}) after some algebra (see Appendix~\ref{appendix:first_markov_derivs} for explicit steps),
\begin{align}
\label{C_formal_soln_first_markov}
&\mathbf{C}(t) = e^{-\mathcal{G}t} \mathbf{C}(0) e^{-\mathcal{G^\dagger}t} \nonumber \\
& +\epsilon^2 \int \frac{d\omega}{2\pi} \left( \frac{1-e^{-(\mathcal{G}+i\omega \mathbb{I})t}}{\mathcal{G}+i\omega \mathbb{I}} \mathbf{F}(\omega) \frac{1-e^{-(\mathcal{G}^\dagger-i\omega \mathbb{I})t}}{\mathcal{G}^\dagger-i\omega \mathbb{I}} \right), 
\end{align}  
where $\mathbb{I}$ is $N$ dimensional identity matrix, 
\begin{align}
\label{def_G}
\mathcal{G}=-i\mathbf{H}_{\rm NH}^*,
\end{align}
with $\mathbf{H}_{\rm NH}^*$ representing complex conjugate of $\mathbf{H}_{\rm NH}$ 
and $\mathbf{F}(\omega)$ is a $N \times N$ matrix defined in Eq.(\ref{def_F}).
Since $\mathbf{F}(\omega)$ is a positive semi-definite diagonal matrix by construction, the structure of Eq.(\ref{C_formal_soln_first_markov}) ensures Hermiticity and positive semi-definiteness of $\mathbf{C}(t)$ at all times. If the NESS is unique, i.e, if the real parts of eigenvalues of $\mathcal{G}$ are positive, $\mathbf{C}(\infty)$ is given by
\begin{align}
\label{NESS_first_markov}
\mathbf{C}(\infty)=\epsilon^2 \int \frac{d\omega}{2\pi} \left( \frac{1}{\mathcal{G}+i\omega \mathbb{I}} \mathbf{F}(\omega) \frac{1}{\mathcal{G}^\dagger-i\omega \mathbb{I}} \right).
\end{align}
This equation could also be obtained by solving Eq.(\ref{QLE_first_Markov}) by a fourier transform and then obtaining the correlation matrix.

While Eq.(\ref{C_formal_soln_first_markov}) can be directly used to numerically obtain the correlation matrix, it does not reveal the relation to the Lyapunov equation. To see this, we write down the equation of motion for the correlation matrix directly from Eq.(\ref{QLE_first_Markov}). After some simplification, this can be written as
\begin{align}
\label{Lyapunov_first_markov}
& \frac{d \mathbf{C}}{dt}= -(\mathcal{G} \mathbf{C} + \mathbf{C} \mathcal{G}^\dagger ) + \epsilon^2 \mathbf{Q}(t), \nonumber \\
& \mathbf{Q}(t) = i\left( \mathbf{C}_\xi (t) -\mathbf{C}_\xi^\dagger (t) \right),
\end{align}
where elements of the matrix $\mathbf{C}_\xi(t)$ are given by
$\mathbf{C}_{\xi_{\ell m}}(t) = \langle \hat{\xi}_\ell^\dagger (t) \hat{c}_m(t)\rangle$. Using Eq.(\ref{formal_soln_first_markov}), $\mathbf{C}_\xi(t)$ is obtained as
\begin{align}
\label{C_xi}
\mathbf{C}_\xi(t) = -i \int_0^t dt^\prime \int \frac{d\omega}{2 \pi} \mathbf{F}(\omega)e^{i\omega t^\prime} e^{-\mathcal{G}^\dagger t^\prime}.
\end{align}
The Eq.(\ref{Lyapunov_first_markov}) is exactly of the Lyapunov equation form, except for a time-dependent inhomogeneous part. It can be checked that Eq.(\ref{C_formal_soln_first_markov}) is the solution of this equation (see Appendix~\ref{appendix:some_proofs}). Therefore, this Lyapunov-like form preserves the positive semi-definiteness of $\mathbf{C}(t)$ at all times. Also note that since at the first-Markov approximation level we treat the noise correlations without further approximations, and the noise correlations are not delta functions in time, the dynamics of the system is actually explicitly non-Markovian in this case. Next, we make approximations on the noise correlations.

\subsection{Second Markov approximation}
It can be seen that the time-dependence of the inhomogeneous part in Eq.(\ref{Lyapunov_first_markov}) stems from the noise correlation not being a delta function. We do the second Markov approximation to do away with this time dependence.  This can be done at two levels.
\subsubsection{Level-I second Markov approximation}
 Let $\tau_{B_2}$ be defined via the condition 
\begin{align}
\label{def_tau_B2}
&\left| \int \frac{d\omega}{2 \pi} \mathfrak{J}_\ell(\omega) \mathfrak{n}_{B_\ell}(\omega)e^{i\omega t}\right|<\textrm{some tolerance, say $O(\epsilon)$, }\nonumber \\
&~\forall~t>\tau_{B_2}
\end{align}
for all $\ell$.
 So, assuming $t \gg \tau_{B_2}$, from Eq.(\ref{def_F}), we see that we can essentially extend the upper limit of the time integral in Eq.(\ref{C_xi}) to infinity, since this only changes the higher order terms. This gives the Lyapunov equation,
\begin{align}
\label{Lyapunov_second_markov_1}
& \frac{d \mathbf{C}}{dt}= -(\mathcal{G} \mathbf{C} + \mathbf{C} \mathcal{G}^\dagger ) + \epsilon^2 \mathbf{Q}_1,  \\
& \mathbf{Q}_1 = \int \frac{d\omega}{2\pi} \left(\mathbf{F}(\omega)\frac{1}{\mathcal{G}^\dagger-i\omega \mathbb{I}}+\frac{1}{\mathcal{G}+i\omega \mathbb{I}}\mathbf{F}(\omega) \right). \label{Q_1}
\end{align} 
The $\mathbf{Q}_1$ above is nothing but $\mathbf{Q}(\infty)$.
This level-I second Markov approximation approximation, by construction, becomes more and more accurate as $t$ is increased. In the $t\rightarrow \infty$ limit, corresponding to NESS, the results from Eq.(\ref{Lyapunov_second_markov_1}) and Eq.(\ref{Lyapunov_first_markov}) match. Thus, $\mathbf{C}(\infty)$ obtained from the algebraic version of above Lyapunov equation (see Eq.(\ref{algebraic_Lyapunov})) is Hermitian and positive semi-definite as required. If the NESS is unique, it is given by Eq.(\ref{NESS_first_markov}) (see Appendix~\ref{appendix:some_proofs}).

However, $\mathbf{Q}_1$ given in the expression in Eq.(\ref{Lyapunov_second_markov_1}) is not guaranteed to be positive semi-definite. Thus, unlike Eq.(\ref{Lyapunov_first_markov}), Eq.(\ref{Lyapunov_second_markov_1}) may not preserve the positive semi-definiteness $\mathbf{C}(t)$ at all times. Nevertheless, the negative eigenvalues of $\mathbf{C}(t)$, if at all they exist, will go towards zero on reducing  $\epsilon$ and with increase in time. So, they should be taken to be zero within the accuracy regime of this equation.

The fact that the NESS obtained from Eq.(\ref{Lyapunov_second_markov_1}) is positive semi-definite even though $\mathbf{Q}_1$ may not be so is presumably quite interesting on the mathematical front. This is because, it means that the algebraic Lyapunov equation, Eq.(\ref{algebraic_Lyapunov}), can provably yield positive semi-definite results even when its inhomogeneous part is not positive semi-definite.

Next, we make one more level of approximation, which, as we will see later, makes the description equivalent to that obtained from the Redfield equation.

\subsubsection{Level-II second Markov approximation}
In this level, a further approximation, using the fact that the inhomogeneous part is already $O(\epsilon^2)$, is made to essentially perform the integration in the definition of $\mathbf{Q}_1$ in Eq.(\ref{Lyapunov_second_markov_1}).  To do this, we define,
\begin{align}
\label{eigenbasis_matrices}
&\mathbf{F}^E(\omega) = \Phi^\dagger \mathbf{F}(\omega) \Phi,~\mathbf{v}^E = \Phi^\dagger \mathbf{v} \Phi, \nonumber \\
&~\mathcal{G}^E = \Phi^\dagger\mathcal{G}\Phi, ~~\mathcal{G}^E = -i\mathbf{D}-\epsilon^2 \mathbf{v}^E.
\end{align} 
From Eq.(\ref{C_xi}) and above definitions, it follows that
\begin{align}
\label{C_xi_level_II_second_markov}
\mathbf{C}_\xi(t) \simeq & -i \Phi \left(\int_0^t dt^\prime \int \frac{d\omega}{2 \pi} \mathbf{F}^E(\omega)e^{i\omega t^\prime} e^{-i\mathbf{D} t^\prime} \right) \Phi^\dagger \nonumber \\
&+ O(\epsilon^2).
\end{align}
Assuming $t \gg \tau_{B_2}$ to extend the upper limit of the time integral to infinity, performing the integral, and neglecting $O(\epsilon^2)$ terms, we obtain,
\begin{align}
\mathbf{C}_{\xi_{\ell m}}(\infty) \simeq -\frac{i}{2}\sum_{\alpha=1}^N  \Phi_{\ell \alpha} \Phi_{m \alpha}^* \Big( \mathbf{F}_{\ell \ell} (\omega_\alpha) - i \mathbf{F}_{\ell \ell}^H (\omega_\alpha) \Big).
\end{align}
The Lyapunov equation is then given by
\begin{align}
\label{Lyapunov_second_markov_2}
& \frac{d \mathbf{C}}{dt}= -(\mathcal{G} \mathbf{C} + \mathbf{C} \mathcal{G}^\dagger ) + \epsilon^2 \mathbf{Q}_2, 
\end{align}
with the elements of the matrix $\mathbf{Q}_2$ given by
\begin{align}
\label{def_Q2}
\mathbf{Q}_{2_{\ell m}} = \frac{1}{2}\sum_{\alpha=1}^N & \Phi_{\ell \alpha} \Phi_{m \alpha}^* \Big( \mathbf{F}_{\ell \ell} (\omega_\alpha) - i \mathbf{F}_{\ell \ell}^H (\omega_\alpha)  \nonumber \\
& + \mathbf{F}_{m m} (\omega_\alpha) + i \mathbf{F}_{m m}^H (\omega_\alpha) \Big).
\end{align}
Thus, $\mathbf{Q}_2$ is given in terms of single-particle eigenvalues and eigenvectors of the system. Even though $\mathbf{Q}_2$ is Hermitian, like $\mathbf{Q}_1$, it is not guaranteed to be positive semi-definite. Moreover, unlike Eq.(\ref{Lyapunov_first_markov}), even the steady state solution of Eq.(\ref{Lyapunov_second_markov_2}) may not strictly be same as that from Eq.(\ref{Lyapunov_second_markov_1}). So there might be violation of positive semi-definiteness of $\mathbf{C}(\infty)$. But, the deviations from the results of  Eq.(\ref{Lyapunov_second_markov_1}) will be small and will decrease on decreasing $\epsilon$. 

The level-II second markov approximation explicitly requires weak system-bath coupling. It requires weak system-bath coupling even for fermionic baths in the wide-band limit (Eq.(\ref{wide-band-limit})) (unless all the baths are either completely full or completely empty, i.e, the chemical potentials are $\pm\infty$). However, the first Markov approximation and the level-I second Markov approximation do not require weak system-bath coupling approximation for fermionic baths at wide-band limit. The first-Markov approximation gives exact dynamics in this case, while the level-I second Markov approximation gives approximate dynamics but exact steady state (even when system-bath coupling is strong, i.e, $\epsilon=1$).

Having derived the Lyapunov equations, in the next section, we find the QMEs which can be associated with them.

\section{Relation to quantum master equations}
\label{sec:deriving_associated_QMEs}

In above, we have derived the Lyapunov equations directly from equations of motion. Usually, Lyapunov equations arise in open quantum systems when calculating correlation matrices of linear systems governed by some QME. The question, then, is what are the QMEs corresponding to the Lyapunov equations derived in the previous section. 

Using standard Born-Markov approximations (without any secular approximation), the Redfield QME corresponding to our set-up can be derived. This is given by \cite{archak_2016}
\begin{widetext}
\begin{align}
\label{RQME}
&\frac{\partial \hat{\rho}}{\partial t} = i[\hat{\rho}, \hat{\mathcal{H}}_S +\hat{\mathcal{H}}_{LS}] + \epsilon^2 \sum_{\ell,m=1}^N \Big[  \big( \mathbf{v}_{\ell m} + \mathbf{v}_{m \ell}^* \mp \mathbf{Q}_{2_{m \ell }}  \big) \left( \hat{c}_m \rho \hat{c}_\ell^\dagger - \frac{1}{2} \{ \hat{c}_\ell^\dagger \hat{c}_m, \hat{\rho} \} \right) + \mathbf{Q}_{2_{m \ell}} \left(\hat{c}_\ell^\dagger \rho \hat{c}_m - \frac{1}{2} \{ \hat{c}_m \hat{c}_\ell^\dagger, \hat{\rho} \} \right) \Big], 
\end{align} 
\end{widetext}
where $-$ sign is for fermions and $+$ sign is for bosons, $\{\hat{A},\hat{B}\}=\hat{A}\hat{B}+\hat{B}\hat{A}$ is the anti-commutator,  $\mathbf{Q}_2$ and $\mathbf{v}$ are as given in Eqs.(\ref{def_Q2}) and (\ref{def_v}), and
\begin{align}
\hat{\mathcal{H}}_{LS} = \sum_{\ell, m, \alpha=1}^N \Phi_{\ell \alpha} \Phi_{m \alpha}^* \left( \frac{\mathfrak{J}_\ell^H(\omega_\alpha) +  \mathfrak{J}_m^H(\omega_\alpha)}{4} \right) \hat{c_\ell}^\dagger \hat{c}_m
\end{align}
is the so called Lamb-shift Hamiltonian.
It can be checked, directly by calculating the correlation matrix $\mathbf{C}_{\ell m}(t) = {\rm Tr}\left(\hat{c}_\ell^\dagger \hat{c}_m \hat{\rho}(t) \right)$, that the corresponding Lyapunov equation is nothing but Eq.(\ref{Lyapunov_second_markov_2}).  This shows equivalence of the equation of motion approach and the Redfield QME approach for Gaussian systems. This is of course expected as they describe the same set-up under same approximations. 

From above, it follows by direct inspection that
\begin{widetext} 
\begin{align}
\label{QME-2-second_markov_1}
&\frac{\partial \hat{\rho}}{\partial t} = i[\hat{\rho}, \hat{\mathcal{H}}_S +\hat{\mathcal{H}}_{LS}] + \epsilon^2 \sum_{\ell,m=1}^N \Big[  \big( \mathbf{v}_{\ell m} + \mathbf{v}_{ m \ell}^* \mp \mathbf{Q}_{1_{m \ell}}  \big) \left( \hat{c}_m \rho \hat{c}_\ell^\dagger - \frac{1}{2} \{ \hat{c}_\ell^\dagger \hat{c}_m, \hat{\rho} \} \right) + \mathbf{Q}_{1_{m\ell}} \left(\hat{c}_\ell^\dagger \rho \hat{c}_m - \frac{1}{2} \{ \hat{c}_m \hat{c}_\ell^\dagger, \hat{\rho} \} \right) \Big] 
\end{align}  
\end{widetext}
is the QME corresponding to the Lyapunov equation after level-I second Markov approximation Eq.(\ref{Lyapunov_second_markov_1}), while,
\begin{widetext}
\begin{align}
\label{QME-3-first_markov}
&\frac{\partial \hat{\rho}}{\partial t} = i[\rho, \hat{\mathcal{H}}_S +\hat{\mathcal{H}}_{LS}] + \epsilon^2 \sum_{\ell,m=1}^N \Big[  \Big( \mathbf{v}_{\ell m} + \mathbf{v}_{m \ell}^* \mp \mathbf{Q}_{{ m \ell}}(t)  \Big) \left( \hat{c}_m \rho \hat{c}_\ell^\dagger - \frac{1}{2} \{ \hat{c}_\ell^\dagger \hat{c}_m, \hat{\rho} \} \right) + \mathbf{Q}_{{m \ell }}(t) \left(\hat{c}_\ell^\dagger \rho \hat{c}_m - \frac{1}{2} \{ \hat{c}_m \hat{c}_\ell^\dagger, \hat{\rho} \} \right) \Big] 
\end{align}  
\end{widetext}
is the QME corresponding to the Lyapunov equation with time-dependent inhomogeneous part after first Markov approximation, Eq.(\ref{Lyapunov_first_markov}). These QMEs would be hard to derive using the standard approaches. In particular, Eq.(\ref{QME-3-first_markov}) has explicitly time-dependent rates, which means it can potentially describe non-Markovian dynamics. Further, for fermionic systems in the wide-band limit, Eq.(\ref{QME-3-first_markov}) is essentially exact, i.e, does not require any more approximation (holds even in strong system-bath coupling, i.e, when $\epsilon=1$). On the other hand, Eq.(\ref{QME-2-second_markov_1}), although makes some approximation on the dynamics, in this case, it gives the correct steady state exactly, without any further approximations. In the next section, we discuss in more detail the positivity, accuracy and additivity issues of the Lyapunov equations and the associated QMEs.

\section{Positivity, accuracy and additivity}
\label{sec:positivity_accuracy_additivity}

The positivity issues of the Lyapunov equation have been mentioned above, while those in QMEs have already been well studied \cite{lindblad1976, GKS1976, Gorini_1978, Breuer_book, Rivas_book}. In the following, we discuss both kinds of positivity issues, the relation between them, their origin from the accuracy issue and how they may be circumvented in practice. We also comment on additivity of the QMEs.
\subsection{Positivity}
All the QMEs given above are in the Gorini-Kossakowski-Sudarshan-Lindblad (GKSL) \cite{lindblad1976, GKS1976, Gorini_1978} form. For time-independent rates, Eqs.(\ref{RQME}), (\ref{QME-2-second_markov_1}), the condition for complete positivity of the density matrix $\hat{\rho}$ is given by \cite{Breuer_book, Rivas_book}
\begin{align}
\label{GKSL_positivity_condition}
& \mathbf{v}^*+\mathbf{v}^T \mp \mathbf{Q}_{1,2} \rightarrow \textrm{positive semi-definite} \nonumber \\
& \mathbf{Q}_{1,2} \rightarrow \textrm{positive semi-definite}.
\end{align} 
On the other hand, as we have discussed before, positive semi-definiteness of the correlation matrix $\mathbf{C}$ requires
\begin{align}
\label{Lyapunov_positivity_condition}
\mathbf{Q}_{1,2} \rightarrow \textrm{positive semi-definite}.
\end{align}
Thus, the first condition in Eq.(\ref{GKSL_positivity_condition}) is not required. It follows that, for Gaussian initial states of the system, only $\mathbf{Q}_{1,2}$ being positive semi-definite is sufficient for positivity of $\hat{\rho}$. To see this, we note that QMEs derived above preserve the Gaussianity of the initial state. This is, of course, consistent, because the actual time evolution, without any approximation, $e^{-i\hat{\mathcal{H}}t} \hat{\rho}_{tot}(0) e^{i\hat{\mathcal{H}}t}$ has this property. Now, for all Gaussian states, the density matrix can be constructed from the correlation matrix (see, for example,\cite{Peschel_2001,Dhar_2012,Peschel_2017}), giving a one-to-one mapping between the two at all times. So, a valid, positive semi-definite correlation matrix, guaranteed by Eq.(\ref{Lyapunov_positivity_condition}), will yield a valid density matrix. If, on the other hand, the initial state is non-Gaussian, this one-to-one mapping breaks. The correlation matrix has no information about the non-Gaussianity of the initial state. To ensure complete positivity of the density matrix at all times in such cases, both conditions in Eq.(\ref{GKSL_positivity_condition}) are required. It immediately follows that if the steady state is unique irrespective of the initial condition (i.e, if the real parts of eigenvalues of $\mathcal{G}$ are positive), then it must be Gaussian. This means that to ensure positivity in the steady state, only Eq.(\ref{Lyapunov_positivity_condition}) is sufficient.

The above discussion leads us to the following two important conclusions. The first is that, for Gaussian initial states, the dynamics obtained from Eq.(\ref{QME-3-first_markov}) is free from positivity issues at all times. The second is that, the NESS of Eq.(\ref{QME-2-second_markov_1}) is free from any positivity issues. This is despite the fact that in general, $\mathbf{Q}_1$ may not be positive semi-definite and therefore Eq.(\ref{QME-2-second_markov_1}) may not be completely positive. In fact, for fermionic wide-band baths, even in cases where Eq.(\ref{QME-2-second_markov_1}) is not completely positive, it always yields exact NESS results. This is contrary to the somewhat popular belief that complete positivity is a necessary requirement for accurately describing the steady state.   

\subsection{Accuracy}
The positivity issues remain in the Redfield QME Eq.(\ref{RQME}), and the corresponding Lyapunov equation Eq.(\ref{Lyapunov_second_markov_2}), even for Gaussian initial states and even in the steady state. But this is related to accuracy of the results. Both of these are differential equations written correct to $O(\epsilon^2)$. But, their solution requires an exponentiation, thereby generating all orders of $\epsilon$ in the result. Clearly, all orders of $\epsilon$ in the result are not correct. For the Redfield equation, it can be shown that  the diagonal elements of $\hat{\rho}$ in the eigenbasis of the system Hamiltonian $\hat{\mathcal{H}}_S$ are given correctly to $O(1)$ (leading order), the error occurring at $O(\epsilon^2)$, while the off-diagonal elements in that basis are given correctly to $O(\epsilon^2)$ (leading order), the error occurring at $O(\epsilon^4)$ \cite{fleming_cummings_accuracy,Archak_2021} (assuming no degeneracy). The correlation matrix in the eigenbasis of the system Hamiltonian is given by,
\begin{align}
\label{eigenbasis_correlations}
\mathbf{C}^E(t) = \Phi^\dagger \mathbf{C}(t) \Phi.
\end{align} 
Following similar arguments, it can be shown that $\mathbf{C}^E_{\alpha \alpha}(t)$ is obtained correct to $O(1)$ while $\mathbf{C}^E_{\alpha \nu}(t)$, $\alpha \neq \nu$ is given correct to $O(\epsilon^2)$. It is exactly this mismatch in orders of accuracy between the diagonal and the off-diagonal elements that leads to positivity violation both for $\hat{\rho}(t)$ and $\mathbf{C}(t)$. It can be checked that, the diagonal elements in any basis carry an error of $O(\epsilon^2)$. So $O(\epsilon^2)$ violation of positivity is related with accuracy limits of the equations. 

If positivity is restored by doing ad-hoc approximations at the level of the differential equations, for example, by neglecting the negative eigenvalues of $\mathbf{Q}_{1,2}$, or making secular approximations, it does not guarantee better accuracy. In fact, accuracy usually become worse \cite{Archak_2021,Davidovic_2022}. On the other hand, the results from the Redfield QME and the corresponding Lyapunov equation can be checked by scaling the required matrix elements with $\epsilon$ \cite{Archak_2020, Archak_2021}. If the scaling observed is higher than that dictated by the accuracy, then that matrix element is to be taken as zero. This is because, it would mean that the leading order contribution is zero while the non-zero values are coming as an artefact of the higher orders present in the solution of the equation.  This gives a completely controlled way of checking and correcting the results. Further, interesting techniques for obtaining the $O(\epsilon^2)$ correction to diagonal elements in the energy eigenbasis at steady state from the Redfield equation have been developed \cite{Juzar_2012,Juzar_2013,Juzar_2017}. These techniques can be used to alleviate positivity issues at steady state \cite{Archak_2020}.

Also, in the first and the second Markov approximations, we have assumed $t \gg \tau_{B_1}$ (Eq.(\ref{def_tau_B1})) and $t \gg \tau_{B_2}$ (Eq.(\ref{def_tau_B2}))  respectively. So results for times smaller than these time scales will not be accurate.

\subsection{Additivity}
\label{subsec:additivity}

A point to note is that in all the forms of Lyapunov equations and the associated QMEs the contribution from each bath comes as an additive term. It is sometimes believed that additive QMEs cannot give accurate NESS, especially at strong system-bath coupling \cite{Chan_2014,Mitchison_2018,Giusteri_2017,Bogna_2018,Maguire_2019}. However, as mentioned before, Eqs.(\ref{QME-3-first_markov}) and (\ref{QME-2-second_markov_1}) give exact NESS for fermionic case with wide-band baths, irrespective of the strenght of system-bath coupling. But clearly they are additive. This prompts a deeper discussion.

Most works exploring inaccuracy of additive QMEs (except Ref.~\cite{Bogna_2018}), refer to additivity of equations in the diagonal Lindblad form,
\begin{align}
\label{Lindblad_form}
& \frac{\partial \hat{\rho}}{\partial t}=i[\hat{\rho},\mathcal{\hat{H}}_S]+\hat{\mathcal{L}}(\hat{\rho}), \nonumber \\
&\hat{\mathcal{L}}(\hat{\rho}) = i[\hat{\rho},\mathcal{\hat{H}}_{LS}]+\sum_{\lambda}\gamma_{\lambda} \Big( \hat{L}_{\lambda} \hat{\rho} \hat{L}_{\lambda}^\dagger - \frac{1}{2} \{ \hat{L}_{\lambda}^\dagger \hat{L}_{\lambda}, \hat{\rho}   \} \Big),
\end{align}
where $\hat{L}_{\lambda}$ are called Lindblad operators, and $\gamma_\lambda$ are called rates. Clearly, for a system with more than one site, none of Eqs.(\ref{RQME}), (\ref{QME-2-second_markov_1}) and (\ref{QME-3-first_markov}) are explicitly in this form. Rather, they are in the so-called off-diagonal GKSL form. They can be cast into the form of Eq.(\ref{Lindblad_form}) by making a change of basis \cite{lindblad1976, GKS1976, Gorini_1978, Breuer_book, Rivas_book}. The Lindblad operators and the rates so obtained will no longer have just additive contribution from each bath, but rather will be a linear combination of contributions from all the baths. Thus, once converted to form of Eq.(\ref{Lindblad_form}), they will be non-additive in this strict sense. Note that, the above statements also hold true for the Redfield equation Eq.(\ref{RQME}), which therefore would be non-additive in this sense, whenever the system has more than one site. 

A single-site system, then, gives an exception. A single fermionic site coupled to two wide-band fermionic baths at different temperatures and chemical potentials is the so-called resonant level model. This is one of the simplest and extremely well-studied open quantum system. Exact results are known from various approaches like NEGF \cite{Jauho_book}. As we will see explicitly later in Sec.\ref{example:resonant_level}, it turns out, for this system, the QME after level-I second Markov approximation, i.e, the analog of Eq.(\ref{QME-2-second_markov_1}), is additive, perfectly of (diagonal) Lindblad form, and yet gives the exact NESS answer. This simple example therefore shows perfectly additive QMEs, even this the strict diagonal Lindblad sense, may be able to give exact results in some cases. This example does not even satisfy the sufficient condition for microscopically derived additive QMEs found in Ref. \cite{Bogna_2018}. 

From above discussion, we see that our equations suggest a non-trivial result: linear fermionic open systems in the wide-band limit are governed by additive QMEs (not in the strict diagonal Lindblad sense) at all strengths of system-bath coupling. Interestingly, the exact QME for linear systems, which can be derived without wide-band limit approximation, is manifestly non-additive \cite{WMZ_2012}. Thus, it seems that, at least for linear systems, additivity of QMEs depends on the validity of our first Markov approximation. This can be done exactly for all coupling strengths for wide-band fermionic baths, which makes the corresponding QME additive. For bosonic baths, it would require a weak system-bath coupling approximation.  
 
With the positivity, accuracy and additivity issues of the Lyapunov equations and the associated QMEs clarified, in the next section, we look at the possibility of having dark states, and also provide perturbative solutions in the regime of very small system-bath coupling.

\section{Dark states, the perturbative solution and thermalization}
\label{sec:dark_states_perturbative_solution_thermalization}

\subsection{Dark states}

Dark states are eigenstates of the system Hamiltonian which are left invariant by presence of the baths (for example, see \cite{Emary_2007, Buca_2012, Keller_2019, Quach_2020}).  Depending on the geometry of the lattice, it may so happen that the eigenfunction of some system mode has nodes at exactly the sites where the baths are attached. In that case, it can be checked by transforming the exact quantum Langevin equation (Eq.\ref{exact_QLE})) to the single-particle eigenbasis that the corresponding mode is completely detached from the baths, and will evolve in the same way as the isolated system. So, if the system was initially prepared in that state, it will remain in that state even in presence of the baths, making it a dark state. This, in turn, means, there is no unique steady state of the system. By transforming the Lyapunov equations to the single-particle eigenbasis, it is easy to show that this property is respected by them. 

\subsection{Perturbative solution: dynamics and NESS}
In the limit of very small system-bath coupling, the  Level-II second Markov approximation is good for single-time correlations, the corresponding Lyapunov equation being Eq.(\ref{Lyapunov_second_markov_2}).
Although this equation can be exactly solved, it gives correct answers only up to the leading order term in $\epsilon$. So, it is useful to find  analytical expressions for $\mathbf{C}(t)$ up to the leading order term in $\epsilon$. In order to do so, we transform Eq.(\ref{Lyapunov_second_markov_2}) to single-particle eigenbasis
\begin{align}
\label{Lyapunov_second_markov_2_eigenbasis}
\frac{d \mathbf{C}^E}{dt}= -(\mathcal{G}^E \mathbf{C}^E + \mathbf{C} {\mathcal{G}^E}^\dagger ) + \epsilon^2 \mathbf{Q}^E_2,
\end{align}
where $\mathcal{G}^E$, $\mathbf{C}^E$ are as defined in Eqs.(\ref{eigenbasis_matrices}), (\ref{eigenbasis_correlations}), and $\mathbf{Q}^E_2=\Phi^\dagger \mathbf{Q}_2 \Phi$ likewise. The explicit expressions for the elements of  
$\mathbf{v}^E$ and $\mathbf{Q}_2^E$ are
\begin{align}
\mathbf{v}^E_{\alpha \nu} &= \frac{1}{2} \left(\mathbf{f}^E_{\alpha \nu}(\omega_\nu) + i {\mathbf{f}^E}^H_{\alpha \nu}(\omega_\nu)\right), \\
\mathbf{Q}^E_{2_{\alpha \nu}} &= \frac{1}{2}\Big( \mathbf{F}_{\alpha \nu}^E (\omega_\nu) - i {\mathbf{F}^E}^H_{\alpha \nu} (\omega_\nu)+ (\alpha \leftrightarrow \nu)^* \Big),
\end{align}
where $(\alpha \leftrightarrow \nu)^*$ notation means that the labels $\alpha$ and $\nu$ are to be interchanged, and the resulting expression is to be complex-conjuagted. The functions $\mathbf{f}^E_{\alpha \nu}(\omega)$  and $\mathbf{F}^E_{\alpha \nu}(\omega)$ are
\begin{align}
&\mathbf{f}^E_{\alpha \nu}(\omega) = \sum_{\ell=1}^N \Phi_{\ell \alpha}^* \Phi_{\ell \nu} \mathfrak{J}_\ell(\omega), \nonumber \\
&\mathbf{F}^E_{\alpha \nu}(\omega) = \sum_{\ell=1}^N \Phi_{\ell \alpha}^* \Phi_{\ell \nu} \mathfrak{J}_\ell(\omega)\mathfrak{n}_{B_\ell}(\omega),
\end{align}
and ${\mathbf{f}^E}^H_{\alpha \nu} (\omega)$ and ${\mathbf{F}^E}^H_{\alpha \nu} (\omega)$ are the corresponding Hilbert transforms. This form of the Lyapunov equation can be used to find perturbative solutions. In the following, we will assume that the NESS is unique, and there is no degeneracy in the system. Results can be easily generalized to cases without these approximations. In particular, we use the condition 
\begin{align}
\label{pert_cond}
|\omega_{\alpha} - \omega_{\nu}| \gg \epsilon^2 \left | \mathbf{v}^E_{\alpha \alpha} + \mathbf{v}^{E*}_{\nu \nu} \right|, ~~~\forall~~ \alpha \neq \nu.
\end{align}
To this end, first we define
\begin{align}
\label{def_w}
w_{\alpha \nu} = i(\omega_{\alpha} - \omega_{\nu}) + \epsilon^2 \left( \mathbf{v}^E_{\alpha \alpha} + \mathbf{v}^{E*}_{\nu \nu} \right).
\end{align}
The perturbative solutions of Eq.(\ref{Lyapunov_second_markov_2_eigenbasis}) up to leading order in $\epsilon$ are given by
\begin{align}
&\mathbf{C}^E_{\alpha \alpha}(t) \simeq \mathbf{C}^E_{\alpha \alpha}(0)e^{-2\epsilon^2\mathbf{f}^E_{\alpha \alpha}(\omega_\alpha)t} \nonumber \\
&+ \frac{\mathbf{F}^E_{\alpha \alpha}(\omega_\alpha)}{\mathbf{f}^E_{\alpha \alpha}(\omega_\alpha)}(1-e^{-2\epsilon^2\mathbf{f}^E_{\alpha \alpha}(\omega_\alpha)t}), \label{pert_sol_occ} \\
&\mathbf{C}^E_{\alpha \nu}(t) \simeq \mathbf{C}^E_{\alpha \nu}(0)e^{-w_{\alpha \nu} t}-\frac{i \epsilon^2 \mathbf{Q}^E_{2_{\alpha \nu}}}{\omega_\alpha - \omega_\nu}(1-e^{-w_{\alpha \nu} t}) \nonumber \\
&+ \frac{i \epsilon^2}{\omega_\alpha - \omega_\nu}\Big[ \mathbf{v}^{E*}_{\nu \alpha} \mathbf{C}^E_{\alpha \alpha}(0)(e^{-2\epsilon^2\mathbf{f}^E_{\alpha \alpha}(\omega_\alpha)t} - e^{-w_{\alpha \nu}t})\nonumber \\
&+\mathbf{v}^E_{\alpha \nu } \mathbf{C}^E_{\nu \nu}(0)(e^{-2\epsilon^2\mathbf{f}^E_{\nu \nu}(\omega_\alpha)t} - e^{-w_{\alpha \nu}t}) \Big] \nonumber \\
& +\frac{i \epsilon^2}{\omega_\alpha - \omega_\nu}\Big[ \mathbf{v}^{E^*}_{\nu \alpha}\frac{\mathbf{F}^E_{\alpha \alpha}(\omega_\alpha)}{\mathbf{f}^E_{\alpha \alpha}(\omega_\alpha)}(1-e^{-2\epsilon^2\mathbf{f}^E_{\alpha \alpha}(\omega_\alpha)t}) \nonumber \\
&+ \mathbf{v}^E_{\alpha \nu }\frac{\mathbf{F}^E_{\nu \nu}(\omega_\nu)}{\mathbf{f}^E_{\nu \nu}(\omega_\nu)}(1-e^{-2\epsilon^2\mathbf{f}^E_{\nu \nu}(\omega_\nu)t})\Big]. \label{pert_sol_off_diag}
\end{align}
Real part of $w_{\alpha \nu}$ is $\mathbf{f}^E_{\alpha \alpha}(\omega_\alpha)+\mathbf{f}^E_{\nu \nu}(\omega_\nu)$, $\mathbf{f}^E_{\alpha \alpha}(\omega_\alpha)>0$ by construction. Thus, for time $t\gg \textrm{max}\{[\epsilon^2 \mathbf{f}^E_{\alpha \alpha}(\omega_\alpha)]^{-1}\}$, the steady state is reached. The perturbative results for the steady state are given by
\begin{align}
& \mathbf{C}^E_{\alpha \alpha}(\infty) = \frac{\mathbf{F}^E_{\alpha \alpha}(\omega_\alpha)}{\mathbf{f}^E_{\alpha \alpha}(\omega_\alpha)} = \frac{\sum_{\ell} |\Phi_{\ell\alpha}|^2\mathfrak{J}_\ell(\omega_\alpha)\mathfrak{n}_{\ell}(\omega_\alpha)}{\sum_{\ell} |\Phi_{\ell\alpha}|^2\mathfrak{J}_\ell(\omega_\alpha)} \label{occ_pert_ss} \\
& \mathbf{C}^E_{\alpha \nu}(\infty) = \frac{i\epsilon^2}{\omega_\alpha - \omega_\nu}\Big[\frac{\mathbf{v}^{E*}_{\nu \alpha}\mathbf{F}^E_{\alpha \alpha}(\omega_\alpha)}{\mathbf{f}^E_{\alpha \alpha}(\omega_\alpha)}\nonumber \\
& \qquad \qquad \qquad \qquad  + \frac{\mathbf{v}^E_{\alpha \nu}\mathbf{F}^E_{\nu \nu}(\omega_\nu)}{\mathbf{f}^E_{\nu \nu}(\omega_\nu)}-\mathbf{Q}^E_{2_{\alpha\nu}} \Big]. \label{off_diag_pert_ss}
\end{align}
We clearly see that the leading order diagonal elements in the single-particle eigenbasis of the system are $O(1)$, while the leading-order off-diagonal elements are $O(\epsilon^2)$, as already mentioned before. 
Given the single-particle eigenvalues and eigenvectors of the system, the equations in this subsection are almost closed-form solutions for dynamics and NESS of our very general set-up under Born-Markov approximation and Eq.(\ref{pert_cond}). Thus, we see that, in this regime, all results for both the transient dynamics and the NESS can be written in terms of the single-particle eigenvalues and eigenvectors of the system. This simplifies the extremely complicated problem of the system in arbitrary dimension connected to an arbitrary number of baths at different temperatures and chemical potentials, to the same level as obtaining the dynamics of the isolated system in absence of the baths. This is a very significant simplification. 
Moreover, as we will see later in Sec.~\ref{example:one_dimension}, if the system is one-dimensional with nearest neighbour hopping, further simplification is possible and  a very general and insightful expression for current at NESS in terms of single-particle eigenvalues and eigenvectors can be obtained.  Below, we use Eqs.(\ref{occ_pert_ss}), (\ref{off_diag_pert_ss}) to discuss thermalization in the arbitrary dimensions and geometry setting.


\subsection{Thermalization}
The Eqs.(\ref{occ_pert_ss}), (\ref{off_diag_pert_ss}) reveal the very important physics of thermalization. In equilibrium all baths have same temperatures and chemical potentials, i.e, the Bose or Fermi distributions of all the baths are exactly the same, $\mathfrak{n}_\ell(\omega)\rightarrow  \mathfrak{n}(\omega)$. Thus, from Eq.(\ref{occ_pert_ss}), $\mathbf{C}^E_{\alpha \alpha}(\infty)=\mathfrak{n}(\omega_\alpha)$. So we get the non-trivial and physically important result that
\begin{align}
\label{thermalization1}
\textrm{in equilibrium},~~~\lim_{\epsilon\rightarrow 0} \left( \lim_{t\rightarrow\infty} \hat{\rho}(t) \right) = \frac{e^{-\beta(\hat{\mathcal{H}}_S -\mu \hat{N}_S)}}{Tr\left(e^{-\beta(\hat{\mathcal{H}}_S -\mu \hat{N}_S)}\right)},
\end{align}
where the order of limits cannot be changed, and $\hat{N}_S=\sum_{\ell=1}^N \hat{c}_\ell^\dagger  \hat{c}_\ell$. Note that the $\epsilon\rightarrow 0$ limit is consistent with Eq.(\ref{pert_cond}). Although Eq.(\ref{occ_pert_ss}) is written for the case where each site is attached to the bath, as mentioned before, it is possible to send an arbitrary number of system-bath couplings to zero to obtain an arbitrary distribution for location of baths. Even if a bath at one site is kept connected, while all others are disconnected, Eq.(\ref{occ_pert_ss}) shows that Eq.(\ref{thermalization1}) holds.  To appreciate the non-triviality of the result, remember that we are working with an extremely general system in arbitrary lattice and geometry with arbitrary number of site attached to baths. What we showed above is that even if a thermal bath is attached to one site of such a system, all the modes attached to it will reach thermal equilibrium, and if the NESS is unique, the system will thermalize in the sense of Eq.(\ref{thermalization1}), irrespective of any further details of the system.

Away from equilibrium, when the temperatures and chemical potentials of the baths are different, there will be non-zero current in NESS. For systems with time-reversal symmetry, $\mathbf{H}$ will be real symmetric and consequently, $\Phi$ can be chosen to be real orthogonal. In this case, the current in NESS depends directly on the imaginary part of the off-diagonal elements $\mathbf{C}^E_{\alpha \nu}$. After some algebra, the imaginary part of $\mathbf{C}^E_{\alpha \nu}$ can be explicitly written as
\begin{align}
\label{imag_off_diag_pert_ss}
&\textrm{Im}\left(\mathbf{C}^E_{\alpha \nu}(\infty)\right) = \frac{\epsilon^2}{\omega_\alpha - \omega_\nu}\Big[ \nonumber \\
& \frac{\sum_{\ell, m} \Phi_{m\alpha}^2 \Phi_{\ell\alpha} \Phi_{\ell\nu} \mathfrak{J}_{\ell}(\omega_\alpha)\mathfrak{J}_{m}(\omega_\alpha)\left( \mathfrak{n}_m(\omega_\alpha)- \mathfrak{n}_\ell(\omega_\alpha)\right)}{\sum_{\ell}^\prime \Phi_{\ell\alpha}^2 \mathfrak{J}_{\ell}(\omega_\alpha)} \nonumber \\
&+(\alpha \leftrightarrow \nu) \Big].
\end{align}
In equilibrium, $\mathfrak{n}_\ell(\omega_\alpha) = \mathfrak{n}_m(\omega_\alpha)=\mathfrak{n}(\omega_\alpha)$, so Im$\left(\mathbf{C}^E_{\alpha \nu}(\infty)\right)=0$, which is consistent with the fact that there is no steady state current in equilibrium for systems with time-reversal symmetry.

Until now, we have discussed only equal time correlations of the system. The understanding in terms of equation of motion gives us a natural way to discuss two-time correlation functions and regression formulas, which we discuss in the next section.


\begin{table*}
\begin{tabular}{|c||c|c|c|}
\hline
& & &\\ 
Approximation level & First Markov,  & Level-I second Markov,  & Level-II second Markov,  \\
& Eqs.(\ref{Lyapunov_first_markov}), (\ref{C_xi}) & Eqs.(\ref{Lyapunov_second_markov_1}),(\ref{Q_1}) & Eqs.(\ref{Lyapunov_second_markov_2}), (\ref{def_Q2})\\
\hline
\hline
\hline
& & & \\
Assumption & $t\gg \tau_{B_1}$   &  $t\gg \tau_{B_1},\tau_{B_2}$   & $t\gg \tau_{B_1},\tau_{B_2}$   \\
& see Eq.(\ref{def_tau_B1}) for $\tau_{B_1}$ & see Eq.(\ref{def_tau_B2}) for $\tau_{B_2}$ & + weak system-bath coupling\\
\hline
& & & \\
Weak system-bath coupling & Holds & Holds & Holds \\
& & &\\
\hline
& & & \\
Strong coupling wide-band limit & Holds and exact & Holds, exact at NESS & Does not hold \\
fermionic case [Eq.(\ref{wide-band-limit})] & & &  \\
& & & \\
\hline
& & & \\
Positivity of Lyapunov equation & Preserved at all times & Preserved at NESS & Violated below accuracy level \\
& & &\\
\hline
& & &\\
 & Eq.(\ref{QME-3-first_markov})   & Eq.(\ref{QME-2-second_markov_1}) & Eq.(\ref{RQME}) \\
Associated QME & time-dependent rates & time-independent rates, & Redfield equation \\
& & may not be completely positive, &  \\
& & but always gives positive NESS &\\
\hline
& & & \\
Additivity & Additive & Additive & Additive \\
(both Lyapunov equation and QME) & & & \\
\hline
& & & \\
Generalized regression formula & Can be derived,   & Can be derived,  & Cannot be derived\\
from equations of motion & Eqs.(\ref{regression_formula_first_Markov}), (\ref{C_xi_two_time}) & Eqs.(\ref{regression_formula_level_I_second_Markov}), (\ref{C_xi_two_time_level_I_second_Markov}) &  \\
\hline
\end{tabular}
\caption{ Summary of the three level of approximations for the three slightly different continuous-time Lyapunov equations. The homogeneous part of the Lyapunov equations, which is associated with the non-Hermitian Hamiltonian governing the dynamics (see Eqs.(\ref{QLE_first_Markov}),  (\ref{def_non_Hermitian_Hamiltonian}), (\ref{def_v}), (\ref{def_G})), is the same in all cases. The approximations change the inhomogeneous part of the Lyapunov equations which embody the  quantum and thermal fluctuations. The level-II second Markov approximation is identical to the Redfield equation and has the same, controlled, accuracy and positivity issues. It can be used to obtain semi-analytical results (Eqs.(\ref{pert_sol_occ}), (\ref{pert_sol_off_diag}), (\ref{occ_pert_ss}), (\ref{off_diag_pert_ss})) in the regime of very small system-bath coupling (Eq.(\ref{pert_cond})), which give open system dynamics in terms of single-particle eigenvalues and eigenvectors of the system.  \label{table1}}  
\end{table*}

\section{Two-time correlations and regression formulas}
\label{sec:regression_formulas}

The matrix $\mathbf{C}(t)$ gives equal time correlations. In this section, we move to calculating two-time correlations
\begin{align}
\mathbf{C}_{\ell m}(t, t^\prime) = \langle \hat{c}_\ell^\dagger (t) \hat{c}_m(t^\prime) \rangle.
\end{align}
These quantities are directly related to NEGF \cite{Jauho_book}. For calculating two-time correlations from QMEs one has to resort to the corresponding quantum regression formulas, which often relies on assumptions with their own set of issues \cite{Talkner_1986,Ford_1996,Giac_2014}. On the other hand, having an operator equation of motion description makes obtaining two-time correlations completely straightforward. 

\subsection{First Markov approximation}
After first Markov approximation, calculating two-time correlation functions from Eq.(\ref{formal_soln_first_markov}) gives (see Appendix~\ref{appendix:first_markov_derivs})
\begin{align}
\label{two_time_correlations_first_markov}
&\mathbf{C}(t,t^\prime) = e^{-\mathcal{G}t} \mathbf{C}(0) e^{-\mathcal{G^\dagger}t^\prime} \nonumber \\
& +\epsilon^2 \int \frac{d\omega}{2\pi} \left( \frac{1-e^{-(\mathcal{G}+i\omega \mathbb{I})t}}{\mathcal{G}+i\omega \mathbb{I}} \mathbf{F}(\omega) \frac{1-e^{-(\mathcal{G}^\dagger-i\omega \mathbb{I})t^\prime}}{\mathcal{G}^\dagger-i\omega \mathbb{I}} \right), 
\end{align}  
which is exactly same as Eq.(\ref{C_formal_soln_first_markov}) except with the two time arguments being different. The regression formula deals with evolution of  $\mathbf{C}(t+\tau, t)$ as a function of $\tau$, $\tau>0$. If we naively used the quantum regression relation on the QMEs \cite{Breuer_book}, we would get,
\begin{align}
\label{regression_from_QME}
\frac{d \mathbf{C}(t+\tau, t)}{d \tau} = -\mathcal{G} \mathbf{C}(t+\tau, t). 
\end{align}
This would be same for all three QMEs Eq.(\ref{QME-3-first_markov}), (\ref{QME-2-second_markov_1}), (\ref{RQME}). However, we can derive the differential equation for evolution of $\mathbf{C}(t+\tau, t)$ as a function of $\tau$ directly from equations of motion, exactly as Eq.(\ref{Lyapunov_first_markov}) was obtained, without further approximations. This gives,
\begin{align}
\label{regression_formula_first_Markov}
\frac{d \mathbf{C}(t+\tau, t)}{d \tau} = -\mathcal{G} \mathbf{C}(t+\tau, t) + i\epsilon^2 \mathbf{C}_{\xi}(t+\tau, t).
\end{align}
Here, $\mathbf{C}_{\xi_{\ell m}}(t+\tau, t)=\langle \hat{\xi}_\ell^\dagger(t+\tau) \hat{c}_m(t) \rangle$, which can be calculated from Eq.(\ref{formal_soln_first_markov}) as
\begin{align}
\label{C_xi_two_time}
\mathbf{C}_\xi(t+\tau, t) = -i \int_0^t dt^\prime \int \frac{d\omega}{2 \pi} \mathbf{F}(\omega)e^{i\omega (t^\prime+\tau)} e^{-\mathcal{G}^\dagger t^\prime}.
\end{align}
Thus, naively using quantum regression at the QMEs would not have given the inhomogeneous part of Eq.(\ref{regression_formula_first_Markov}). 
Such deviations from the quantum regression formula at the level of QMEs has been associated with non-Markovian behavior \cite{Giac_2014}. We call equations of the form of Eq.(\ref{regression_formula_first_Markov}) generalized regression formulas. The Eq.(\ref{regression_formula_first_Markov}) is under the first Markov approximation, and requires $t+\tau \gg \tau_{B_1}$ with $\tau_{B_1}$ defined in Eq.(\ref{def_tau_B1}). So, if $\tau \gg \tau_{B_1}$, only then $t=0$ is allowed, otherwise not. If we put $t=0$, the inhomogeneous part becomes zero. Thus, only in this case the equation becomes same as the one expected by using quantum regression from the QME. 

The formal solution of Eq.(\ref{regression_formula_first_Markov}) can be written as
\begin{align}
&\mathbf{C}(t+\tau, t) = e^{-\mathcal{G}\tau}\mathbf{C}(t) \nonumber \\
& +\epsilon^2 \int \frac{d\omega}{2\pi} \left( \frac{e^{i\omega \tau}-e^{-\mathcal{G}\tau}}{\mathcal{G}+i\omega \mathbb{I}} \mathbf{F}(\omega) \frac{1-e^{-(\mathcal{G}^\dagger-i\omega \mathbb{I})t}}{\mathcal{G}^\dagger-i\omega \mathbb{I}} \right).
\end{align}
If real parts of eigenvalues of $\mathcal{G}$ are positive, i.e, if there is unique NESS, then noting that $\mathbf{C}(\infty)$ is given by Eq.(\ref{NESS_first_markov}), we can write the above equation in the following suggestive form, 
\begin{align}
\label{first_Markov_regression_suggestive_form}
&\mathbf{C}(t+\tau, t) = e^{-\mathcal{G}\tau}\left[\mathbf{C}(t) - \mathbf{C}(\infty)\right] \nonumber \\
&+ \epsilon^2 \int \frac{d\omega}{2\pi} e^{i\omega \tau}\left( \frac{1}{\mathcal{G}+i\omega \mathbb{I}} \mathbf{F}(\omega) \frac{1}{\mathcal{G}^\dagger-i\omega \mathbb{I}} \right) \\
& -\epsilon^2 \int \frac{d\omega}{2\pi} \left( \frac{e^{i\omega \tau}-e^{-\mathcal{G}\tau}}{\mathcal{G}+i\omega \mathbb{I}} \mathbf{F}(\omega) \frac{e^{-(\mathcal{G}^\dagger-i\omega \mathbb{I})t}}{\mathcal{G}^\dagger-i\omega \mathbb{I}} \right).\nonumber
\end{align}
So the contribution from the term that would be obtained via naive application of quantum regression at the level of the QME actually goes to zero at NESS. We get the following result for two-time correlations at NESS,
\begin{align}
\label{long-time_regression}
\lim_{t\rightarrow\infty}\mathbf{C}(t+\tau,t)=\epsilon^2 \int \frac{d\omega}{2\pi} e^{i\omega \tau}\left( \frac{1}{\mathcal{G}+i\omega \mathbb{I}} \mathbf{F}(\omega) \frac{1}{\mathcal{G}^\dagger-i\omega \mathbb{I}} \right).
\end{align}
This is exactly the same as would be obtained by solving Eq.(\ref{QLE_first_Markov}) by a Fourier transform and then obtaining the two-time correlation functions. Clearly, quantum regression at the level of the QME would not give this result.

\subsection{Second Markov approximation}
The level-I second Markov approximation would assume $t+\tau \gg \tau_{B_2}$, with $\tau_{B_2}$ defined in Eq.(\ref{def_tau_B2}),  and take $t\rightarrow \infty$ in Eq.(\ref{C_xi_two_time}). This yields,
\begin{align}
\label{regression_formula_level_I_second_Markov}
\frac{d \mathbf{C}(t+\tau, t)}{d \tau} = -\mathcal{G} \mathbf{C}(t+\tau, t) + i\epsilon^2 \widetilde{\mathbf{C}}_{\xi}(\tau), 
\end{align}
where 
\begin{align}
\label{C_xi_two_time_level_I_second_Markov}
\widetilde{\mathbf{C}}_\xi(\tau) = -i \int \frac{d\omega}{2 \pi} e^{i\omega \tau}\mathbf{F}(\omega)\frac{1}{\mathcal{G}^\dagger-i\omega \mathbb{I}} .
\end{align}
The formal solution at for the above equations is
\begin{align}
\label{second_Markov_regression_suggestive_form}
&\mathbf{C}(t+\tau, t) = e^{-\mathcal{G}\tau}\mathbf{C}(t) \nonumber \\
& +\epsilon^2 \int \frac{d\omega}{2\pi} \left( \frac{e^{i\omega \tau}-e^{-\mathcal{G}\tau}}{\mathcal{G}+i\omega \mathbb{I}} \mathbf{F}(\omega) \frac{1}{\mathcal{G}^\dagger-i\omega \mathbb{I}} \right).
\end{align}
If real parts of eigenvalues of $\mathcal{G}$ are positive, the above expression can be written in the same form as Eq.(\ref{first_Markov_regression_suggestive_form}) except the last line being set to zero. So, in the long time limit, the expression again reduces to Eq.(\ref{long-time_regression}).

A level-II second Markov approximation would further approximate Eq.(\ref{C_xi_two_time_level_I_second_Markov}) and carry out an analogous procedure of Eq.(\ref{C_xi_level_II_second_markov}). However, carrying out this approximation gives inconsistent results for two-time correlations. This is because, $\widetilde{\mathbf{C}}_\xi(\tau)$ typically would decay with $\tau$. Setting $\epsilon=0$ in the expression with finite $\tau$ would not give this behavior and therefore would be inconsistent. As a result, the  level-II second Markov approximation cannot be performed here. So, to obtain two-time correlations corresponding to the Redfield equation, Eq.(\ref{RQME}), one has to resort to the regression relation at the level of QMEs Eq.(\ref{regression_from_QME}), which is likely to impose additional restrictions \cite{Giac_2014}. In particular, we see from Eq.(\ref{second_Markov_regression_suggestive_form}) and Eq.(\ref{NESS_first_markov}) that only if, due to some additional approximations over and above the level-I second Markov approximation,  the following relation approximately holds
\begin{align}
\int \frac{d\omega}{2\pi} e^{i\omega \tau}\left( \frac{1}{\mathcal{G}+i\omega \mathbb{I}} \mathbf{F}(\omega) \frac{1}{\mathcal{G}^\dagger-i\omega \mathbb{I}} \right)\approx e^{-\mathcal{G}\tau}\mathbf{C}(\infty),
\end{align}
will Eq.(\ref{regression_from_QME}) be satisfied at all times. Whether these additional approximations are same as those required for deriving Redfield equation is not clear and requires further work \cite{Bijay_2021}.

It is usually believed that if the NESS is given by a QME of Lindblad form, the quantum regression formula must be valid. However, as we will see below in Sec.~\ref{example:resonant_level}, the simple example of the resonant level model with wide-band baths shows that it is not so. Although its NESS can be obtained from a Lindblad equation for arbitrary strengths of system-bath coupling, the exact two-time correlations at NESS are given by the analog of Eq.(\ref{long-time_regression}), which is different from what is obtained via analog of Eq.(\ref{regression_from_QME}). Only if further approximations, like weak system-bath coupling, or high temperatures, are made, can the two results be reduced to the same form. So, clearly, regression relations require further set of approximations, which can sometimes be different from those required to obtain Lindblad descriptions.

The formulas given in this section allow calculation of  the $\mathbf{C}(t+\tau, t)$, $\tau>0$, knowing $\mathbf{C}(t)$. If $\mathbf{C}(t, t+\tau)$ is desired instead, it can be obtained by simply noting that $\mathbf{C}^\dagger(t+\tau, t)=\mathbf{C}(t, t+\tau)$.

This concludes our main general results. A summary of our results is given in Table~\ref{table1}. In the next section, we explicitly discuss the two examples we have referred to before, viz., the resonant level model, and a one-dimensional system with nearest neighbour hopping.

\section{Insightful examples}
\label{sec:examples}

\subsection{Resonant level model}
\label{example:resonant_level}
\begin{figure}
\includegraphics[width=0.7\columnwidth]{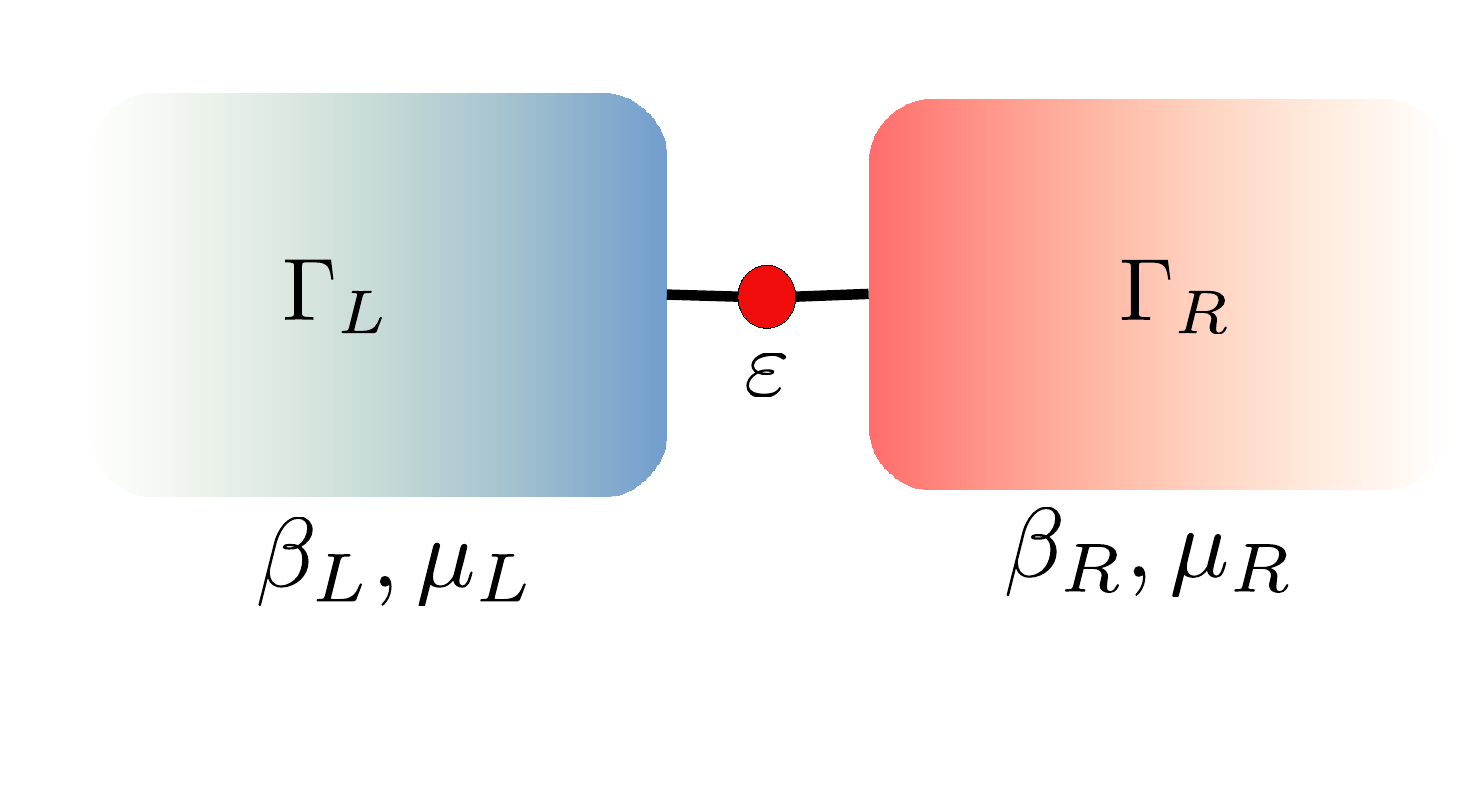}
\caption{Schematic of the resonant level model. \label{fig:resonant_level}}
\end{figure}

In this section, we work out the extremely simple example of a resonant level model, which is a single fermionic site, coupled to two wide-band fermionic reservoirs at different temperatures and chemical potentials, see Fig.~\ref{fig:resonant_level}. The Hamiltonian is given by
\begin{align}
& \hat{\mathcal{H}}_{SB} = \varepsilon \hat{c}^\dagger \hat{c}, \nonumber \\
&\hat{\mathcal{H}}_{SB}=\sum_{\ell=L,R}\hat{\mathcal{H}}_{SB_\ell},~\hat{\mathcal{H}}_{SB_\ell}= \sum_{r=1}^\infty \left(\kappa_{rl}\hat{c}^\dagger \hat{B}_{r\ell} + \kappa_{rl}^*\hat{B}_{r\ell}^\dagger \hat{c}\right) \nonumber \\
& \hat{\mathcal{H}}_{B}=\sum_{\ell=L,R}\hat{\mathcal{H}}_{B_\ell},~ \hat{\mathcal{H}}_{B_\ell}=\sum_{r=1}^\infty \Omega_{rl}\hat{B}_{r\ell}^\dagger \hat{B}_{r\ell}.
\end{align}
In above, $\ell=L,R$ labels the left and the right baths attached to the system site. The system-bath coupling and the baths are of the same form as Eq.(\ref{system_bath_coupling}) except the following differences. Since we will consider the wide-band limit fermionic case, 
\begin{align}
\mathfrak{J}_{\ell}(\omega)=\Gamma_\ell,~~\ell=L,R,
\end{align}
we have set $\epsilon=1$. Further, unlike in previous sections, there are two baths attached to the same single site. As mentioned before, since all the equations are additive this hardly complicates anything. We will specifically focus on the level-I second Markov approximation. For each bath, we have an analog of $\mathbf{v}$ and $\mathbf{Q}_{1}$ matrices, and they are just summed over. 
\begin{align}
\mathbf{v}=\mathbf{v}^{(L)} + \mathbf{v}^{(R)},~~\mathbf{Q}_{1} = \mathbf{Q}_{1}^{(L)} + \mathbf{Q}_{1}^{(R)},  
\end{align}
where  $\mathbf{v}^{(L,R)}$ and $\mathbf{Q}_{1}^{(L,R)}$ are calculated by using Eq.(\ref{wide_band_v}) and Eq.(\ref{Q_1}). For a single site, they are just scalars, $\mathbf{v}^{(L,R)}$ being given by
\begin{align}
\mathbf{v}^{(\ell)}= \frac{\Gamma_{\ell}}{2},~~\ell=L,R,
\end{align}
see Eq.(\ref{wide_band_v}). The non-Hermitian Hamiltonian (Eq.(\ref{def_non_Hermitian_Hamiltonian})) and $\mathcal{G}$ (Eq.(\ref{def_G})) are then
\begin{align}
\mathbf{H}_{\rm NH}=\varepsilon-i\frac{\Gamma_{L}+\Gamma_{R}}{2},~\mathcal{G}=-i\varepsilon+\frac{\Gamma_{L}+\Gamma_{R}}{2}.
\end{align}
Using Eq.(\ref{Q_1}), we get the following expressions for $\mathbf{Q}_{2}^{(L,R)}$,
\begin{align}
\label{resonant_level_Q}
\mathbf{Q}_{1}^{(\ell)} = \int \frac{d\omega}{2 \pi} \frac{(\Gamma_L+\Gamma_R)\Gamma_{\ell}\mathfrak{n}_{B_{\ell}}(\omega)}{(\omega-\varepsilon)^2+\left(\frac{\Gamma_L+\Gamma_R}{2}\right)^2}~~\ell=L,R,
\end{align}
where $\mathfrak{n}_{B_{L}}(\omega)$ ($\mathfrak{n}_{B_{R}}(\omega)$) is the Fermi distribution corresponding to the left (right) bath.
For a single-site, the correlation matrix is also a scalar, $\langle\hat{n}\rangle=\langle\hat{c}^\dagger \hat{c}\rangle $ being the only element. The Lyapunov equation then becomes
\begin{align}
\label{resonant_level_Lyapunov}
\frac{d \langle \hat{n} \rangle}{dt}= - (\Gamma_L+\Gamma_R)\langle \hat{n} \rangle + \mathbf{Q}_{1}^{(L)}+\mathbf{Q}_{1}^{(R)}.
\end{align}
The corresponding QME is
\begin{align}
\label{resonant_level_QME}
\frac{\partial \hat{\rho}}{\partial t}& = i[\hat{\rho}, \varepsilon \hat{n}] + \sum_{\ell=L,R} \Big[  \big( \Gamma_\ell-\mathbf{Q}_{1}^{(\ell)} \big) \left( \hat{c} \rho \hat{c}^\dagger - \frac{1}{2} \{ \hat{c}^\dagger \hat{c}, \hat{\rho} \} \right) \nonumber \\
&+ \mathbf{Q}_{1}^{(\ell)} \left(\hat{c}^\dagger \rho \hat{c} - \frac{1}{2} \{ \hat{c} \hat{c}^\dagger, \hat{\rho} \} \right) \Big]. 
\end{align}  
This is clearly additive and of Lindblad form. It can be checked explicitly that above QME gives Eq.(\ref{resonant_level_Lyapunov}).

The NESS is obtained by setting the left-hand-side of Eq.(\ref{resonant_level_Lyapunov}) to zero, which, gives
\begin{align}
\label{resonant_level_occ_1}
\langle \hat{n} \rangle = \frac{\mathbf{Q}_{1}^{(L)}+\mathbf{Q}_{1}^{(R)}}{\Gamma_L+\Gamma_R}.
\end{align}
This, along with Eq.(\ref{resonant_level_Q}), gives,
\begin{align}
\langle \hat{n} \rangle = \int \frac{d\omega}{2\pi} \frac{\Gamma_L \mathfrak{n}_{B_L}(\omega)+\Gamma_R \mathfrak{n}_{B_R}(\omega)}{(\omega-\varepsilon)^2+\left(\frac{\Gamma_L+\Gamma_R}{2}\right)^2}.
\end{align}
This is the well-known correct expression for occupation at NESS for the resonant level model \cite{Jauho_book}. 

Let us now calculate the current at NESS. The Eq.(\ref{resonant_level_Lyapunov}) is essentially a continuity equation for occupation, and currents from the left and the right baths, $I_L$ and $I_R$, can be identified as
$
I_\ell = -\Gamma_\ell \langle \hat{n} \rangle + \mathbf{Q}_1^{(\ell)},~~\ell=L,R.
$
At NESS, since the left-hand-side of Eq.(\ref{resonant_level_Lyapunov}) is zero, we have $I_L = -I_R = I$.
Using Eq.(\ref{resonant_level_occ_1}), the expression for current $I$ is
\begin{align}
I = \frac{\Gamma_R \mathbf{Q}_1^{(L)}-\Gamma_L \mathbf{Q}_1^{(R)} }{\Gamma_L + \Gamma_R}.
\end{align}
Substituting the expression for $\mathbf{Q}_{1}^{(\ell)}$ from Eq.(\ref{resonant_level_Q}) gives
\begin{align}
I = \int \frac{d\omega}{2\pi} \frac{\Gamma_L \Gamma_R [\mathfrak{n}_{B_L}(\omega)-\mathfrak{n}_{B_R}(\omega)]}{(\omega-\varepsilon)^2+\left(\frac{\Gamma_L+\Gamma_R}{2}\right)^2},
\end{align}
which is the well-known correct expression for current in NESS resonant level model with wide-band baths \cite{Jauho_book}. Note that these results are true irrespective of strength of system-bath coupling. Therefore the perfectly additive Lindblad equation, Eq.(\ref{resonant_level_QME}) gives completely exact results for NESS of the resonant level model (Refer to discussion in Sec.~\ref{subsec:additivity}). Note that, the system-bath coupling Hamiltonians neither commute with each other, nor with the system Hamiltonian. So it does not satisfy the sufficient condition for accurate additive dynamics given in Ref.~\cite{Bogna_2018}. 

It is interesting to note that even though the NESS is given exactly by a QME of perfectly Lindblad form, Eq.(\ref{resonant_level_QME}), the two-time correlations at NESS cannot be obtained by using quantum regression relation at the level of the QME (Refer to discussions in Sec.~\ref{sec:regression_formulas}). For simplicity, here we will assume
$\Gamma_L = \Gamma_R= \Gamma$.
Let us define the notation
\begin{align}
\langle \hat{c}^\dagger(\tau) \hat{c} \rangle = \lim_{t\rightarrow\infty}\langle \hat{c}^\dagger(t+\tau) \hat{c}(t) \rangle.
\end{align}
Using quantum regression at the level of the QME gives
\begin{align}
\label{resonant_level_QME_regression}
&\langle \hat{c}^\dagger(\tau) \hat{c} \rangle = e^{(i\varepsilon-\Gamma)\tau}\langle \hat{n} \rangle \nonumber \\
&= e^{(i\varepsilon-\Gamma)\tau}\int \frac{d\omega}{2\pi} \frac{\Gamma\left[ \mathfrak{n}_{B_L}(\omega)+\mathfrak{n}_{B_R}(\omega)\right]}{(\omega-\varepsilon)^2+\Gamma^2}.
\end{align}
On the other hand, using Eq.(\ref{long-time_regression}) gives 
\begin{align}
\label{resonant_level_regression}
\langle \hat{c}^\dagger(\tau) \hat{c} \rangle = \int \frac{d\omega}{2\pi} e^{i\omega \tau}\frac{\Gamma\left[ \mathfrak{n}_{B_L}(\omega)+\mathfrak{n}_{B_R}(\omega)\right]}{(\omega-\varepsilon)^2+\Gamma^2} ,
\end{align}
which is the correct expression obtained from NEGF \cite{Jauho_book}. Clearly this is not the same as Eq.(\ref{resonant_level_QME_regression}). So, despite the NESS being exactly given by a Lindblad equation, the regression relation does not hold.

The Eq.(\ref{resonant_level_regression}) has the form of Fourier transform of sum of Fermi distributions weighted by a Lorentzian function. This can approximately reduce to Eq.(\ref{resonant_level_QME_regression}) if the Fermi distributions are reasonably flat within the width of the Lorentzian. Since Fermi distributions vary in a scale of $1/\beta$, this can be satisfied if
\begin{align}
\beta_\ell \Gamma \ll 1,~~\ell=L,R.
\end{align}
The above condition can either be satisfied at weak coupling, or at high temperatures. In either case, it will be justified to approximately set $\mathfrak{n}_{L,R}(\omega)\approx \mathfrak{n}_{L,R}(\varepsilon)$ in the integrations in Eqs.(\ref{resonant_level_QME_regression}) and (\ref{resonant_level_regression}). Due to properties of Lorentzian functions, both integrations will now yield
\begin{align}
\langle \hat{c}^\dagger(\tau) \hat{c} \rangle \approx e^{(i\varepsilon-\Gamma)\tau} \frac{\mathfrak{n}_{B_L}(\varepsilon)+\mathfrak{n}_{B_R}(\varepsilon)}{2}.
\end{align}
This same result could be obtained by using quantum regression on the Redfield equation (which will also be of Lindblad form in this case), deriving which will explicitly require weak system-bath coupling approximation. Clearly, validity of quantum regression requires additional approximations.

\subsection{Simple expression for current and dimensionless conductance in one-dimensional nearest neighbour systems}
\label{example:one_dimension}

\begin{figure}
\includegraphics[width=\columnwidth]{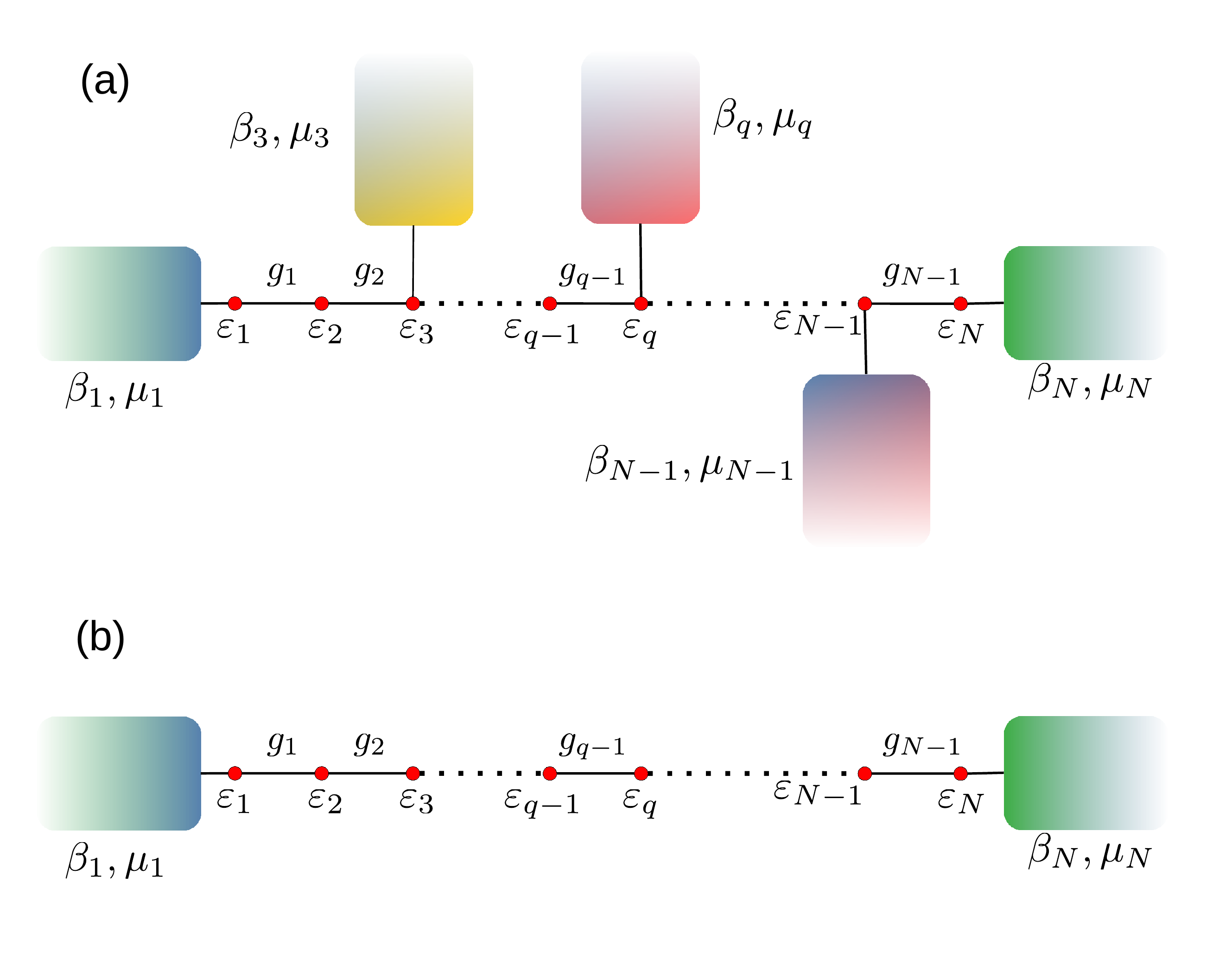}
\caption{(a) General linear one-dimensional nearest neighbour system coupled to multiple thermal baths. (b) The same system but with baths attached at only the two ends.\label{fig:one_dimension_nearest_neighbour}}
\end{figure}

 Here we consider a one-dimensional chain with nearest-neighbour hopping and derive a simple and insightful expression for the particle current using the result in Eq.(\ref{off_diag_pert_ss}). The system Hamiltonian we will be considering now is given by
\begin{align}
\label{1D_nearest_neighbour}
\hat{\mathcal{H}}_S = \sum_{\ell=1}^N \varepsilon_{\ell} \hat{c}_\ell^\dagger\hat{c}_\ell 
+ \sum_{\ell=1}^{N-1} g_\ell \left(\hat{c}_\ell^\dagger\hat{c}_{\ell+1}  +\hat{c}_{\ell+1}^\dagger\hat{c}_\ell \right). 
\end{align}
In other words, $\mathbf{H}$ in Eq.(\ref{general_non_int}) is now the tridiagonal matrix
\begin{align}
\label{tridiag_Hs}
\mathbf{H}=
\begin{bmatrix}
\varepsilon_{1} & g_1 &0 &\ldots &\ldots  \\
g_1 & \varepsilon_{2} & g_2 &0 &\ldots  \\
0 & g_2 & \varepsilon_{3} & g_3 &0  \\
\vdots &\ddots &\ddots &\ddots &\ddots  \\
\ldots  &\ldots  &0 & g_{N-1} & \varepsilon_{N}
\end{bmatrix}.
\end{align} 
This system is coupled at an arbitrary number of sites to different baths which are all at different temperatures and chemical potentials, see Fig.~\ref{fig:one_dimension_nearest_neighbour}(a). We will find an expression for particle current at a system bond in terms of the single-particle eigenvalues and eigenfunctions of the system. The particle current $I_p$ at the $p$th bond of the one-dimensional system is given by
\begin{align}
I_p = g_p \textrm{Im}(\langle \hat{c}_p^\dagger\hat{c}_{p+1} \rangle)
= g_p\sum_{\alpha,\nu=1}^N \Phi_{p \alpha} \Phi_{p \nu} \textrm{Im}(\mathbf{C}^E_{\alpha \nu}),
\end{align}
where ${\rm Im}(A)$ refers to imaginary part of $A$. Using Eq.(\ref{off_diag_pert_ss}) and simplifying utilizing some properties of a  tridiagonal matrix, we obtain the following expression for current at the $p$th bond (see Appendix~\ref{appendix:current_derivation}),
\begin{align}
\label{tridiag_I_2}
&I_p = \epsilon^2 \sum_{\alpha=1}^N\frac{1}{\sum_{\ell}^\prime \Phi_{\ell\alpha}^2 \mathfrak{J}_{\ell}(\omega_\alpha)} \Big[ \nonumber \\ 
&\sum_{\ell,m}^\prime \sum_{k=1}^p  \delta_{\ell k}\Phi_{m\alpha}^2 \Phi_{\ell\alpha}^2 \mathfrak{J}_{\ell}(\omega_\alpha)\mathfrak{J}_{m}(\omega_\alpha)\left[ \mathfrak{n}_\ell(\omega_\alpha)- \mathfrak{n}_m(\omega_\alpha)\right] \Big].
\end{align}
Here $\sum^\prime$ refers to sum over the points where the baths are attached. So, we see that current in the limit of very small system-bath couplings (Eq.(\ref{pert_cond})) is governed by the amplitudes of the single-particle eigenfunctions at the sites where the baths are attached.

Now let us further specialize to the case where the only two baths are attached, which are at the first and last sites of the system, see Fig.~\ref{fig:one_dimension_nearest_neighbour}(b). In this case, the current should be independent of the bond where it is calculated, $I_p=I$. Indeed it is so, as is seen from carrying out the sum over $k$ in Eq.(\ref{tridiag_I_2}), which gives
\begin{align}
\label{tridiag_I_3}
&I =\epsilon^2 \sum_{\alpha=1}^N\frac{\Phi_{1\alpha}^2 \Phi_{N\alpha}^2 \mathfrak{J}_{1}(\omega_\alpha)\mathfrak{J}_{N}(\omega_\alpha)\left[ \mathfrak{n}_1(\omega_\alpha)- \mathfrak{n}_N(\omega_\alpha)\right]}{\Phi_{1\alpha}^2 \mathfrak{J}_{1}(\omega_\alpha)+\Phi_{N\alpha}^2 \mathfrak{J}_{N}(\omega_\alpha)}.
\end{align}

The above expressions for current are valid for both bosonic and fermionic cases. We can now obtain an insightful expression for particle conductance of a fermionic system. Let set-up be fermionic with the two baths having same temperature, $\beta_1=\beta_N=\beta$, but different chemical potentials, $\mu_1=\mu+\Delta \mu, \mu_N=\mu$ and being described by wide-band baths coupled at same strength, $\mathfrak{J}_{1}(\omega_\alpha)=\mathfrak{J}_{N}(\omega_\alpha)=\Gamma$. Then particle conductance is given by,
\begin{align}
G =& \lim_{\Delta\mu_\rightarrow 0} \frac{dI}{d\mu} \nonumber \\
&= \epsilon^2 \beta \Gamma \sum_{\alpha=1}^N  \left[ \frac{\Phi_{1\alpha}^2 \Phi_{N\alpha}^2 }{ \Phi_{1\alpha}^2 +\Phi_{N\alpha}^2}  \mathfrak{n}^2(\omega_\alpha)e^{\beta(\omega_\alpha-\mu)}\right]
\end{align}
Note, here we have already imposed very small system-bath coupling condition before, so we are not in the regime of Eq.(\ref{Lyapunov_second_markov_1}), even though we are using wide-band fermionic baths now. In the high temperature limit, $\mathfrak{n}^2(\omega_\alpha)e^{\beta(\omega_\alpha-\mu)}\simeq1/4$. So particle conductance of the fermionic system in wide-band and high temperature limit is given by
\begin{align}
\label{dimensionless_conductance}
G = \epsilon^2 \frac{\Gamma \beta}{4} W(1,N),~{\rm where}~W(r,s)=\sum_{\alpha=1}^N  \left[ \frac{\Phi_{r\alpha}^2 \Phi_{s\alpha}^2 }{ \Phi_{r\alpha}^2 +\Phi_{s\alpha}^2}\right].
\end{align}
Thus, in this limit, $W(1,N)$ is proportional to conductance. It can be termed a dimensionless conductance. It is interesting to note that $W(1,N)$ depends only on system eigenfunctions and is independent of the baths. It is essentially an isolated system quantity, but to derive Eq.(\ref{dimensionless_conductance}), we required to consider an open system. This expression for particle conductance was used in Ref.~\cite{Archak_AAH_2018}, without proof, to explain the origin of sub-diffusive scaling of conductance with system size at the critical point of the Aubry-Andr{\'e}-Harper model. The results were also checked against exact calculations.

\section{Summary and outlook}
\label{sec:summary_and_outlook}

\noindent
{\it Summary and distinguishing features from previous works ---}
In this work, we have derived the Lyapunov equation for describing the dynamics of number conserving linear systems (quadratic Hamiltonians) in a lattice of arbitrary dimension and geometry, coupled to an arbitrary number of macroscopic thermal baths which can all be at different temperatures and chemical potentials. Three slightly different forms of the Lyapunov equation are derived. Table~\ref{table1} gives the summary of our results. The following points distinguish our work from previous works involving Lyapunov equations.  We have given a detailed, systematic, controlled derivation starting from a fully microscopic Hamiltonian (Hermitian) model of the system coupled with the baths. Unlike most previous works \cite{Prosen_2008, Prosen_2010, Prosen_2012, Koga_2012, Nicacio_2012, Landi_2013, Nicacio_2016, Landi_2017, Landi_2019, Mcdonald_2021, Bernal_Garcia_2021}),  we have done so without referring to any QME. Exact approaches require inverse Laplace transformation \cite{WMZ_2008,Jin_2010,WMZ_2012,Martinez_2013,WMZ_2013}, inverting which can become challenging depending on number of sites in the system and the spectrum of the system. The Lyapunov equations greatly simplify the problem of obtaining dynamics of such open quantum systems via bypassing the need for Laplace transform.  In particular, for fermionic reservoirs in the so-called wide-band limit, our results do not require a weak system-bath coupling approximation. Our microscopic derivation makes the validity regimes of the Lyapunov equations clear.  We have then found the associated QMEs which yields the corresponding Lyapunov equations. For Gaussian initial states of the system, two of the associated QMEs also allow us to resolve the positivity problem of the Redfield equation. These two QMEs would have been difficult to obtain otherwise. On the other hand, the third QME is the Redfield equation, which shows the equivalence of QME and equation of motion approaches. In the limit of very small coupling to the baths, we have found semi-analytical results for the system which are written only in terms of the single-particle eigenvalues and eigenvectors of the system. This therefore reduces the problem of obtaining open system dynamics to the same difficulty level as solving the isolated system dynamics. Finally, we have given the generalized regression formulas from the correlation matrix, which allows calculation of all two-time correlation functions for a Gaussian system. These formulas would not be possible to derive by naively applying quantum regression at the level of the QMEs. Finally, we have worked out two insightful examples which highlight several features of our results.   
\\

\noindent
{\it Implications for non-Hermitian quantum physics ---}
One of most immediate implications of our work is that our microscopic derivation explicitly separates the non-Hermitian Hamiltonian governing the dynamics of the system from the quantum and thermal fluctuations due to the presence of the baths. Therefore, it shows how dynamics governed by effective non-Hermitian Hamiltonians can microscopically arise out of Hermitian quantum mechanics, and gives an unified way to treat quantum and thermal fluctuations on such systems. Since our formulation is very general, all kinds of non-Hermitian Hamiltonians can be microscopically designed in this way, as long as there are no sources of gain.  The homogeneous part of the Lyapunov equation is associated with the non-Hermitian Hamiltonian, while the inhomogeneous part of the Lyapunov equation is associated with the quantum and thermal fluctuations. Interestingly, the three levels of approximation only change the inhomogeneous part. They therefore give various accuracy levels of treating the quantum and thermal fluctuations. While Lyapunov equations from Lindblad QMEs have been used to explore non-Hermitian physics in few works  \cite{Mcdonald_2021,Arkhipov_2021,Roccati_2021}, our formulation goes beyond the validity regimes of such Lindblad QMEs deriving which usually require  further approximations over the Redfield equation \cite{Breuer_book,Rivas_book,Archak_2021}.

Sources of gain typically require a non-linear coupling with bath, which is beyond the scope of the present paper. However, Lyapunov equations can be obtained from microscopic considerations, completely out of Hermitian quantum mechanics,  in such cases also, see \cite{Archak_2020_PT}. But, such Lyapunov equations, which give an effective linearized description, may be unstable and may not be valid up to long times. Nevertheless, at least up to some finite time (which can be estimated), our present work, in combination with \cite{Archak_2020_PT}, suggests that systems described by all kinds of effective non-Hermitian Hamiltonians can be obtained from standard quantum mechanics, as long as quantum and thermal fluctuations are properly accounted for via a Lyapunov equation.  However, in presence of gain, the non-linear coupling can make the state non-Gaussian. So, the Lyapunov equation in presence of gain may not describe the full state of the system. 
\\

\noindent
{\it Possible implications for dissipative quantum many-body systems ---}
Systems governed by quadratic Hamiltonians form the starting point for much of our understanding of physics, especially, in higher than one dimension. Much of analytical techniques in physics deal with formulating sophisticated methods to obtain corrections above such quadratic descriptions \cite{Mahan_book, Jauho_book, Kamenev_book}. These techniques may be combined with our general Lyapunov equations and regression formulas to treat many-body interactions (i.e higher than quadratic terms). The simplest of these is the Hartree-Fock mean field approximation, which is very often made for realistic three-dimensional systems. At such mean-field level, the dynamics and NESS can be obtained by solving the Lyapunov equation self-consistently. The Lyapunov equations may therefore be used as a natural and simple way to include dissipation into the mean-field description of realistic quantum many-body systems. This also gives a microscopic meaning, consistent with standard quantum mechanics, to band structures of non-Hermitian Hamiltonians, which have been explored in various works (for example, \cite{Kawabata_2019,Gong_2018, Wang2021}). 
\\

\noindent  
{\it Relevance to quantum thermodynamics of Gaussian systems ---}
Lyapunov equations have been recently used to introduce and describe concepts related to quantum thermodynamics in Gaussian systems like Wigner entropy production \cite{Landi_2017,Landi_2019}, and non-linear Onsager relations \cite{Landi_2020}. Once again, this was done from the viewpoint of Markovian Lindblad equations. Our results therefore suggest possible generalizations of those results beyond their present validity regimes. 
\\

\noindent
{\it Further generalizations and future works ---}
Here we have considered a number conserving quadratic Hamiltonian. But most of our results can be readily generalized to the case of number non-conserving systems, such as superconductors and squeezed harmonic oscillators, using a slightly different definition for the correlation matrix \cite{Ribeiro_2015}. They may also be possible generalize to the case of athermal Gaussian baths, instead of thermal baths. Another interesting direction is generalization of the Lyapunov equation to the case where the temperatures and the chemical potentials of the baths are time-dependent. At the level of the Redfield equation and the associated Lyapunov equation, this has already been achieved \cite{Archak_2017}. It will be interesting to see if a generalization beyond the validity regime of Redfield equation would be possible. Moreover, the entire rich mathematical understanding of Lyapunov equations \cite{Control_theory_book1,Control_theory_book2} can now be carried over to open quantum systems and non-Hermitian physics. Investigations in these directions, as well as in the direction of quantum thermodynamics, will be taken up in future works. Overall, we find it fascinating that an equation used in daily life for control of macroscopic objects \cite{Control_theory_book1} can be shown to play such a fundamental role in describing physics of microscopic quantum systems.
\\

\noindent
{\it Acknowledgements ---}
The author acknowledges funding from the European Union's Horizon 2020 research and innovation programme under the Marie Sklodowska-Curie Grant Agreement No. 890884. The author also acknowledges funding from the Centre for complex quantum systems, Aarhus University, Denmark.

\appendix
\section*{Appendix}
\section{Derivation of Eq.(\ref{C_formal_soln_first_markov}) from Eq.(\ref{formal_soln_first_markov})}
\label{appendix:first_markov_derivs}

In order to derive Eq.(\ref{C_formal_soln_first_markov}) from Eq.(\ref{formal_soln_first_markov}), first we take the transpose of Eq.(\ref{formal_soln_first_markov}),
\begin{align}
\label{formal_soln_first_markov_transpose}
c_{vec}^T(t) =  c_{vec}^T(0) e^{-i\mathbf{H}_{\rm NH}^T t} + i\epsilon \int_0^t dt^\prime \xi_{vec}^T(t^\prime) e^{-i\mathbf{H}_{\rm NH}^T (t-t^\prime)} .
\end{align}
Note that with the transpose, the column vectors $c_{vec}(t)$ and $\xi_{vec}(t)$ with elements $\{\hat{c}_\ell(t)\}$ and $\{\hat{\xi}_\ell(t)\}$ have now become row vectors. Next, we take the Hermitian conjugate of the above equation. This gives,
\begin{align}
\label{formal_soln_first_markov_dagger}
c^\dagger_{vec}(t) = e^{i\mathbf{H}_{\rm NH}^* t} c_{vec}^\dagger(0)  - i\epsilon \int_0^t dt^\prime e^{i\mathbf{H}_{\rm NH}^* (t-t^\prime)}\xi_{vec}^\dagger(t^\prime),
\end{align}
where $c^\dagger_{vec}(t)$ and $\xi_{vec}^\dagger(t)$ are column vectors with elements $\{\hat{c}_\ell^\dagger(t)\}$ and $\{\hat{\xi}_\ell^\dagger(t)\}$. Multiplying Eqs.(\ref{formal_soln_first_markov_dagger}) and (\ref{formal_soln_first_markov_transpose}) and taking expectation values with respect to the initial state Eq.(\ref{initial_state}), we get
\begin{align}
&\mathbf{C}(t)= e^{-\mathcal{G}t} \mathbf{C}(0) e^{-\mathcal{G^\dagger}t} \nonumber \\
& +\epsilon^2 \int_0^t dt_1 \int_0^t dt_2 e^{-i(t-t_1)\mathcal{G} }\langle\xi_{vec}^\dagger(t_1)\xi_{vec}^T(t_2)\rangle e^{-i(t-t_2)\mathcal{G}^\dagger },
\end{align}
where $\mathcal{G}=-i\mathbf{H}_{\rm NH}^*$, as defined in Eq.(\ref{def_G}), and $\langle\xi_{vec}^\dagger(t_1)\xi_{vec}^T(t_2)\rangle$ is a matrix with elements $\langle \hat{\xi}_{\ell}^\dagger(t_1)\hat{\xi}_{m}(t_2)\rangle$. Using the expression for $\langle \hat{\xi}_{\ell}^\dagger(t_1)\hat{\xi}_{m}(t_2)\rangle$ from Eqs.(\ref{noise_corrs}) and (\ref{def_F}), and carrying out the integrations over time, we obtain Eq.(\ref{C_formal_soln_first_markov}).

Note that Eq.(\ref{two_time_correlations_first_markov}) can also be obtained  as above, simply by making the time arguments in Eqs.(\ref{formal_soln_first_markov_transpose}) and (\ref{formal_soln_first_markov_dagger}) different.

\section{Some explicit proofs}
\label{appendix:some_proofs}

In the main text, it was mentioned that Eq.(\ref{C_formal_soln_first_markov}) is the solution for Eq.(\ref{Lyapunov_first_markov}). Further, it was mentioned that although Eq.(\ref{C_formal_soln_first_markov}) is not the solution of Eq.(\ref{Lyapunov_second_markov_1}), the long-time solution of Eq.(\ref{Lyapunov_second_markov_1}) is given by Eq.(\ref{NESS_first_markov}), and is hence positive semi-definite. Here we explicitly show the steps for proving these statements.

\begin{figure}[b!]
\includegraphics[width=\columnwidth]{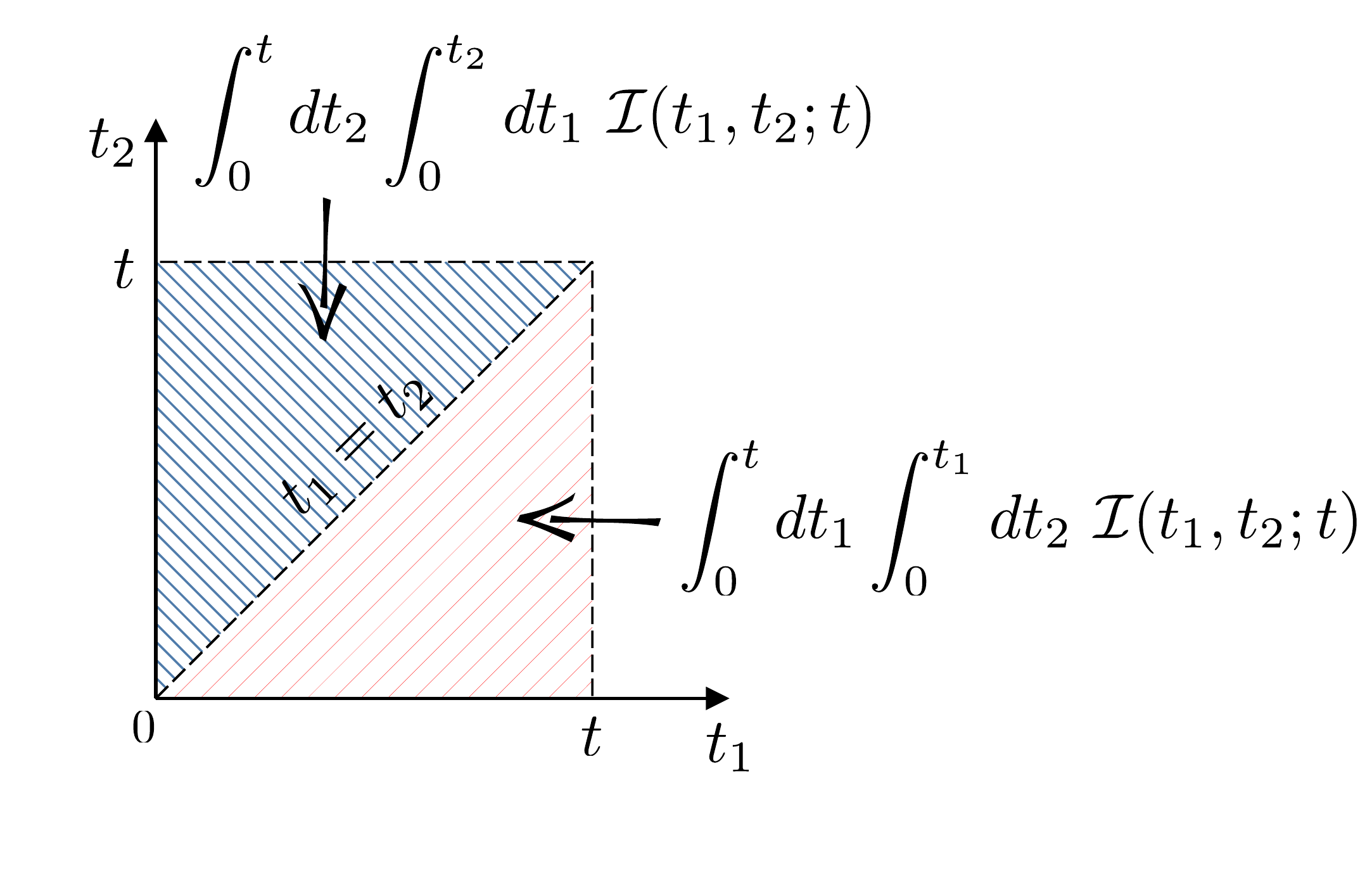}
\caption{The domain of the integrations in Eq.(\ref{R_simplified}). \label{fig:for_integration}}
\end{figure}

The formal solution of Eq.(\ref{Lyapunov_first_markov}) is
\begin{align}
\mathbf{C}(t) = e^{-\mathcal{G}t} \mathbf{C}(0) e^{-\mathcal{G}^\dagger t} + \epsilon^2\mathbf{R}(t),
\end{align}
where 
\begin{align}
\mathbf{R}(t)=\int_0^t dt^\prime e^{-(t-t^\prime)\mathcal{G}} \mathbf{Q}(t^\prime) e^{-(t-t^\prime)\mathcal{G}^\dagger }.
\end{align}
Then, we have to show that $\mathbf{R}(t)$ simplifies to the second term in Eq.(\ref{C_formal_soln_first_markov}), when $\mathbf{Q}(t)$ is as defined in Eqs.(\ref{Lyapunov_first_markov}), (\ref{C_xi}). To show this, first, we use the expression for $\mathbf{Q}(t)$ in $\mathbf{R}$, and after some algebra, it can be written in the form
\begin{align}
\label{R_simplified}
\mathbf{R}(t)=& \int_0^t dt_1 \int_0^{t_1} dt_2~\mathcal{I}(t_1,t_2;t) \nonumber \\
&+ \int_0^t dt_2 \int_0^{t_2} dt_1~\mathcal{I}(t_1,t_2;t),
\end{align}
with
\begin{align}
\mathcal{I}(t_1,t_2;t) = \int \frac{d\omega}{2\pi}e^{-(t-t_1)\mathcal{G}} e^{i(t_1-t_2)\omega} \mathbf{F}(\omega) e^{-(t-t_1)\mathcal{G}^\dagger}.
\end{align}
At this point, visualizing the domain of the integrations, shown in Fig.(\ref{fig:for_integration}) becomes useful. Since the two terms in Eq.(\ref{R_simplified}) correspond to exactly the same function getting integrated over the lower triangular and the upper triangular parts of square shown in Fig.(\ref{fig:for_integration}), their sum exactly corresponds to integration of the function over the entire square. Thus, we have
\begin{align}
\mathbf{R}(t)=\int_0^t dt_1 \int_0^{t} dt_2~\mathcal{I}(t_1,t_2;t).
\end{align} 
Now, explicitly carrying out the integrations over time gives exactly second term in Eq.(\ref{C_formal_soln_first_markov}). This therefore proves that Eq.(\ref{C_formal_soln_first_markov}) is indeed the solution for Eq.(\ref{Lyapunov_first_markov}).

The formal solution of Eq.(\ref{Lyapunov_second_markov_1}) is
\begin{align}
\mathbf{C}(t) = e^{-\mathcal{G}t} \mathbf{C}(0) e^{-\mathcal{G}^\dagger t} + \epsilon^2\mathbf{R}_1(t),
\end{align}
where
\begin{align}
\mathbf{R}_1(t)=\int_0^t dt^\prime e^{-\mathcal{G}t^\prime} \mathbf{Q}_1 e^{-\mathcal{G}^\dagger t^\prime}.
\end{align}
Noting that $\mathbf{Q}_1$ (Eq.(\ref{Q_1})) is nothing but $\mathbf{Q}(\infty)$, similar algebraic simplifications as done to obtain Eq.(\ref{R_simplified}) gives
\begin{align}
\mathbf{R}_1(t) = & \int_0^t dt_1 \int_0^{\infty} dt_2~\mathcal{I}(t_1,t_2;t) \nonumber \\
&+ \int_0^t dt_2 \int_0^{\infty} dt_1~\mathcal{I}(t_1,t_2;t).
\end{align}
Clearly, using the geometric visualization of domain to combine the two terms in the sum into one integration is no longer possible for finite $t$. But for $t\rightarrow \infty$, this can be done and it gives exactly same result as $\mathbf{R}(\infty)$. By carrying out the time integration, it can be checked that the result is exactly Eq.(\ref{NESS_first_markov}), which is manifestly positive semi-definite.

\section{Derivation of Eq.(\ref{tridiag_I_2})}
\label{appendix:current_derivation}

In this section, we give the derivation of Eq.(\ref{tridiag_I_2}).
In going to the single-particle eigenbasis we have to diagonalise $\mathbf{H}$ as in Eq.(\ref{diag_Hs}). Writing out Eq.(\ref{diag_Hs}) explicitly for Eq.(\ref{tridiag_Hs}), we have the following set of equations
\begin{align}
\label{tridiag_Phi}
&(\omega_\alpha - \varepsilon_{\ell}) \Phi_{\ell \alpha} = g_\ell \Phi_{\ell+1 \alpha} + g_{\ell-1} \Phi_{\ell-1 \alpha}~~~\forall~~~\ell \neq 1,N; \nonumber \\
&(\omega_\alpha - \varepsilon_{1}) \Phi_{1 \alpha}  = g_1 \Phi_{2 \alpha}, ~~
(\omega_\alpha - \varepsilon_{N}) \Phi_{N \alpha}  = g_{N-1} \Phi_{N-1 \alpha}.
\end{align}
Using Eq.(\ref{off_diag_pert_ss}), and simplifying, the expression for NESS current becomes,
\begin{align}
\label{tridiag_I_1}
&I = \epsilon^2 g_p  \sum_{\alpha,\nu=1}^N \frac{\Phi_{p \alpha}\Phi_{p+1 \nu} -\Phi_{p \nu}\Phi_{p+1 \alpha} }{\omega_\alpha -\omega_\nu}\Big[ \nonumber \\
& \frac{\sum_{\ell, m}^\prime \Phi_{m\alpha}^2 \Phi_{\ell\alpha} \Phi_{\ell\nu} \mathfrak{J}_{\ell}(\omega_\alpha)\mathfrak{J}_{m}(\omega_\alpha)\left[ \mathfrak{n}_m(\omega_\alpha)- \mathfrak{n}_\ell(\omega_\alpha)\right]}{\sum_{\ell}^\prime \Phi_{\ell\alpha}^2 \mathfrak{J}_{\ell}(\omega_\alpha)} \Big],
\end{align}
where $\sum^\prime$ refers to sum over the points where the baths are attached.
Remember that since we have used Eq.(\ref{imag_off_diag_pert_ss}), above expression for current is only valid when Eq.(\ref{pert_cond}) is satisfied.
This expression can be further simplified using an interesting result which we find for tridiagonal matrices using Eq.(\ref{tridiag_Phi}),
\begin{align}
\label{tridiag_Phi_cond}
\Phi_{p \alpha}\Phi_{p+1 \nu} -\Phi_{p\nu}\Phi_{p+1 \alpha} = \frac{\omega_\nu-\omega_\alpha}{g_p}\sum_{k=1}^p \Phi_{k \alpha}\Phi_{k \nu}.
\end{align}
Also, the eigenvectors are orthogonal, so
\begin{align}
\label{Phi_orthogonal}
\sum_{\nu=1}^N \Phi_{\ell \nu} \Phi_{m \nu}= \delta_{\ell m}.
\end{align}
Using Eqs.(\ref{tridiag_Phi_cond}), (\ref{Phi_orthogonal}) in Eq.(\ref{tridiag_I_1}), we obtain Eq.(\ref{tridiag_I_2}).

\bibliography{ref_lyapunov}

\begin{thebibliography}{123}%
\makeatletter
\providecommand \@ifxundefined [1]{%
 \@ifx{#1\undefined}
}%
\providecommand \@ifnum [1]{%
 \ifnum #1\expandafter \@firstoftwo
 \else \expandafter \@secondoftwo
 \fi
}%
\providecommand \@ifx [1]{%
 \ifx #1\expandafter \@firstoftwo
 \else \expandafter \@secondoftwo
 \fi
}%
\providecommand \natexlab [1]{#1}%
\providecommand \enquote  [1]{``#1''}%
\providecommand \bibnamefont  [1]{#1}%
\providecommand \bibfnamefont [1]{#1}%
\providecommand \citenamefont [1]{#1}%
\providecommand \href@noop [0]{\@secondoftwo}%
\providecommand \href [0]{\begingroup \@sanitize@url \@href}%
\providecommand \@href[1]{\@@startlink{#1}\@@href}%
\providecommand \@@href[1]{\endgroup#1\@@endlink}%
\providecommand \@sanitize@url [0]{\catcode `\\12\catcode `\$12\catcode
  `\&12\catcode `\#12\catcode `\^12\catcode `\_12\catcode `\%12\relax}%
\providecommand \@@startlink[1]{}%
\providecommand \@@endlink[0]{}%
\providecommand \url  [0]{\begingroup\@sanitize@url \@url }%
\providecommand \@url [1]{\endgroup\@href {#1}{\urlprefix }}%
\providecommand \urlprefix  [0]{URL }%
\providecommand \Eprint [0]{\href }%
\providecommand \doibase [0]{http://dx.doi.org/}%
\providecommand \selectlanguage [0]{\@gobble}%
\providecommand \bibinfo  [0]{\@secondoftwo}%
\providecommand \bibfield  [0]{\@secondoftwo}%
\providecommand \translation [1]{[#1]}%
\providecommand \BibitemOpen [0]{}%
\providecommand \bibitemStop [0]{}%
\providecommand \bibitemNoStop [0]{.\EOS\space}%
\providecommand \EOS [0]{\spacefactor3000\relax}%
\providecommand \BibitemShut  [1]{\csname bibitem#1\endcsname}%
\let\auto@bib@innerbib\@empty
\bibitem [{\citenamefont {Hur}\ \emph {et~al.}(2016)\citenamefont {Hur},
  \citenamefont {Henriet}, \citenamefont {Petrescu}, \citenamefont {Plekhanov},
  \citenamefont {Roux},\ and\ \citenamefont {Schir{\'{o}}}}]{LeHur2016}%
  \BibitemOpen
  \bibfield  {author} {\bibinfo {author} {\bibfnamefont {K.~L.}\ \bibnamefont
  {Hur}}, \bibinfo {author} {\bibfnamefont {L.}~\bibnamefont {Henriet}},
  \bibinfo {author} {\bibfnamefont {A.}~\bibnamefont {Petrescu}}, \bibinfo
  {author} {\bibfnamefont {K.}~\bibnamefont {Plekhanov}}, \bibinfo {author}
  {\bibfnamefont {G.}~\bibnamefont {Roux}}, \ and\ \bibinfo {author}
  {\bibfnamefont {M.}~\bibnamefont {Schir{\'{o}}}},\ }\href {\doibase
  10.1016/j.crhy.2016.05.003} {\bibfield  {journal} {\bibinfo  {journal}
  {Comptes Rendus Physique}\ }\textbf {\bibinfo {volume} {17}},\ \bibinfo
  {pages} {808} (\bibinfo {year} {2016})}\BibitemShut {NoStop}%
\bibitem [{\citenamefont {Binder}\ \emph {et~al.}(2018)\citenamefont {Binder},
  \citenamefont {Correa}, \citenamefont {Gogolin}, \citenamefont {Anders},\
  and\ \citenamefont {Adesso}}]{quantum_thermodynamics}%
  \BibitemOpen
  \bibinfo {editor} {\bibfnamefont {F.}~\bibnamefont {Binder}}, \bibinfo
  {editor} {\bibfnamefont {L.~A.}\ \bibnamefont {Correa}}, \bibinfo {editor}
  {\bibfnamefont {C.}~\bibnamefont {Gogolin}}, \bibinfo {editor} {\bibfnamefont
  {J.}~\bibnamefont {Anders}}, \ and\ \bibinfo {editor} {\bibfnamefont
  {G.}~\bibnamefont {Adesso}},\ eds.,\ \href@noop {} {\emph {\bibinfo {title}
  {Thermodynamics in the Quantum Regime}}}\ (\bibinfo  {publisher} {Springer,
  Cham},\ \bibinfo {year} {2018})\BibitemShut {NoStop}%
\bibitem [{\citenamefont {Cao}\ \emph {et~al.}(2019)\citenamefont {Cao},
  \citenamefont {Romero}, \citenamefont {Olson}, \citenamefont {Degroote},
  \citenamefont {Johnson}, \citenamefont {Kieferová}, \citenamefont
  {Kivlichan}, \citenamefont {Menke}, \citenamefont {Peropadre}, \citenamefont
  {Sawaya},\ and\ \citenamefont {et~al.}}]{quantum_chemistry}%
  \BibitemOpen
  \bibfield  {author} {\bibinfo {author} {\bibfnamefont {Y.}~\bibnamefont
  {Cao}}, \bibinfo {author} {\bibfnamefont {J.}~\bibnamefont {Romero}},
  \bibinfo {author} {\bibfnamefont {J.~P.}\ \bibnamefont {Olson}}, \bibinfo
  {author} {\bibfnamefont {M.}~\bibnamefont {Degroote}}, \bibinfo {author}
  {\bibfnamefont {P.~D.}\ \bibnamefont {Johnson}}, \bibinfo {author}
  {\bibfnamefont {M.}~\bibnamefont {Kieferová}}, \bibinfo {author}
  {\bibfnamefont {I.~D.}\ \bibnamefont {Kivlichan}}, \bibinfo {author}
  {\bibfnamefont {T.}~\bibnamefont {Menke}}, \bibinfo {author} {\bibfnamefont
  {B.}~\bibnamefont {Peropadre}}, \bibinfo {author} {\bibfnamefont {N.~P.~D.}\
  \bibnamefont {Sawaya}}, \ and\ \bibinfo {author} {\bibnamefont {et~al.}},\
  }\href {http://dx.doi.org/10.1021/acs.chemrev.8b00803} {\bibfield  {journal}
  {\bibinfo  {journal} {Chemical Reviews}\ }\textbf {\bibinfo {volume} {119}},\
  \bibinfo {pages} {10856–10915} (\bibinfo {year} {2019})}\BibitemShut
  {NoStop}%
\bibitem [{\citenamefont {Cao}\ \emph {et~al.}(2020)\citenamefont {Cao},
  \citenamefont {Cogdell}, \citenamefont {Coker}, \citenamefont {Duan},
  \citenamefont {Hauer}, \citenamefont {Kleinekath{\"o}fer}, \citenamefont
  {Jansen}, \citenamefont {Man{\v c}al}, \citenamefont {Miller}, \citenamefont
  {Ogilvie}, \citenamefont {Prokhorenko}, \citenamefont {Renger}, \citenamefont
  {Tan}, \citenamefont {Tempelaar}, \citenamefont {Thorwart}, \citenamefont
  {Thyrhaug}, \citenamefont {Westenhoff},\ and\ \citenamefont
  {Zigmantas}}]{quantum_biology}%
  \BibitemOpen
  \bibfield  {author} {\bibinfo {author} {\bibfnamefont {J.}~\bibnamefont
  {Cao}}, \bibinfo {author} {\bibfnamefont {R.~J.}\ \bibnamefont {Cogdell}},
  \bibinfo {author} {\bibfnamefont {D.~F.}\ \bibnamefont {Coker}}, \bibinfo
  {author} {\bibfnamefont {H.-G.}\ \bibnamefont {Duan}}, \bibinfo {author}
  {\bibfnamefont {J.}~\bibnamefont {Hauer}}, \bibinfo {author} {\bibfnamefont
  {U.}~\bibnamefont {Kleinekath{\"o}fer}}, \bibinfo {author} {\bibfnamefont
  {T.~L.~C.}\ \bibnamefont {Jansen}}, \bibinfo {author} {\bibfnamefont
  {T.}~\bibnamefont {Man{\v c}al}}, \bibinfo {author} {\bibfnamefont
  {R.~J.~D.}\ \bibnamefont {Miller}}, \bibinfo {author} {\bibfnamefont {J.~P.}\
  \bibnamefont {Ogilvie}}, \bibinfo {author} {\bibfnamefont {V.~I.}\
  \bibnamefont {Prokhorenko}}, \bibinfo {author} {\bibfnamefont
  {T.}~\bibnamefont {Renger}}, \bibinfo {author} {\bibfnamefont {H.-S.}\
  \bibnamefont {Tan}}, \bibinfo {author} {\bibfnamefont {R.}~\bibnamefont
  {Tempelaar}}, \bibinfo {author} {\bibfnamefont {M.}~\bibnamefont {Thorwart}},
  \bibinfo {author} {\bibfnamefont {E.}~\bibnamefont {Thyrhaug}}, \bibinfo
  {author} {\bibfnamefont {S.}~\bibnamefont {Westenhoff}}, \ and\ \bibinfo
  {author} {\bibfnamefont {D.}~\bibnamefont {Zigmantas}},\ }\href
  {https://advances.sciencemag.org/content/6/14/eaaz4888} {\bibfield  {journal}
  {\bibinfo  {journal} {Science Advances}\ }\textbf {\bibinfo {volume} {6}}
  (\bibinfo {year} {2020})}\BibitemShut {NoStop}%
\bibitem [{\citenamefont {Awschalom}\ \emph {et~al.}(2021)\citenamefont
  {Awschalom}, \citenamefont {Du}, \citenamefont {He}, \citenamefont
  {Heremans}, \citenamefont {Hoffmann}, \citenamefont {Hou}, \citenamefont
  {Kurebayashi}, \citenamefont {Li}, \citenamefont {Liu}, \citenamefont
  {Novosad}, \citenamefont {Sklenar}, \citenamefont {Sullivan}, \citenamefont
  {Sun}, \citenamefont {Tang}, \citenamefont {Tyberkevych}, \citenamefont
  {Trevillian}, \citenamefont {Tsen}, \citenamefont {Weiss}, \citenamefont
  {Zhang}, \citenamefont {Zhang}, \citenamefont {Zhao},\ and\ \citenamefont
  {Zollitsch}}]{quantum_engineering}%
  \BibitemOpen
  \bibfield  {author} {\bibinfo {author} {\bibfnamefont {D.~D.}\ \bibnamefont
  {Awschalom}}, \bibinfo {author} {\bibfnamefont {C.~H.~R.}\ \bibnamefont
  {Du}}, \bibinfo {author} {\bibfnamefont {R.}~\bibnamefont {He}}, \bibinfo
  {author} {\bibfnamefont {J.}~\bibnamefont {Heremans}}, \bibinfo {author}
  {\bibfnamefont {A.}~\bibnamefont {Hoffmann}}, \bibinfo {author}
  {\bibfnamefont {J.}~\bibnamefont {Hou}}, \bibinfo {author} {\bibfnamefont
  {H.}~\bibnamefont {Kurebayashi}}, \bibinfo {author} {\bibfnamefont
  {Y.}~\bibnamefont {Li}}, \bibinfo {author} {\bibfnamefont {L.}~\bibnamefont
  {Liu}}, \bibinfo {author} {\bibfnamefont {V.}~\bibnamefont {Novosad}},
  \bibinfo {author} {\bibfnamefont {J.}~\bibnamefont {Sklenar}}, \bibinfo
  {author} {\bibfnamefont {S.}~\bibnamefont {Sullivan}}, \bibinfo {author}
  {\bibfnamefont {D.}~\bibnamefont {Sun}}, \bibinfo {author} {\bibfnamefont
  {H.}~\bibnamefont {Tang}}, \bibinfo {author} {\bibfnamefont {V.}~\bibnamefont
  {Tyberkevych}}, \bibinfo {author} {\bibfnamefont {C.}~\bibnamefont
  {Trevillian}}, \bibinfo {author} {\bibfnamefont {A.~W.}\ \bibnamefont
  {Tsen}}, \bibinfo {author} {\bibfnamefont {L.}~\bibnamefont {Weiss}},
  \bibinfo {author} {\bibfnamefont {W.}~\bibnamefont {Zhang}}, \bibinfo
  {author} {\bibfnamefont {X.}~\bibnamefont {Zhang}}, \bibinfo {author}
  {\bibfnamefont {L.}~\bibnamefont {Zhao}}, \ and\ \bibinfo {author}
  {\bibfnamefont {C.~W.}\ \bibnamefont {Zollitsch}},\ }\href {\doibase
  10.1109/TQE.2021.3057799} {\bibfield  {journal} {\bibinfo  {journal} {IEEE
  Transactions on Quantum Engineering}\ ,\ \bibinfo {pages} {1}} (\bibinfo
  {year} {2021})}\BibitemShut {NoStop}%
\bibitem [{\citenamefont {Mahan}(2000)}]{Mahan_book}%
  \BibitemOpen
  \bibfield  {author} {\bibinfo {author} {\bibfnamefont {G.~D.}\ \bibnamefont
  {Mahan}},\ }\href {\doibase https://doi.org/10.1007/978-1-4757-5714-9} {\emph
  {\bibinfo {title} {Many-particle physics}}}\ (\bibinfo  {publisher}
  {Springer, Boston, MA},\ \bibinfo {year} {2000})\BibitemShut {NoStop}%
\bibitem [{\citenamefont {Haug}\ and\ \citenamefont
  {Jauho}(2008)}]{Jauho_book}%
  \BibitemOpen
  \bibfield  {author} {\bibinfo {author} {\bibfnamefont {H.}~\bibnamefont
  {Haug}}\ and\ \bibinfo {author} {\bibfnamefont {A.-P.}\ \bibnamefont
  {Jauho}},\ }\href {\doibase 10.1007/978-3-540-73564-9} {\emph {\bibinfo
  {title} {Quantum Kinetics in Transport and Optics of Semiconductors}}}\
  (\bibinfo  {publisher} {Springer-Verlag Berlin Heidelberg},\ \bibinfo {year}
  {2008})\BibitemShut {NoStop}%
\bibitem [{\citenamefont {Wang}\ \emph {et~al.}(2014)\citenamefont {Wang},
  \citenamefont {Agarwalla}, \citenamefont {Li},\ and\ \citenamefont
  {Thingna}}]{Wang_2014}%
  \BibitemOpen
  \bibfield  {author} {\bibinfo {author} {\bibfnamefont {J.-S.}\ \bibnamefont
  {Wang}}, \bibinfo {author} {\bibfnamefont {B.~K.}\ \bibnamefont {Agarwalla}},
  \bibinfo {author} {\bibfnamefont {H.}~\bibnamefont {Li}}, \ and\ \bibinfo
  {author} {\bibfnamefont {J.}~\bibnamefont {Thingna}},\ }\href {\doibase
  10.1007/s11467-013-0340-x} {\bibfield  {journal} {\bibinfo  {journal}
  {Frontiers of Physics}\ }\textbf {\bibinfo {volume} {9}},\ \bibinfo {pages}
  {673} (\bibinfo {year} {2014})}\BibitemShut {NoStop}%
\bibitem [{\citenamefont {Kamenev}(2011)}]{Kamenev_book}%
  \BibitemOpen
  \bibfield  {author} {\bibinfo {author} {\bibfnamefont {A.}~\bibnamefont
  {Kamenev}},\ }\href {\doibase 10.1017/CBO9781139003667} {\emph {\bibinfo
  {title} {Field Theory of Non-Equilibrium Systems}}}\ (\bibinfo  {publisher}
  {Cambridge University Press, Cambridge},\ \bibinfo {year} {2011})\BibitemShut
  {NoStop}%
\bibitem [{\citenamefont {Ford}\ \emph {et~al.}(1965)\citenamefont {Ford},
  \citenamefont {Kac},\ and\ \citenamefont {Mazur}}]{Ford_Kac_Mazur_1965}%
  \BibitemOpen
  \bibfield  {author} {\bibinfo {author} {\bibfnamefont {G.~W.}\ \bibnamefont
  {Ford}}, \bibinfo {author} {\bibfnamefont {M.}~\bibnamefont {Kac}}, \ and\
  \bibinfo {author} {\bibfnamefont {P.}~\bibnamefont {Mazur}},\ }\href
  {\doibase 10.1063/1.1704304} {\bibfield  {journal} {\bibinfo  {journal}
  {Journal of Mathematical Physics}\ }\textbf {\bibinfo {volume} {6}},\
  \bibinfo {pages} {504} (\bibinfo {year} {1965})}\BibitemShut {NoStop}%
\bibitem [{\citenamefont {Benguria}\ and\ \citenamefont
  {Kac}(1981)}]{Kac_1981}%
  \BibitemOpen
  \bibfield  {author} {\bibinfo {author} {\bibfnamefont {R.}~\bibnamefont
  {Benguria}}\ and\ \bibinfo {author} {\bibfnamefont {M.}~\bibnamefont {Kac}},\
  }\href {\doibase 10.1103/PhysRevLett.46.1} {\bibfield  {journal} {\bibinfo
  {journal} {Phys. Rev. Lett.}\ }\textbf {\bibinfo {volume} {46}},\ \bibinfo
  {pages} {1} (\bibinfo {year} {1981})}\BibitemShut {NoStop}%
\bibitem [{\citenamefont {Cort\'es}\ \emph {et~al.}(1985)\citenamefont
  {Cort\'es}, \citenamefont {West},\ and\ \citenamefont
  {Lindenberg}}]{Cortes_1985}%
  \BibitemOpen
  \bibfield  {author} {\bibinfo {author} {\bibfnamefont {E.}~\bibnamefont
  {Cort\'es}}, \bibinfo {author} {\bibfnamefont {B.~J.}\ \bibnamefont {West}},
  \ and\ \bibinfo {author} {\bibfnamefont {K.}~\bibnamefont {Lindenberg}},\
  }\href {\doibase 10.1063/1.448268} {\bibfield  {journal} {\bibinfo  {journal}
  {The Journal of Chemical Physics}\ }\textbf {\bibinfo {volume} {82}},\
  \bibinfo {pages} {2708} (\bibinfo {year} {1985})}\BibitemShut {NoStop}%
\bibitem [{\citenamefont {Ford}\ \emph {et~al.}(1988)\citenamefont {Ford},
  \citenamefont {Lewis},\ and\ \citenamefont {O'Connell}}]{Ford_1988}%
  \BibitemOpen
  \bibfield  {author} {\bibinfo {author} {\bibfnamefont {G.~W.}\ \bibnamefont
  {Ford}}, \bibinfo {author} {\bibfnamefont {J.~T.}\ \bibnamefont {Lewis}}, \
  and\ \bibinfo {author} {\bibfnamefont {R.~F.}\ \bibnamefont {O'Connell}},\
  }\href {\doibase 10.1103/PhysRevA.37.4419} {\bibfield  {journal} {\bibinfo
  {journal} {Phys. Rev. A}\ }\textbf {\bibinfo {volume} {37}},\ \bibinfo
  {pages} {4419} (\bibinfo {year} {1988})}\BibitemShut {NoStop}%
\bibitem [{\citenamefont {Dhar}\ and\ \citenamefont
  {Sriram~Shastry}(2003)}]{Dhar_2003}%
  \BibitemOpen
  \bibfield  {author} {\bibinfo {author} {\bibfnamefont {A.}~\bibnamefont
  {Dhar}}\ and\ \bibinfo {author} {\bibfnamefont {B.}~\bibnamefont
  {Sriram~Shastry}},\ }\href {\doibase 10.1103/PhysRevB.67.195405} {\bibfield
  {journal} {\bibinfo  {journal} {Phys. Rev. B}\ }\textbf {\bibinfo {volume}
  {67}},\ \bibinfo {pages} {195405} (\bibinfo {year} {2003})}\BibitemShut
  {NoStop}%
\bibitem [{\citenamefont {Banerjee}\ \emph {et~al.}(2004)\citenamefont
  {Banerjee}, \citenamefont {Bag}, \citenamefont {Banik},\ and\ \citenamefont
  {Ray}}]{Dhruba_2004}%
  \BibitemOpen
  \bibfield  {author} {\bibinfo {author} {\bibfnamefont {D.}~\bibnamefont
  {Banerjee}}, \bibinfo {author} {\bibfnamefont {B.~C.}\ \bibnamefont {Bag}},
  \bibinfo {author} {\bibfnamefont {S.~K.}\ \bibnamefont {Banik}}, \ and\
  \bibinfo {author} {\bibfnamefont {D.~S.}\ \bibnamefont {Ray}},\ }\href
  {\doibase 10.1063/1.1711593} {\bibfield  {journal} {\bibinfo  {journal} {The
  Journal of Chemical Physics}\ }\textbf {\bibinfo {volume} {120}},\ \bibinfo
  {pages} {8960} (\bibinfo {year} {2004})}\BibitemShut {NoStop}%
\bibitem [{\citenamefont {Dhar}\ and\ \citenamefont
  {Sen}(2006)}]{Dhar_Sen_2006}%
  \BibitemOpen
  \bibfield  {author} {\bibinfo {author} {\bibfnamefont {A.}~\bibnamefont
  {Dhar}}\ and\ \bibinfo {author} {\bibfnamefont {D.}~\bibnamefont {Sen}},\
  }\href {\doibase 10.1103/PhysRevB.73.085119} {\bibfield  {journal} {\bibinfo
  {journal} {Phys. Rev. B}\ }\textbf {\bibinfo {volume} {73}},\ \bibinfo
  {pages} {085119} (\bibinfo {year} {2006})}\BibitemShut {NoStop}%
\bibitem [{\citenamefont {Dhar}\ and\ \citenamefont
  {Roy}(2006)}]{Dhar_Roy_2006}%
  \BibitemOpen
  \bibfield  {author} {\bibinfo {author} {\bibfnamefont {A.}~\bibnamefont
  {Dhar}}\ and\ \bibinfo {author} {\bibfnamefont {D.}~\bibnamefont {Roy}},\
  }\href {\doibase 10.1007/s10955-006-9235-3} {\bibfield  {journal} {\bibinfo
  {journal} {Journal of Statistical Physics}\ }\textbf {\bibinfo {volume}
  {125}},\ \bibinfo {pages} {801} (\bibinfo {year} {2006})}\BibitemShut
  {NoStop}%
\bibitem [{\citenamefont {Breuer}\ and\ \citenamefont
  {Petruccione}(2007)}]{Breuer_book}%
  \BibitemOpen
  \bibfield  {author} {\bibinfo {author} {\bibfnamefont {H.-P.}\ \bibnamefont
  {Breuer}}\ and\ \bibinfo {author} {\bibfnamefont {F.}~\bibnamefont
  {Petruccione}},\ }\href {\doibase 10.1093/acprof:oso/9780199213900.001.0001}
  {\emph {\bibinfo {title} {The Theory of Open Quantum Systems}}}\ (\bibinfo
  {publisher} {Oxford University Press, Oxford},\ \bibinfo {year}
  {2007})\BibitemShut {NoStop}%
\bibitem [{\citenamefont {Rivas}\ and\ \citenamefont
  {Huelga}(2012)}]{Rivas_book}%
  \BibitemOpen
  \bibfield  {author} {\bibinfo {author} {\bibfnamefont {A.}~\bibnamefont
  {Rivas}}\ and\ \bibinfo {author} {\bibfnamefont {S.~F.}\ \bibnamefont
  {Huelga}},\ }\href {\doibase 10.1007/978-3-642-23354-8} {\emph {\bibinfo
  {title} {Open Quantum Systems}}}\ (\bibinfo  {publisher} {Springer Berlin
  Heidelberg},\ \bibinfo {year} {2012})\BibitemShut {NoStop}%
\bibitem [{\citenamefont {Tu}\ and\ \citenamefont {Zhang}(2008)}]{WMZ_2008}%
  \BibitemOpen
  \bibfield  {author} {\bibinfo {author} {\bibfnamefont {M.~W.~Y.}\
  \bibnamefont {Tu}}\ and\ \bibinfo {author} {\bibfnamefont {W.-M.}\
  \bibnamefont {Zhang}},\ }\href {\doibase 10.1103/PhysRevB.78.235311}
  {\bibfield  {journal} {\bibinfo  {journal} {Phys. Rev. B}\ }\textbf {\bibinfo
  {volume} {78}},\ \bibinfo {pages} {235311} (\bibinfo {year}
  {2008})}\BibitemShut {NoStop}%
\bibitem [{\citenamefont {Jin}\ \emph {et~al.}(2010)\citenamefont {Jin},
  \citenamefont {Tu}, \citenamefont {Zhang},\ and\ \citenamefont
  {Yan}}]{Jin_2010}%
  \BibitemOpen
  \bibfield  {author} {\bibinfo {author} {\bibfnamefont {J.}~\bibnamefont
  {Jin}}, \bibinfo {author} {\bibfnamefont {M.~W.-Y.}\ \bibnamefont {Tu}},
  \bibinfo {author} {\bibfnamefont {W.-M.}\ \bibnamefont {Zhang}}, \ and\
  \bibinfo {author} {\bibfnamefont {Y.}~\bibnamefont {Yan}},\ }\href {\doibase
  10.1088/1367-2630/12/8/083013} {\bibfield  {journal} {\bibinfo  {journal}
  {New Journal of Physics}\ }\textbf {\bibinfo {volume} {12}},\ \bibinfo
  {pages} {083013} (\bibinfo {year} {2010})}\BibitemShut {NoStop}%
\bibitem [{\citenamefont {Zhang}\ \emph {et~al.}(2012)\citenamefont {Zhang},
  \citenamefont {Lo}, \citenamefont {Xiong}, \citenamefont {Tu},\ and\
  \citenamefont {Nori}}]{WMZ_2012}%
  \BibitemOpen
  \bibfield  {author} {\bibinfo {author} {\bibfnamefont {W.-M.}\ \bibnamefont
  {Zhang}}, \bibinfo {author} {\bibfnamefont {P.-Y.}\ \bibnamefont {Lo}},
  \bibinfo {author} {\bibfnamefont {H.-N.}\ \bibnamefont {Xiong}}, \bibinfo
  {author} {\bibfnamefont {M.~W.-Y.}\ \bibnamefont {Tu}}, \ and\ \bibinfo
  {author} {\bibfnamefont {F.}~\bibnamefont {Nori}},\ }\href {\doibase
  10.1103/PhysRevLett.109.170402} {\bibfield  {journal} {\bibinfo  {journal}
  {Phys. Rev. Lett.}\ }\textbf {\bibinfo {volume} {109}},\ \bibinfo {pages}
  {170402} (\bibinfo {year} {2012})}\BibitemShut {NoStop}%
\bibitem [{\citenamefont {McCutcheon}\ \emph {et~al.}(2015)\citenamefont
  {McCutcheon}, \citenamefont {Paz},\ and\ \citenamefont
  {Roncaglia}}]{WMZ_2012_comment}%
  \BibitemOpen
  \bibfield  {author} {\bibinfo {author} {\bibfnamefont {D.~P.~S.}\
  \bibnamefont {McCutcheon}}, \bibinfo {author} {\bibfnamefont {J.~P.}\
  \bibnamefont {Paz}}, \ and\ \bibinfo {author} {\bibfnamefont {A.~J.}\
  \bibnamefont {Roncaglia}},\ }\href {\doibase 10.1103/PhysRevLett.115.168901}
  {\bibfield  {journal} {\bibinfo  {journal} {Phys. Rev. Lett.}\ }\textbf
  {\bibinfo {volume} {115}},\ \bibinfo {pages} {168901} (\bibinfo {year}
  {2015})}\BibitemShut {NoStop}%
\bibitem [{\citenamefont {Martinez}\ and\ \citenamefont
  {Paz}(2013)}]{Martinez_2013}%
  \BibitemOpen
  \bibfield  {author} {\bibinfo {author} {\bibfnamefont {E.~A.}\ \bibnamefont
  {Martinez}}\ and\ \bibinfo {author} {\bibfnamefont {J.~P.}\ \bibnamefont
  {Paz}},\ }\href {\doibase 10.1103/PhysRevLett.110.130406} {\bibfield
  {journal} {\bibinfo  {journal} {Phys. Rev. Lett.}\ }\textbf {\bibinfo
  {volume} {110}},\ \bibinfo {pages} {130406} (\bibinfo {year}
  {2013})}\BibitemShut {NoStop}%
\bibitem [{\citenamefont {Yang}\ \emph {et~al.}(2013)\citenamefont {Yang},
  \citenamefont {Cai}, \citenamefont {Xu}, \citenamefont {Zhang},\ and\
  \citenamefont {Sun}}]{WMZ_2013}%
  \BibitemOpen
  \bibfield  {author} {\bibinfo {author} {\bibfnamefont {L.-P.}\ \bibnamefont
  {Yang}}, \bibinfo {author} {\bibfnamefont {C.~Y.}\ \bibnamefont {Cai}},
  \bibinfo {author} {\bibfnamefont {D.~Z.}\ \bibnamefont {Xu}}, \bibinfo
  {author} {\bibfnamefont {W.-M.}\ \bibnamefont {Zhang}}, \ and\ \bibinfo
  {author} {\bibfnamefont {C.~P.}\ \bibnamefont {Sun}},\ }\href {\doibase
  10.1103/PhysRevA.87.012110} {\bibfield  {journal} {\bibinfo  {journal} {Phys.
  Rev. A}\ }\textbf {\bibinfo {volume} {87}},\ \bibinfo {pages} {012110}
  (\bibinfo {year} {2013})}\BibitemShut {NoStop}%
\bibitem [{\citenamefont {Ribeiro}\ and\ \citenamefont
  {Vieira}(2015)}]{Ribeiro_2015}%
  \BibitemOpen
  \bibfield  {author} {\bibinfo {author} {\bibfnamefont {P.}~\bibnamefont
  {Ribeiro}}\ and\ \bibinfo {author} {\bibfnamefont {V.~R.}\ \bibnamefont
  {Vieira}},\ }\href {\doibase 10.1103/PhysRevB.92.100302} {\bibfield
  {journal} {\bibinfo  {journal} {Phys. Rev. B}\ }\textbf {\bibinfo {volume}
  {92}},\ \bibinfo {pages} {100302} (\bibinfo {year} {2015})}\BibitemShut
  {NoStop}%
\bibitem [{\citenamefont {Walls}(1970)}]{Walls1970}%
  \BibitemOpen
  \bibfield  {author} {\bibinfo {author} {\bibfnamefont {D.~F.}\ \bibnamefont
  {Walls}},\ }\href {\doibase 10.1007/bf01396784} {\bibfield  {journal}
  {\bibinfo  {journal} {Zeitschrift f\"{u}r Physik A Hadrons and nuclei}\
  }\textbf {\bibinfo {volume} {234}},\ \bibinfo {pages} {231} (\bibinfo {year}
  {1970})}\BibitemShut {NoStop}%
\bibitem [{\citenamefont {Novotn{\'{y}}}(2002)}]{Novotny_2002}%
  \BibitemOpen
  \bibfield  {author} {\bibinfo {author} {\bibfnamefont {T.}~\bibnamefont
  {Novotn{\'{y}}}},\ }\href {\doibase 10.1209/epl/i2002-00174-3} {\bibfield
  {journal} {\bibinfo  {journal} {Europhysics Letters ({EPL})}\ }\textbf
  {\bibinfo {volume} {59}},\ \bibinfo {pages} {648} (\bibinfo {year}
  {2002})}\BibitemShut {NoStop}%
\bibitem [{\citenamefont {Wichterich}\ \emph {et~al.}(2007)\citenamefont
  {Wichterich}, \citenamefont {Henrich}, \citenamefont {Breuer}, \citenamefont
  {Gemmer},\ and\ \citenamefont {Michel}}]{Wichterich_2007}%
  \BibitemOpen
  \bibfield  {author} {\bibinfo {author} {\bibfnamefont {H.}~\bibnamefont
  {Wichterich}}, \bibinfo {author} {\bibfnamefont {M.~J.}\ \bibnamefont
  {Henrich}}, \bibinfo {author} {\bibfnamefont {H.-P.}\ \bibnamefont {Breuer}},
  \bibinfo {author} {\bibfnamefont {J.}~\bibnamefont {Gemmer}}, \ and\ \bibinfo
  {author} {\bibfnamefont {M.}~\bibnamefont {Michel}},\ }\href {\doibase
  10.1103/PhysRevE.76.031115} {\bibfield  {journal} {\bibinfo  {journal} {Phys.
  Rev. E}\ }\textbf {\bibinfo {volume} {76}},\ \bibinfo {pages} {031115}
  (\bibinfo {year} {2007})}\BibitemShut {NoStop}%
\bibitem [{\citenamefont {Prachar}\ and\ \citenamefont
  {Novotn{\'{y}}}(2010)}]{Novotny_2010}%
  \BibitemOpen
  \bibfield  {author} {\bibinfo {author} {\bibfnamefont {J.}~\bibnamefont
  {Prachar}}\ and\ \bibinfo {author} {\bibfnamefont {T.}~\bibnamefont
  {Novotn{\'{y}}}},\ }\href {\doibase
  https://doi.org/10.1016/j.physe.2009.06.026} {\bibfield  {journal} {\bibinfo
  {journal} {Physica E: Low-dimensional Systems and Nanostructures}\ }\textbf
  {\bibinfo {volume} {42}},\ \bibinfo {pages} {565} (\bibinfo {year}
  {2010})}\BibitemShut {NoStop}%
\bibitem [{\citenamefont {Rivas}\ \emph {et~al.}(2010)\citenamefont {Rivas},
  \citenamefont {Plato}, \citenamefont {Huelga},\ and\ \citenamefont
  {Plenio}}]{Rivas_2010}%
  \BibitemOpen
  \bibfield  {author} {\bibinfo {author} {\bibfnamefont {{\'{A}}.}~\bibnamefont
  {Rivas}}, \bibinfo {author} {\bibfnamefont {A.~D.~K.}\ \bibnamefont {Plato}},
  \bibinfo {author} {\bibfnamefont {S.~F.}\ \bibnamefont {Huelga}}, \ and\
  \bibinfo {author} {\bibfnamefont {M.~B.}\ \bibnamefont {Plenio}},\ }\href
  {\doibase 10.1088/1367-2630/12/11/113032} {\bibfield  {journal} {\bibinfo
  {journal} {New Journal of Physics}\ }\textbf {\bibinfo {volume} {12}},\
  \bibinfo {pages} {113032} (\bibinfo {year} {2010})}\BibitemShut {NoStop}%
\bibitem [{\citenamefont {Deçordi}\ and\ \citenamefont
  {Vidiella-Barranco}(2017)}]{barranco_2014}%
  \BibitemOpen
  \bibfield  {author} {\bibinfo {author} {\bibfnamefont {G.}~\bibnamefont
  {Deçordi}}\ and\ \bibinfo {author} {\bibfnamefont {A.}~\bibnamefont
  {Vidiella-Barranco}},\ }\href {\doibase
  https://doi.org/10.1016/j.optcom.2016.10.017} {\bibfield  {journal} {\bibinfo
   {journal} {Optics Communications}\ }\textbf {\bibinfo {volume} {387}},\
  \bibinfo {pages} {366} (\bibinfo {year} {2017})}\BibitemShut {NoStop}%
\bibitem [{\citenamefont {Levy}\ and\ \citenamefont
  {Kosloff}(2014)}]{Levy2014}%
  \BibitemOpen
  \bibfield  {author} {\bibinfo {author} {\bibfnamefont {A.}~\bibnamefont
  {Levy}}\ and\ \bibinfo {author} {\bibfnamefont {R.}~\bibnamefont {Kosloff}},\
  }\href {\doibase 10.1209/0295-5075/107/20004} {\bibfield  {journal} {\bibinfo
   {journal} {{EPL} (Europhysics Letters)}\ }\textbf {\bibinfo {volume}
  {107}},\ \bibinfo {pages} {20004} (\bibinfo {year} {2014})}\BibitemShut
  {NoStop}%
\bibitem [{\citenamefont {Purkayastha}\ \emph {et~al.}(2016)\citenamefont
  {Purkayastha}, \citenamefont {Dhar},\ and\ \citenamefont
  {Kulkarni}}]{archak_2016}%
  \BibitemOpen
  \bibfield  {author} {\bibinfo {author} {\bibfnamefont {A.}~\bibnamefont
  {Purkayastha}}, \bibinfo {author} {\bibfnamefont {A.}~\bibnamefont {Dhar}}, \
  and\ \bibinfo {author} {\bibfnamefont {M.}~\bibnamefont {Kulkarni}},\ }\href
  {\doibase 10.1103/PhysRevA.93.062114} {\bibfield  {journal} {\bibinfo
  {journal} {Phys. Rev. A}\ }\textbf {\bibinfo {volume} {93}},\ \bibinfo
  {pages} {062114} (\bibinfo {year} {2016})}\BibitemShut {NoStop}%
\bibitem [{\citenamefont {Trushechkin}\ and\ \citenamefont
  {Volovich}(2016)}]{Trushechkin_2016}%
  \BibitemOpen
  \bibfield  {author} {\bibinfo {author} {\bibfnamefont {A.~S.}\ \bibnamefont
  {Trushechkin}}\ and\ \bibinfo {author} {\bibfnamefont {I.~V.}\ \bibnamefont
  {Volovich}},\ }\href {\doibase 10.1209/0295-5075/113/30005} {\bibfield
  {journal} {\bibinfo  {journal} {{EPL} (Europhysics Letters)}\ }\textbf
  {\bibinfo {volume} {113}},\ \bibinfo {pages} {30005} (\bibinfo {year}
  {2016})}\BibitemShut {NoStop}%
\bibitem [{\citenamefont {Eastham}\ \emph {et~al.}(2016)\citenamefont
  {Eastham}, \citenamefont {Kirton}, \citenamefont {Cammack}, \citenamefont
  {Lovett},\ and\ \citenamefont {Keeling}}]{Eastham_2016}%
  \BibitemOpen
  \bibfield  {author} {\bibinfo {author} {\bibfnamefont {P.~R.}\ \bibnamefont
  {Eastham}}, \bibinfo {author} {\bibfnamefont {P.}~\bibnamefont {Kirton}},
  \bibinfo {author} {\bibfnamefont {H.~M.}\ \bibnamefont {Cammack}}, \bibinfo
  {author} {\bibfnamefont {B.~W.}\ \bibnamefont {Lovett}}, \ and\ \bibinfo
  {author} {\bibfnamefont {J.}~\bibnamefont {Keeling}},\ }\href
  {http://dx.doi.org/10.1103/PhysRevA.94.012110} {\bibfield  {journal}
  {\bibinfo  {journal} {Physical Review A}\ }\textbf {\bibinfo {volume} {94}}
  (\bibinfo {year} {2016})}\BibitemShut {NoStop}%
\bibitem [{\citenamefont {Hofer}\ \emph {et~al.}(2017)\citenamefont {Hofer},
  \citenamefont {Perarnau-Llobet}, \citenamefont {Miranda}, \citenamefont
  {Haack}, \citenamefont {Silva}, \citenamefont {Brask},\ and\ \citenamefont
  {Brunner}}]{Hofer_2017}%
  \BibitemOpen
  \bibfield  {author} {\bibinfo {author} {\bibfnamefont {P.~P.}\ \bibnamefont
  {Hofer}}, \bibinfo {author} {\bibfnamefont {M.}~\bibnamefont
  {Perarnau-Llobet}}, \bibinfo {author} {\bibfnamefont {L.~D.~M.}\ \bibnamefont
  {Miranda}}, \bibinfo {author} {\bibfnamefont {G.}~\bibnamefont {Haack}},
  \bibinfo {author} {\bibfnamefont {R.}~\bibnamefont {Silva}}, \bibinfo
  {author} {\bibfnamefont {J.~B.}\ \bibnamefont {Brask}}, \ and\ \bibinfo
  {author} {\bibfnamefont {N.}~\bibnamefont {Brunner}},\ }\href {\doibase
  10.1088/1367-2630/aa964f} {\bibfield  {journal} {\bibinfo  {journal} {New
  Journal of Physics}\ }\textbf {\bibinfo {volume} {19}},\ \bibinfo {pages}
  {123037} (\bibinfo {year} {2017})}\BibitemShut {NoStop}%
\bibitem [{\citenamefont {González}\ \emph {et~al.}(2017)\citenamefont
  {González}, \citenamefont {Correa}, \citenamefont {Nocerino}, \citenamefont
  {Palao}, \citenamefont {Alonso},\ and\ \citenamefont
  {Adesso}}]{Gonzalez_2017}%
  \BibitemOpen
  \bibfield  {author} {\bibinfo {author} {\bibfnamefont {J.~O.}\ \bibnamefont
  {González}}, \bibinfo {author} {\bibfnamefont {L.~A.}\ \bibnamefont
  {Correa}}, \bibinfo {author} {\bibfnamefont {G.}~\bibnamefont {Nocerino}},
  \bibinfo {author} {\bibfnamefont {J.~P.}\ \bibnamefont {Palao}}, \bibinfo
  {author} {\bibfnamefont {D.}~\bibnamefont {Alonso}}, \ and\ \bibinfo {author}
  {\bibfnamefont {G.}~\bibnamefont {Adesso}},\ }\href {\doibase
  10.1142/S1230161217400108} {\bibfield  {journal} {\bibinfo  {journal} {Open
  Systems \& Information Dynamics}\ }\textbf {\bibinfo {volume} {24}},\
  \bibinfo {pages} {1740010} (\bibinfo {year} {2017})}\BibitemShut {NoStop}%
\bibitem [{\citenamefont {Mitchison}\ and\ \citenamefont
  {Plenio}(2018)}]{Mitchison_2018}%
  \BibitemOpen
  \bibfield  {author} {\bibinfo {author} {\bibfnamefont {M.~T.}\ \bibnamefont
  {Mitchison}}\ and\ \bibinfo {author} {\bibfnamefont {M.~B.}\ \bibnamefont
  {Plenio}},\ }\href {\doibase 10.1088/1367-2630/aa9f70} {\bibfield  {journal}
  {\bibinfo  {journal} {New Journal of Physics}\ }\textbf {\bibinfo {volume}
  {20}},\ \bibinfo {pages} {033005} (\bibinfo {year} {2018})}\BibitemShut
  {NoStop}%
\bibitem [{\citenamefont {Cattaneo}\ \emph {et~al.}(2019)\citenamefont
  {Cattaneo}, \citenamefont {Giorgi}, \citenamefont {Maniscalco},\ and\
  \citenamefont {Zambrini}}]{Cattaneo_2019}%
  \BibitemOpen
  \bibfield  {author} {\bibinfo {author} {\bibfnamefont {M.}~\bibnamefont
  {Cattaneo}}, \bibinfo {author} {\bibfnamefont {G.~L.}\ \bibnamefont
  {Giorgi}}, \bibinfo {author} {\bibfnamefont {S.}~\bibnamefont {Maniscalco}},
  \ and\ \bibinfo {author} {\bibfnamefont {R.}~\bibnamefont {Zambrini}},\
  }\href {\doibase 10.1088/1367-2630/ab54ac} {\bibfield  {journal} {\bibinfo
  {journal} {New Journal of Physics}\ }\textbf {\bibinfo {volume} {21}},\
  \bibinfo {pages} {113045} (\bibinfo {year} {2019})}\BibitemShut {NoStop}%
\bibitem [{\citenamefont {Hartmann}\ and\ \citenamefont
  {Strunz}(2020)}]{Hartmann_2020_1}%
  \BibitemOpen
  \bibfield  {author} {\bibinfo {author} {\bibfnamefont {R.}~\bibnamefont
  {Hartmann}}\ and\ \bibinfo {author} {\bibfnamefont {W.~T.}\ \bibnamefont
  {Strunz}},\ }\href {\doibase 10.1103/PhysRevA.101.012103} {\bibfield
  {journal} {\bibinfo  {journal} {Phys. Rev. A}\ }\textbf {\bibinfo {volume}
  {101}},\ \bibinfo {pages} {012103} (\bibinfo {year} {2020})}\BibitemShut
  {NoStop}%
\bibitem [{\citenamefont {Benatti}\ \emph {et~al.}(2020)\citenamefont
  {Benatti}, \citenamefont {Floreanini},\ and\ \citenamefont
  {Memarzadeh}}]{Benatti_2020}%
  \BibitemOpen
  \bibfield  {author} {\bibinfo {author} {\bibfnamefont {F.}~\bibnamefont
  {Benatti}}, \bibinfo {author} {\bibfnamefont {R.}~\bibnamefont {Floreanini}},
  \ and\ \bibinfo {author} {\bibfnamefont {L.}~\bibnamefont {Memarzadeh}},\
  }\href {\doibase 10.1103/PhysRevA.102.042219} {\bibfield  {journal} {\bibinfo
   {journal} {Phys. Rev. A}\ }\textbf {\bibinfo {volume} {102}},\ \bibinfo
  {pages} {042219} (\bibinfo {year} {2020})}\BibitemShut {NoStop}%
\bibitem [{\citenamefont {Konopik}\ and\ \citenamefont
  {Lutz}(2022)}]{konopik_2020local}%
  \BibitemOpen
  \bibfield  {author} {\bibinfo {author} {\bibfnamefont {M.}~\bibnamefont
  {Konopik}}\ and\ \bibinfo {author} {\bibfnamefont {E.}~\bibnamefont {Lutz}},\
  }\href {\doibase 10.1103/PhysRevResearch.4.013171} {\bibfield  {journal}
  {\bibinfo  {journal} {Phys. Rev. Research}\ }\textbf {\bibinfo {volume}
  {4}},\ \bibinfo {pages} {013171} (\bibinfo {year} {2022})}\BibitemShut
  {NoStop}%
\bibitem [{\citenamefont {Scali}\ \emph {et~al.}(2021)\citenamefont {Scali},
  \citenamefont {Anders},\ and\ \citenamefont {Correa}}]{Scali_2021}%
  \BibitemOpen
  \bibfield  {author} {\bibinfo {author} {\bibfnamefont {S.}~\bibnamefont
  {Scali}}, \bibinfo {author} {\bibfnamefont {J.}~\bibnamefont {Anders}}, \
  and\ \bibinfo {author} {\bibfnamefont {L.~A.}\ \bibnamefont {Correa}},\
  }\href {http://dx.doi.org/10.22331/q-2021-05-01-451} {\bibfield  {journal}
  {\bibinfo  {journal} {Quantum}\ }\textbf {\bibinfo {volume} {5}},\ \bibinfo
  {pages} {451} (\bibinfo {year} {2021})}\BibitemShut {NoStop}%
\bibitem [{\citenamefont {Benatti}\ \emph {et~al.}(2021)\citenamefont
  {Benatti}, \citenamefont {Floreanini},\ and\ \citenamefont
  {Memarzadeh}}]{Floreanini_2021}%
  \BibitemOpen
  \bibfield  {author} {\bibinfo {author} {\bibfnamefont {F.}~\bibnamefont
  {Benatti}}, \bibinfo {author} {\bibfnamefont {R.}~\bibnamefont {Floreanini}},
  \ and\ \bibinfo {author} {\bibfnamefont {L.}~\bibnamefont {Memarzadeh}},\
  }\href {\doibase 10.1103/PRXQuantum.2.030344} {\bibfield  {journal} {\bibinfo
   {journal} {PRX Quantum}\ }\textbf {\bibinfo {volume} {2}},\ \bibinfo {pages}
  {030344} (\bibinfo {year} {2021})}\BibitemShut {NoStop}%
\bibitem [{\citenamefont {Trushechkin}(2021)}]{trushechkin2021}%
  \BibitemOpen
  \bibfield  {author} {\bibinfo {author} {\bibfnamefont {A.}~\bibnamefont
  {Trushechkin}},\ }\href {\doibase 10.1103/PhysRevA.103.062226} {\bibfield
  {journal} {\bibinfo  {journal} {Phys. Rev. A}\ }\textbf {\bibinfo {volume}
  {103}},\ \bibinfo {pages} {062226} (\bibinfo {year} {2021})}\BibitemShut
  {NoStop}%
\bibitem [{\citenamefont {Tupkary}\ \emph {et~al.}(2022)\citenamefont
  {Tupkary}, \citenamefont {Dhar}, \citenamefont {Kulkarni},\ and\
  \citenamefont {Purkayastha}}]{Archak_2021}%
  \BibitemOpen
  \bibfield  {author} {\bibinfo {author} {\bibfnamefont {D.}~\bibnamefont
  {Tupkary}}, \bibinfo {author} {\bibfnamefont {A.}~\bibnamefont {Dhar}},
  \bibinfo {author} {\bibfnamefont {M.}~\bibnamefont {Kulkarni}}, \ and\
  \bibinfo {author} {\bibfnamefont {A.}~\bibnamefont {Purkayastha}},\ }\href
  {\doibase 10.1103/PhysRevA.105.032208} {\bibfield  {journal} {\bibinfo
  {journal} {Phys. Rev. A}\ }\textbf {\bibinfo {volume} {105}},\ \bibinfo
  {pages} {032208} (\bibinfo {year} {2022})}\BibitemShut {NoStop}%
\bibitem [{\citenamefont {Redfield}(1965)}]{redfield1965}%
  \BibitemOpen
  \bibfield  {author} {\bibinfo {author} {\bibfnamefont {A.}~\bibnamefont
  {Redfield}},\ }in\ \href {\doibase 10.1016/b978-1-4832-3114-3.50007-6} {\emph
  {\bibinfo {booktitle} {Advances in Magnetic Resonance}}}\ (\bibinfo
  {publisher} {Elsevier},\ \bibinfo {year} {1965})\ pp.\ \bibinfo {pages}
  {1--32}\BibitemShut {NoStop}%
\bibitem [{\citenamefont {Davidovic}(2021)}]{Davidovic_2022}%
  \BibitemOpen
  \bibfield  {author} {\bibinfo {author} {\bibfnamefont {D.}~\bibnamefont
  {Davidovic}},\ }\href@noop {} {\  (\bibinfo {year} {2021})},\ \Eprint
  {http://arxiv.org/abs/2112.07863} {arXiv:2112.07863 [quant-ph]} \BibitemShut
  {NoStop}%
\bibitem [{\citenamefont {Purkayastha}\ \emph
  {et~al.}(2020{\natexlab{a}})\citenamefont {Purkayastha}, \citenamefont
  {Guarnieri}, \citenamefont {Mitchison}, \citenamefont {Filip},\ and\
  \citenamefont {Goold}}]{Archak_2020}%
  \BibitemOpen
  \bibfield  {author} {\bibinfo {author} {\bibfnamefont {A.}~\bibnamefont
  {Purkayastha}}, \bibinfo {author} {\bibfnamefont {G.}~\bibnamefont
  {Guarnieri}}, \bibinfo {author} {\bibfnamefont {M.~T.}\ \bibnamefont
  {Mitchison}}, \bibinfo {author} {\bibfnamefont {R.}~\bibnamefont {Filip}}, \
  and\ \bibinfo {author} {\bibfnamefont {J.}~\bibnamefont {Goold}},\ }\href
  {http://dx.doi.org/10.1038/s41534-020-0256-6} {\bibfield  {journal} {\bibinfo
   {journal} {npj Quantum Information}\ }\textbf {\bibinfo {volume} {6}}
  (\bibinfo {year} {2020}{\natexlab{a}})}\BibitemShut {NoStop}%
\bibitem [{\citenamefont {Anderloni}\ \emph {et~al.}(2007)\citenamefont
  {Anderloni}, \citenamefont {Benatti},\ and\ \citenamefont
  {Floreanini}}]{anderloni_2007}%
  \BibitemOpen
  \bibfield  {author} {\bibinfo {author} {\bibfnamefont {S.}~\bibnamefont
  {Anderloni}}, \bibinfo {author} {\bibfnamefont {F.}~\bibnamefont {Benatti}},
  \ and\ \bibinfo {author} {\bibfnamefont {R.}~\bibnamefont {Floreanini}},\
  }\href {\doibase 10.1088/1751-8113/40/7/013} {\bibfield  {journal} {\bibinfo
  {journal} {Journal of Physics A: Mathematical and Theoretical}\ }\textbf
  {\bibinfo {volume} {40}},\ \bibinfo {pages} {1625} (\bibinfo {year}
  {2007})}\BibitemShut {NoStop}%
\bibitem [{\citenamefont {Gaspard}\ and\ \citenamefont
  {Nagaoka}(1999)}]{Gaspard_Nagaoka_1999}%
  \BibitemOpen
  \bibfield  {author} {\bibinfo {author} {\bibfnamefont {P.}~\bibnamefont
  {Gaspard}}\ and\ \bibinfo {author} {\bibfnamefont {M.}~\bibnamefont
  {Nagaoka}},\ }\href {\doibase 10.1063/1.479867} {\bibfield  {journal}
  {\bibinfo  {journal} {The Journal of Chemical Physics}\ }\textbf {\bibinfo
  {volume} {111}},\ \bibinfo {pages} {5668} (\bibinfo {year}
  {1999})}\BibitemShut {NoStop}%
\bibitem [{\citenamefont {Kohen}\ \emph {et~al.}(1997)\citenamefont {Kohen},
  \citenamefont {Marston},\ and\ \citenamefont {Tannor}}]{Kohen_1997}%
  \BibitemOpen
  \bibfield  {author} {\bibinfo {author} {\bibfnamefont {D.}~\bibnamefont
  {Kohen}}, \bibinfo {author} {\bibfnamefont {C.~C.}\ \bibnamefont {Marston}},
  \ and\ \bibinfo {author} {\bibfnamefont {D.~J.}\ \bibnamefont {Tannor}},\
  }\href {\doibase 10.1063/1.474887} {\bibfield  {journal} {\bibinfo  {journal}
  {The Journal of Chemical Physics}\ }\textbf {\bibinfo {volume} {107}},\
  \bibinfo {pages} {5236} (\bibinfo {year} {1997})}\BibitemShut {NoStop}%
\bibitem [{\citenamefont {Gnutzmann}\ and\ \citenamefont
  {Haake}(1996)}]{Gnutzmann_1996}%
  \BibitemOpen
  \bibfield  {author} {\bibinfo {author} {\bibfnamefont {S.}~\bibnamefont
  {Gnutzmann}}\ and\ \bibinfo {author} {\bibfnamefont {F.}~\bibnamefont
  {Haake}},\ }\href {\doibase 10.1007/s002570050208} {\bibfield  {journal}
  {\bibinfo  {journal} {Zeitschrift f{\"u}r Physik B Condensed Matter}\
  }\textbf {\bibinfo {volume} {101}},\ \bibinfo {pages} {263} (\bibinfo {year}
  {1996})}\BibitemShut {NoStop}%
\bibitem [{\citenamefont {Suárez}\ \emph {et~al.}(1992)\citenamefont
  {Suárez}, \citenamefont {Silbey},\ and\ \citenamefont
  {Oppenheim}}]{Suarez_1992}%
  \BibitemOpen
  \bibfield  {author} {\bibinfo {author} {\bibfnamefont {A.}~\bibnamefont
  {Suárez}}, \bibinfo {author} {\bibfnamefont {R.}~\bibnamefont {Silbey}}, \
  and\ \bibinfo {author} {\bibfnamefont {I.}~\bibnamefont {Oppenheim}},\ }\href
  {\doibase 10.1063/1.463831} {\bibfield  {journal} {\bibinfo  {journal} {The
  Journal of Chemical Physics}\ }\textbf {\bibinfo {volume} {97}},\ \bibinfo
  {pages} {5101} (\bibinfo {year} {1992})}\BibitemShut {NoStop}%
\bibitem [{\citenamefont {Fleming}\ and\ \citenamefont
  {Cummings}(2011)}]{fleming_cummings_accuracy}%
  \BibitemOpen
  \bibfield  {author} {\bibinfo {author} {\bibfnamefont {C.~H.}\ \bibnamefont
  {Fleming}}\ and\ \bibinfo {author} {\bibfnamefont {N.~I.}\ \bibnamefont
  {Cummings}},\ }\href {\doibase 10.1103/PhysRevE.83.031117} {\bibfield
  {journal} {\bibinfo  {journal} {Phys. Rev. E}\ }\textbf {\bibinfo {volume}
  {83}},\ \bibinfo {pages} {031117} (\bibinfo {year} {2011})}\BibitemShut
  {NoStop}%
\bibitem [{\citenamefont {Nathan}\ and\ \citenamefont {Rudner}(2020)}]{ule}%
  \BibitemOpen
  \bibfield  {author} {\bibinfo {author} {\bibfnamefont {F.}~\bibnamefont
  {Nathan}}\ and\ \bibinfo {author} {\bibfnamefont {M.~S.}\ \bibnamefont
  {Rudner}},\ }\href {\doibase 10.1103/PhysRevB.102.115109} {\bibfield
  {journal} {\bibinfo  {journal} {Phys. Rev. B}\ }\textbf {\bibinfo {volume}
  {102}},\ \bibinfo {pages} {115109} (\bibinfo {year} {2020})}\BibitemShut
  {NoStop}%
\bibitem [{\citenamefont {Kleinherbers}\ \emph {et~al.}(2020)\citenamefont
  {Kleinherbers}, \citenamefont {Szpak}, \citenamefont {K\"onig},\ and\
  \citenamefont {Sch\"utzhold}}]{Kleinherbers_2020}%
  \BibitemOpen
  \bibfield  {author} {\bibinfo {author} {\bibfnamefont {E.}~\bibnamefont
  {Kleinherbers}}, \bibinfo {author} {\bibfnamefont {N.}~\bibnamefont {Szpak}},
  \bibinfo {author} {\bibfnamefont {J.}~\bibnamefont {K\"onig}}, \ and\
  \bibinfo {author} {\bibfnamefont {R.}~\bibnamefont {Sch\"utzhold}},\ }\href
  {\doibase 10.1103/PhysRevB.101.125131} {\bibfield  {journal} {\bibinfo
  {journal} {Phys. Rev. B}\ }\textbf {\bibinfo {volume} {101}},\ \bibinfo
  {pages} {125131} (\bibinfo {year} {2020})}\BibitemShut {NoStop}%
\bibitem [{\citenamefont {Davidovi{\'c}}(2020)}]{Davidovic_2020}%
  \BibitemOpen
  \bibfield  {author} {\bibinfo {author} {\bibfnamefont {D.}~\bibnamefont
  {Davidovi{\'c}}},\ }\href {\doibase 10.22331/q-2020-09-21-326} {\bibfield
  {journal} {\bibinfo  {journal} {Quantum}\ }\textbf {\bibinfo {volume} {4}},\
  \bibinfo {pages} {326} (\bibinfo {year} {2020})}\BibitemShut {NoStop}%
\bibitem [{\citenamefont {Mozgunov}\ and\ \citenamefont
  {Lidar}(2020)}]{mozgunov2020}%
  \BibitemOpen
  \bibfield  {author} {\bibinfo {author} {\bibfnamefont {E.}~\bibnamefont
  {Mozgunov}}\ and\ \bibinfo {author} {\bibfnamefont {D.}~\bibnamefont
  {Lidar}},\ }\href {http://dx.doi.org/10.22331/q-2020-02-06-227} {\bibfield
  {journal} {\bibinfo  {journal} {Quantum}\ }\textbf {\bibinfo {volume} {4}},\
  \bibinfo {pages} {227} (\bibinfo {year} {2020})}\BibitemShut {NoStop}%
\bibitem [{\citenamefont {McCauley}\ \emph {et~al.}(2020)\citenamefont
  {McCauley}, \citenamefont {Cruikshank}, \citenamefont {Bondar},\ and\
  \citenamefont {Jacobs}}]{mccauley2020}%
  \BibitemOpen
  \bibfield  {author} {\bibinfo {author} {\bibfnamefont {G.}~\bibnamefont
  {McCauley}}, \bibinfo {author} {\bibfnamefont {B.}~\bibnamefont
  {Cruikshank}}, \bibinfo {author} {\bibfnamefont {D.~I.}\ \bibnamefont
  {Bondar}}, \ and\ \bibinfo {author} {\bibfnamefont {K.}~\bibnamefont
  {Jacobs}},\ }\href {http://dx.doi.org/10.1038/s41534-020-00299-6} {\bibfield
  {journal} {\bibinfo  {journal} {npj Quantum Information}\ }\textbf {\bibinfo
  {volume} {6}} (\bibinfo {year} {2020})}\BibitemShut {NoStop}%
\bibitem [{\citenamefont {Kir\ifmmode~\check{s}\else \v{s}\fi{}anskas}\ \emph
  {et~al.}(2018)\citenamefont {Kir\ifmmode~\check{s}\else \v{s}\fi{}anskas},
  \citenamefont {Francki\'e},\ and\ \citenamefont {Wacker}}]{kirvsanskas2018}%
  \BibitemOpen
  \bibfield  {author} {\bibinfo {author} {\bibfnamefont {G.}~\bibnamefont
  {Kir\ifmmode~\check{s}\else \v{s}\fi{}anskas}}, \bibinfo {author}
  {\bibfnamefont {M.}~\bibnamefont {Francki\'e}}, \ and\ \bibinfo {author}
  {\bibfnamefont {A.}~\bibnamefont {Wacker}},\ }\href {\doibase
  10.1103/PhysRevB.97.035432} {\bibfield  {journal} {\bibinfo  {journal} {Phys.
  Rev. B}\ }\textbf {\bibinfo {volume} {97}},\ \bibinfo {pages} {035432}
  (\bibinfo {year} {2018})}\BibitemShut {NoStop}%
\bibitem [{\citenamefont {Prosen}(2008)}]{Prosen_2008}%
  \BibitemOpen
  \bibfield  {author} {\bibinfo {author} {\bibfnamefont {T.}~\bibnamefont
  {Prosen}},\ }\href {\doibase 10.1088/1367-2630/10/4/043026} {\bibfield
  {journal} {\bibinfo  {journal} {New Journal of Physics}\ }\textbf {\bibinfo
  {volume} {10}},\ \bibinfo {pages} {043026} (\bibinfo {year}
  {2008})}\BibitemShut {NoStop}%
\bibitem [{\citenamefont {Prosen}\ and\ \citenamefont
  {{\v{Z}}unkovi{\v{c}}}(2010)}]{Prosen_2010}%
  \BibitemOpen
  \bibfield  {author} {\bibinfo {author} {\bibfnamefont {T.}~\bibnamefont
  {Prosen}}\ and\ \bibinfo {author} {\bibfnamefont {B.}~\bibnamefont
  {{\v{Z}}unkovi{\v{c}}}},\ }\href {\doibase 10.1088/1367-2630/12/2/025016}
  {\bibfield  {journal} {\bibinfo  {journal} {New Journal of Physics}\ }\textbf
  {\bibinfo {volume} {12}},\ \bibinfo {pages} {025016} (\bibinfo {year}
  {2010})}\BibitemShut {NoStop}%
\bibitem [{\citenamefont {Žunkovič}\ and\ \citenamefont
  {Prosen}(2012)}]{Prosen_2012}%
  \BibitemOpen
  \bibfield  {author} {\bibinfo {author} {\bibfnamefont {B.}~\bibnamefont
  {Žunkovič}}\ and\ \bibinfo {author} {\bibfnamefont {T.}~\bibnamefont
  {Prosen}},\ }\href {\doibase 10.1063/1.4745594} {\bibfield  {journal}
  {\bibinfo  {journal} {AIP Conference Proceedings}\ }\textbf {\bibinfo
  {volume} {1468}},\ \bibinfo {pages} {350} (\bibinfo {year}
  {2012})}\BibitemShut {NoStop}%
\bibitem [{\citenamefont {Koga}\ and\ \citenamefont
  {Yamamoto}(2012)}]{Koga_2012}%
  \BibitemOpen
  \bibfield  {author} {\bibinfo {author} {\bibfnamefont {K.}~\bibnamefont
  {Koga}}\ and\ \bibinfo {author} {\bibfnamefont {N.}~\bibnamefont
  {Yamamoto}},\ }\href {\doibase 10.1103/PhysRevA.85.022103} {\bibfield
  {journal} {\bibinfo  {journal} {Phys. Rev. A}\ }\textbf {\bibinfo {volume}
  {85}},\ \bibinfo {pages} {022103} (\bibinfo {year} {2012})}\BibitemShut
  {NoStop}%
\bibitem [{\citenamefont {Nicacio}\ \emph {et~al.}(2015)\citenamefont
  {Nicacio}, \citenamefont {Ferraro}, \citenamefont {Imparato}, \citenamefont
  {Paternostro},\ and\ \citenamefont {Semi\~ao}}]{Nicacio_2012}%
  \BibitemOpen
  \bibfield  {author} {\bibinfo {author} {\bibfnamefont {F.}~\bibnamefont
  {Nicacio}}, \bibinfo {author} {\bibfnamefont {A.}~\bibnamefont {Ferraro}},
  \bibinfo {author} {\bibfnamefont {A.}~\bibnamefont {Imparato}}, \bibinfo
  {author} {\bibfnamefont {M.}~\bibnamefont {Paternostro}}, \ and\ \bibinfo
  {author} {\bibfnamefont {F.~L.}\ \bibnamefont {Semi\~ao}},\ }\href {\doibase
  10.1103/PhysRevE.91.042116} {\bibfield  {journal} {\bibinfo  {journal} {Phys.
  Rev. E}\ }\textbf {\bibinfo {volume} {91}},\ \bibinfo {pages} {042116}
  (\bibinfo {year} {2015})}\BibitemShut {NoStop}%
\bibitem [{\citenamefont {Landi}\ \emph {et~al.}(2013)\citenamefont {Landi},
  \citenamefont {Tom{\'{e}}},\ and\ \citenamefont {de~Oliveira}}]{Landi_2013}%
  \BibitemOpen
  \bibfield  {author} {\bibinfo {author} {\bibfnamefont {G.~T.}\ \bibnamefont
  {Landi}}, \bibinfo {author} {\bibfnamefont {T.}~\bibnamefont {Tom{\'{e}}}}, \
  and\ \bibinfo {author} {\bibfnamefont {M.~J.}\ \bibnamefont {de~Oliveira}},\
  }\href {\doibase 10.1088/1751-8113/46/39/395001} {\bibfield  {journal}
  {\bibinfo  {journal} {Journal of Physics A: Mathematical and Theoretical}\
  }\textbf {\bibinfo {volume} {46}},\ \bibinfo {pages} {395001} (\bibinfo
  {year} {2013})}\BibitemShut {NoStop}%
\bibitem [{\citenamefont {Nicacio}\ \emph {et~al.}(2016)\citenamefont
  {Nicacio}, \citenamefont {Paternostro},\ and\ \citenamefont
  {Ferraro}}]{Nicacio_2016}%
  \BibitemOpen
  \bibfield  {author} {\bibinfo {author} {\bibfnamefont {F.}~\bibnamefont
  {Nicacio}}, \bibinfo {author} {\bibfnamefont {M.}~\bibnamefont
  {Paternostro}}, \ and\ \bibinfo {author} {\bibfnamefont {A.}~\bibnamefont
  {Ferraro}},\ }\href {\doibase 10.1103/PhysRevA.94.052129} {\bibfield
  {journal} {\bibinfo  {journal} {Phys. Rev. A}\ }\textbf {\bibinfo {volume}
  {94}},\ \bibinfo {pages} {052129} (\bibinfo {year} {2016})}\BibitemShut
  {NoStop}%
\bibitem [{\citenamefont {Santos}\ \emph {et~al.}(2017)\citenamefont {Santos},
  \citenamefont {Landi},\ and\ \citenamefont {Paternostro}}]{Landi_2017}%
  \BibitemOpen
  \bibfield  {author} {\bibinfo {author} {\bibfnamefont {J.~P.}\ \bibnamefont
  {Santos}}, \bibinfo {author} {\bibfnamefont {G.~T.}\ \bibnamefont {Landi}}, \
  and\ \bibinfo {author} {\bibfnamefont {M.}~\bibnamefont {Paternostro}},\
  }\href {\doibase 10.1103/PhysRevLett.118.220601} {\bibfield  {journal}
  {\bibinfo  {journal} {Phys. Rev. Lett.}\ }\textbf {\bibinfo {volume} {118}},\
  \bibinfo {pages} {220601} (\bibinfo {year} {2017})}\BibitemShut {NoStop}%
\bibitem [{\citenamefont {Malouf}\ \emph {et~al.}(2019)\citenamefont {Malouf},
  \citenamefont {Santos}, \citenamefont {Correa}, \citenamefont {Paternostro},\
  and\ \citenamefont {Landi}}]{Landi_2019}%
  \BibitemOpen
  \bibfield  {author} {\bibinfo {author} {\bibfnamefont {W.~T.~B.}\
  \bibnamefont {Malouf}}, \bibinfo {author} {\bibfnamefont {J.~P.}\
  \bibnamefont {Santos}}, \bibinfo {author} {\bibfnamefont {L.~A.}\
  \bibnamefont {Correa}}, \bibinfo {author} {\bibfnamefont {M.}~\bibnamefont
  {Paternostro}}, \ and\ \bibinfo {author} {\bibfnamefont {G.~T.}\ \bibnamefont
  {Landi}},\ }\href {\doibase 10.1103/PhysRevA.99.052104} {\bibfield  {journal}
  {\bibinfo  {journal} {Phys. Rev. A}\ }\textbf {\bibinfo {volume} {99}},\
  \bibinfo {pages} {052104} (\bibinfo {year} {2019})}\BibitemShut {NoStop}%
\bibitem [{\citenamefont {McDonald}\ \emph {et~al.}(2022)\citenamefont
  {McDonald}, \citenamefont {Hanai},\ and\ \citenamefont
  {Clerk}}]{Mcdonald_2021}%
  \BibitemOpen
  \bibfield  {author} {\bibinfo {author} {\bibfnamefont {A.}~\bibnamefont
  {McDonald}}, \bibinfo {author} {\bibfnamefont {R.}~\bibnamefont {Hanai}}, \
  and\ \bibinfo {author} {\bibfnamefont {A.~A.}\ \bibnamefont {Clerk}},\ }\href
  {\doibase 10.1103/PhysRevB.105.064302} {\bibfield  {journal} {\bibinfo
  {journal} {Phys. Rev. B}\ }\textbf {\bibinfo {volume} {105}},\ \bibinfo
  {pages} {064302} (\bibinfo {year} {2022})}\BibitemShut {NoStop}%
\bibitem [{\citenamefont {Bernal-García}\ \emph {et~al.}(2021)\citenamefont
  {Bernal-García}, \citenamefont {Huang}, \citenamefont {Miroshnichenko},\
  and\ \citenamefont {Woolley}}]{Bernal_Garcia_2021}%
  \BibitemOpen
  \bibfield  {author} {\bibinfo {author} {\bibfnamefont {D.~N.}\ \bibnamefont
  {Bernal-García}}, \bibinfo {author} {\bibfnamefont {L.}~\bibnamefont
  {Huang}}, \bibinfo {author} {\bibfnamefont {A.~E.}\ \bibnamefont
  {Miroshnichenko}}, \ and\ \bibinfo {author} {\bibfnamefont {M.~J.}\
  \bibnamefont {Woolley}},\ }\href@noop {} {\  (\bibinfo {year} {2021})},\
  \Eprint {http://arxiv.org/abs/2111.04041} {arXiv:2111.04041 [quant-ph]}
  \BibitemShut {NoStop}%
\bibitem [{\citenamefont {Roccati}\ \emph {et~al.}(2021)\citenamefont
  {Roccati}, \citenamefont {Lorenzo}, \citenamefont {Palma}, \citenamefont
  {Landi}, \citenamefont {Brunelli},\ and\ \citenamefont
  {Ciccarello}}]{Roccati_2021}%
  \BibitemOpen
  \bibfield  {author} {\bibinfo {author} {\bibfnamefont {F.}~\bibnamefont
  {Roccati}}, \bibinfo {author} {\bibfnamefont {S.}~\bibnamefont {Lorenzo}},
  \bibinfo {author} {\bibfnamefont {G.~M.}\ \bibnamefont {Palma}}, \bibinfo
  {author} {\bibfnamefont {G.~T.}\ \bibnamefont {Landi}}, \bibinfo {author}
  {\bibfnamefont {M.}~\bibnamefont {Brunelli}}, \ and\ \bibinfo {author}
  {\bibfnamefont {F.}~\bibnamefont {Ciccarello}},\ }\href {\doibase
  10.1088/2058-9565/abcfcc} {\bibfield  {journal} {\bibinfo  {journal} {Quantum
  Science and Technology}\ }\textbf {\bibinfo {volume} {6}},\ \bibinfo {pages}
  {025005} (\bibinfo {year} {2021})}\BibitemShut {NoStop}%
\bibitem [{\citenamefont {Gaji\'c}\ and\ \citenamefont
  {Qureshi}(1995)}]{Control_theory_book1}%
  \BibitemOpen
  \bibfield  {author} {\bibinfo {author} {\bibfnamefont {Z.}~\bibnamefont
  {Gaji\'c}}\ and\ \bibinfo {author} {\bibfnamefont {M.~T.~J.}\ \bibnamefont
  {Qureshi}},\ }\href@noop {} {\emph {\bibinfo {title} {Lyapunov matrix
  equation in system stability and control}}}\ (\bibinfo  {publisher} {Academic
  Press, Inc. San Diego, California},\ \bibinfo {year} {1995})\BibitemShut
  {NoStop}%
\bibitem [{\citenamefont {Sun}\ and\ \citenamefont
  {Yong}(2020)}]{Control_theory_book2}%
  \BibitemOpen
  \bibfield  {author} {\bibinfo {author} {\bibfnamefont {J.}~\bibnamefont
  {Sun}}\ and\ \bibinfo {author} {\bibfnamefont {J.}~\bibnamefont {Yong}},\
  }\href {\doibase https://doi.org/10.1007/978-3-030-20922-3} {\emph {\bibinfo
  {title} {Stochastic linear-quadratic optimal control theory: Open-loop and
  closed-loop solutions}}}\ (\bibinfo  {publisher} {Springer Nature,
  Switzerland},\ \bibinfo {year} {2020})\BibitemShut {NoStop}%
\bibitem [{\citenamefont {Yuan}\ \emph {et~al.}(2021)\citenamefont {Yuan},
  \citenamefont {Lv}, \citenamefont {Baldi},\ and\ \citenamefont
  {Zhang}}]{Yuan_2021_lyapunov_control}%
  \BibitemOpen
  \bibfield  {author} {\bibinfo {author} {\bibfnamefont {S.}~\bibnamefont
  {Yuan}}, \bibinfo {author} {\bibfnamefont {M.}~\bibnamefont {Lv}}, \bibinfo
  {author} {\bibfnamefont {S.}~\bibnamefont {Baldi}}, \ and\ \bibinfo {author}
  {\bibfnamefont {L.}~\bibnamefont {Zhang}},\ }\href {\doibase
  10.1109/TAC.2020.3003647} {\bibfield  {journal} {\bibinfo  {journal} {IEEE
  Transactions on Automatic Control}\ }\textbf {\bibinfo {volume} {66}},\
  \bibinfo {pages} {2250} (\bibinfo {year} {2021})}\BibitemShut {NoStop}%
\bibitem [{\citenamefont {Bergholtz}\ \emph {et~al.}(2021)\citenamefont
  {Bergholtz}, \citenamefont {Budich},\ and\ \citenamefont
  {Kunst}}]{Bergholtz_2021}%
  \BibitemOpen
  \bibfield  {author} {\bibinfo {author} {\bibfnamefont {E.~J.}\ \bibnamefont
  {Bergholtz}}, \bibinfo {author} {\bibfnamefont {J.~C.}\ \bibnamefont
  {Budich}}, \ and\ \bibinfo {author} {\bibfnamefont {F.~K.}\ \bibnamefont
  {Kunst}},\ }\href {\doibase 10.1103/RevModPhys.93.015005} {\bibfield
  {journal} {\bibinfo  {journal} {Rev. Mod. Phys.}\ }\textbf {\bibinfo {volume}
  {93}},\ \bibinfo {pages} {015005} (\bibinfo {year} {2021})}\BibitemShut
  {NoStop}%
\bibitem [{\citenamefont {Ashida}\ \emph {et~al.}(2020)\citenamefont {Ashida},
  \citenamefont {Gong},\ and\ \citenamefont {Ueda}}]{Ashida_2020}%
  \BibitemOpen
  \bibfield  {author} {\bibinfo {author} {\bibfnamefont {Y.}~\bibnamefont
  {Ashida}}, \bibinfo {author} {\bibfnamefont {Z.}~\bibnamefont {Gong}}, \ and\
  \bibinfo {author} {\bibfnamefont {M.}~\bibnamefont {Ueda}},\ }\href {\doibase
  10.1080/00018732.2021.1876991} {\bibfield  {journal} {\bibinfo  {journal}
  {Advances in Physics}\ }\textbf {\bibinfo {volume} {69}},\ \bibinfo {pages}
  {249} (\bibinfo {year} {2020})}\BibitemShut {NoStop}%
\bibitem [{\citenamefont {El-Ganainy}\ \emph {et~al.}(2018)\citenamefont
  {El-Ganainy}, \citenamefont {Makris}, \citenamefont {Khajavikhan},
  \citenamefont {Musslimani}, \citenamefont {Rotter},\ and\ \citenamefont
  {Christodoulides}}]{El_Ganainy2018}%
  \BibitemOpen
  \bibfield  {author} {\bibinfo {author} {\bibfnamefont {R.}~\bibnamefont
  {El-Ganainy}}, \bibinfo {author} {\bibfnamefont {K.~G.}\ \bibnamefont
  {Makris}}, \bibinfo {author} {\bibfnamefont {M.}~\bibnamefont {Khajavikhan}},
  \bibinfo {author} {\bibfnamefont {Z.~H.}\ \bibnamefont {Musslimani}},
  \bibinfo {author} {\bibfnamefont {S.}~\bibnamefont {Rotter}}, \ and\ \bibinfo
  {author} {\bibfnamefont {D.~N.}\ \bibnamefont {Christodoulides}},\ }\href
  {\doibase 10.1038/nphys4323} {\bibfield  {journal} {\bibinfo  {journal}
  {Nature Physics}\ }\textbf {\bibinfo {volume} {14}},\ \bibinfo {pages} {11}
  (\bibinfo {year} {2018})}\BibitemShut {NoStop}%
\bibitem [{\citenamefont {El-Ganainy}\ \emph {et~al.}(2019)\citenamefont
  {El-Ganainy}, \citenamefont {Khajavikhan}, \citenamefont {Christodoulides},\
  and\ \citenamefont {Ozdemir}}]{El_Ganainy2019}%
  \BibitemOpen
  \bibfield  {author} {\bibinfo {author} {\bibfnamefont {R.}~\bibnamefont
  {El-Ganainy}}, \bibinfo {author} {\bibfnamefont {M.}~\bibnamefont
  {Khajavikhan}}, \bibinfo {author} {\bibfnamefont {D.~N.}\ \bibnamefont
  {Christodoulides}}, \ and\ \bibinfo {author} {\bibfnamefont {S.~K.}\
  \bibnamefont {Ozdemir}},\ }\href {\doibase 10.1038/s42005-019-0130-z}
  {\bibfield  {journal} {\bibinfo  {journal} {Communications Physics}\ }\textbf
  {\bibinfo {volume} {2}},\ \bibinfo {pages} {37} (\bibinfo {year}
  {2019})}\BibitemShut {NoStop}%
\bibitem [{\citenamefont {Feng}\ \emph {et~al.}(2017)\citenamefont {Feng},
  \citenamefont {El-Ganainy},\ and\ \citenamefont {Ge}}]{Feng2017}%
  \BibitemOpen
  \bibfield  {author} {\bibinfo {author} {\bibfnamefont {L.}~\bibnamefont
  {Feng}}, \bibinfo {author} {\bibfnamefont {R.}~\bibnamefont {El-Ganainy}}, \
  and\ \bibinfo {author} {\bibfnamefont {L.}~\bibnamefont {Ge}},\ }\href
  {\doibase 10.1038/s41566-017-0031-1} {\bibfield  {journal} {\bibinfo
  {journal} {Nature Photonics}\ }\textbf {\bibinfo {volume} {11}},\ \bibinfo
  {pages} {752} (\bibinfo {year} {2017})}\BibitemShut {NoStop}%
\bibitem [{\citenamefont {Quijandr\'{\i}a}\ \emph {et~al.}(2018)\citenamefont
  {Quijandr\'{\i}a}, \citenamefont {Naether}, \citenamefont {\"Ozdemir},
  \citenamefont {Nori},\ and\ \citenamefont {Zueco}}]{Nori_2018}%
  \BibitemOpen
  \bibfield  {author} {\bibinfo {author} {\bibfnamefont {F.}~\bibnamefont
  {Quijandr\'{\i}a}}, \bibinfo {author} {\bibfnamefont {U.}~\bibnamefont
  {Naether}}, \bibinfo {author} {\bibfnamefont {S.~K.}\ \bibnamefont
  {\"Ozdemir}}, \bibinfo {author} {\bibfnamefont {F.}~\bibnamefont {Nori}}, \
  and\ \bibinfo {author} {\bibfnamefont {D.}~\bibnamefont {Zueco}},\ }\href
  {\doibase 10.1103/PhysRevA.97.053846} {\bibfield  {journal} {\bibinfo
  {journal} {Phys. Rev. A}\ }\textbf {\bibinfo {volume} {97}},\ \bibinfo
  {pages} {053846} (\bibinfo {year} {2018})}\BibitemShut {NoStop}%
\bibitem [{\citenamefont {Purkayastha}\ \emph
  {et~al.}(2020{\natexlab{b}})\citenamefont {Purkayastha}, \citenamefont
  {Kulkarni},\ and\ \citenamefont {Joglekar}}]{Archak_2020_PT}%
  \BibitemOpen
  \bibfield  {author} {\bibinfo {author} {\bibfnamefont {A.}~\bibnamefont
  {Purkayastha}}, \bibinfo {author} {\bibfnamefont {M.}~\bibnamefont
  {Kulkarni}}, \ and\ \bibinfo {author} {\bibfnamefont {Y.~N.}\ \bibnamefont
  {Joglekar}},\ }\href {\doibase 10.1103/PhysRevResearch.2.043075} {\bibfield
  {journal} {\bibinfo  {journal} {Phys. Rev. Research}\ }\textbf {\bibinfo
  {volume} {2}},\ \bibinfo {pages} {043075} (\bibinfo {year}
  {2020}{\natexlab{b}})}\BibitemShut {NoStop}%
\bibitem [{\citenamefont {Huber}\ \emph
  {et~al.}(2020{\natexlab{a}})\citenamefont {Huber}, \citenamefont {Kirton},
  \citenamefont {Rotter},\ and\ \citenamefont {Rabl}}]{Huber_2020}%
  \BibitemOpen
  \bibfield  {author} {\bibinfo {author} {\bibfnamefont {J.}~\bibnamefont
  {Huber}}, \bibinfo {author} {\bibfnamefont {P.}~\bibnamefont {Kirton}},
  \bibinfo {author} {\bibfnamefont {S.}~\bibnamefont {Rotter}}, \ and\ \bibinfo
  {author} {\bibfnamefont {P.}~\bibnamefont {Rabl}},\ }\href {\doibase
  10.21468/scipostphys.9.4.052} {\bibfield  {journal} {\bibinfo  {journal}
  {SciPost Physics}\ }\textbf {\bibinfo {volume} {9}} (\bibinfo {year}
  {2020}{\natexlab{a}}),\ 10.21468/scipostphys.9.4.052}\BibitemShut {NoStop}%
\bibitem [{\citenamefont {Huber}\ \emph
  {et~al.}(2020{\natexlab{b}})\citenamefont {Huber}, \citenamefont {Kirton},\
  and\ \citenamefont {Rabl}}]{Huber_2020_PRA}%
  \BibitemOpen
  \bibfield  {author} {\bibinfo {author} {\bibfnamefont {J.}~\bibnamefont
  {Huber}}, \bibinfo {author} {\bibfnamefont {P.}~\bibnamefont {Kirton}}, \
  and\ \bibinfo {author} {\bibfnamefont {P.}~\bibnamefont {Rabl}},\ }\href
  {\doibase 10.1103/PhysRevA.102.012219} {\bibfield  {journal} {\bibinfo
  {journal} {Phys. Rev. A}\ }\textbf {\bibinfo {volume} {102}},\ \bibinfo
  {pages} {012219} (\bibinfo {year} {2020}{\natexlab{b}})}\BibitemShut
  {NoStop}%
\bibitem [{\citenamefont {Jaramillo~{\'A}vila}\ \emph
  {et~al.}(2020)\citenamefont {Jaramillo~{\'A}vila}, \citenamefont
  {Ventura-Vel{\'a}zquez}, \citenamefont {Le{\'o}n-Montiel}, \citenamefont
  {Joglekar},\ and\ \citenamefont {Rodr{\'i}guez-Lara}}]{Avila_2020}%
  \BibitemOpen
  \bibfield  {author} {\bibinfo {author} {\bibfnamefont {B.}~\bibnamefont
  {Jaramillo~{\'A}vila}}, \bibinfo {author} {\bibfnamefont {C.}~\bibnamefont
  {Ventura-Vel{\'a}zquez}}, \bibinfo {author} {\bibfnamefont {R.~d.~J.}\
  \bibnamefont {Le{\'o}n-Montiel}}, \bibinfo {author} {\bibfnamefont {Y.~N.}\
  \bibnamefont {Joglekar}}, \ and\ \bibinfo {author} {\bibfnamefont {B.~M.}\
  \bibnamefont {Rodr{\'i}guez-Lara}},\ }\href {\doibase
  10.1038/s41598-020-58582-7} {\bibfield  {journal} {\bibinfo  {journal}
  {Scientific Reports}\ }\textbf {\bibinfo {volume} {10}},\ \bibinfo {pages}
  {1761} (\bibinfo {year} {2020})}\BibitemShut {NoStop}%
\bibitem [{\citenamefont {Álvaro Gómez-León}\ \emph
  {et~al.}(2021)\citenamefont {Álvaro Gómez-León}, \citenamefont {Ramos},
  \citenamefont {González-Tudela},\ and\ \citenamefont
  {Porras}}]{Gomez_Leon_2021}%
  \BibitemOpen
  \bibfield  {author} {\bibinfo {author} {\bibnamefont {Álvaro Gómez-León}},
  \bibinfo {author} {\bibfnamefont {T.}~\bibnamefont {Ramos}}, \bibinfo
  {author} {\bibfnamefont {A.}~\bibnamefont {González-Tudela}}, \ and\
  \bibinfo {author} {\bibfnamefont {D.}~\bibnamefont {Porras}},\ }\href@noop {}
  {\  (\bibinfo {year} {2021})},\ \Eprint {http://arxiv.org/abs/2109.10930}
  {arXiv:2109.10930 [quant-ph]} \BibitemShut {NoStop}%
\bibitem [{\citenamefont {Arkhipov}\ \emph {et~al.}(2020)\citenamefont
  {Arkhipov}, \citenamefont {Miranowicz}, \citenamefont {Minganti},\ and\
  \citenamefont {Nori}}]{Arkhipov_2020}%
  \BibitemOpen
  \bibfield  {author} {\bibinfo {author} {\bibfnamefont {I.~I.}\ \bibnamefont
  {Arkhipov}}, \bibinfo {author} {\bibfnamefont {A.}~\bibnamefont
  {Miranowicz}}, \bibinfo {author} {\bibfnamefont {F.}~\bibnamefont
  {Minganti}}, \ and\ \bibinfo {author} {\bibfnamefont {F.}~\bibnamefont
  {Nori}},\ }\href {\doibase 10.1103/PhysRevA.102.033715} {\bibfield  {journal}
  {\bibinfo  {journal} {Phys. Rev. A}\ }\textbf {\bibinfo {volume} {102}},\
  \bibinfo {pages} {033715} (\bibinfo {year} {2020})}\BibitemShut {NoStop}%
\bibitem [{\citenamefont {Arkhipov}\ \emph {et~al.}(2021)\citenamefont
  {Arkhipov}, \citenamefont {Minganti}, \citenamefont {Miranowicz},\ and\
  \citenamefont {Nori}}]{Arkhipov_2021}%
  \BibitemOpen
  \bibfield  {author} {\bibinfo {author} {\bibfnamefont {I.~I.}\ \bibnamefont
  {Arkhipov}}, \bibinfo {author} {\bibfnamefont {F.}~\bibnamefont {Minganti}},
  \bibinfo {author} {\bibfnamefont {A.}~\bibnamefont {Miranowicz}}, \ and\
  \bibinfo {author} {\bibfnamefont {F.}~\bibnamefont {Nori}},\ }\href {\doibase
  10.1103/PhysRevA.104.012205} {\bibfield  {journal} {\bibinfo  {journal}
  {Phys. Rev. A}\ }\textbf {\bibinfo {volume} {104}},\ \bibinfo {pages}
  {012205} (\bibinfo {year} {2021})}\BibitemShut {NoStop}%
\bibitem [{\citenamefont {Wang}\ \emph {et~al.}(2021)\citenamefont {Wang},
  \citenamefont {Dutt}, \citenamefont {Wojcik},\ and\ \citenamefont
  {Fan}}]{Wang2021}%
  \BibitemOpen
  \bibfield  {author} {\bibinfo {author} {\bibfnamefont {K.}~\bibnamefont
  {Wang}}, \bibinfo {author} {\bibfnamefont {A.}~\bibnamefont {Dutt}}, \bibinfo
  {author} {\bibfnamefont {C.~C.}\ \bibnamefont {Wojcik}}, \ and\ \bibinfo
  {author} {\bibfnamefont {S.}~\bibnamefont {Fan}},\ }\href {\doibase
  10.1038/s41586-021-03848-x} {\bibfield  {journal} {\bibinfo  {journal}
  {Nature}\ }\textbf {\bibinfo {volume} {598}},\ \bibinfo {pages} {59}
  (\bibinfo {year} {2021})}\BibitemShut {NoStop}%
\bibitem [{\citenamefont {Kawabata}\ \emph {et~al.}(2019)\citenamefont
  {Kawabata}, \citenamefont {Shiozaki}, \citenamefont {Ueda},\ and\
  \citenamefont {Sato}}]{Kawabata_2019}%
  \BibitemOpen
  \bibfield  {author} {\bibinfo {author} {\bibfnamefont {K.}~\bibnamefont
  {Kawabata}}, \bibinfo {author} {\bibfnamefont {K.}~\bibnamefont {Shiozaki}},
  \bibinfo {author} {\bibfnamefont {M.}~\bibnamefont {Ueda}}, \ and\ \bibinfo
  {author} {\bibfnamefont {M.}~\bibnamefont {Sato}},\ }\href {\doibase
  10.1103/PhysRevX.9.041015} {\bibfield  {journal} {\bibinfo  {journal} {Phys.
  Rev. X}\ }\textbf {\bibinfo {volume} {9}},\ \bibinfo {pages} {041015}
  (\bibinfo {year} {2019})}\BibitemShut {NoStop}%
\bibitem [{\citenamefont {Torres}(2019)}]{Foa_Torres_2019}%
  \BibitemOpen
  \bibfield  {author} {\bibinfo {author} {\bibfnamefont {L.~E. F.~F.}\
  \bibnamefont {Torres}},\ }\href {\doibase 10.1088/2515-7639/ab4092}
  {\bibfield  {journal} {\bibinfo  {journal} {Journal of Physics: Materials}\
  }\textbf {\bibinfo {volume} {3}},\ \bibinfo {pages} {014002} (\bibinfo {year}
  {2019})}\BibitemShut {NoStop}%
\bibitem [{\citenamefont {Gong}\ \emph {et~al.}(2018)\citenamefont {Gong},
  \citenamefont {Ashida}, \citenamefont {Kawabata}, \citenamefont {Takasan},
  \citenamefont {Higashikawa},\ and\ \citenamefont {Ueda}}]{Gong_2018}%
  \BibitemOpen
  \bibfield  {author} {\bibinfo {author} {\bibfnamefont {Z.}~\bibnamefont
  {Gong}}, \bibinfo {author} {\bibfnamefont {Y.}~\bibnamefont {Ashida}},
  \bibinfo {author} {\bibfnamefont {K.}~\bibnamefont {Kawabata}}, \bibinfo
  {author} {\bibfnamefont {K.}~\bibnamefont {Takasan}}, \bibinfo {author}
  {\bibfnamefont {S.}~\bibnamefont {Higashikawa}}, \ and\ \bibinfo {author}
  {\bibfnamefont {M.}~\bibnamefont {Ueda}},\ }\href {\doibase
  10.1103/PhysRevX.8.031079} {\bibfield  {journal} {\bibinfo  {journal} {Phys.
  Rev. X}\ }\textbf {\bibinfo {volume} {8}},\ \bibinfo {pages} {031079}
  (\bibinfo {year} {2018})}\BibitemShut {NoStop}%
\bibitem [{\citenamefont {Harari}\ \emph {et~al.}(2018)\citenamefont {Harari},
  \citenamefont {Bandres}, \citenamefont {Lumer}, \citenamefont {Rechtsman},
  \citenamefont {Chong}, \citenamefont {Khajavikhan}, \citenamefont
  {Christodoulides},\ and\ \citenamefont {Segev}}]{Harari_2019}%
  \BibitemOpen
  \bibfield  {author} {\bibinfo {author} {\bibfnamefont {G.}~\bibnamefont
  {Harari}}, \bibinfo {author} {\bibfnamefont {M.~A.}\ \bibnamefont {Bandres}},
  \bibinfo {author} {\bibfnamefont {Y.}~\bibnamefont {Lumer}}, \bibinfo
  {author} {\bibfnamefont {M.~C.}\ \bibnamefont {Rechtsman}}, \bibinfo {author}
  {\bibfnamefont {Y.~D.}\ \bibnamefont {Chong}}, \bibinfo {author}
  {\bibfnamefont {M.}~\bibnamefont {Khajavikhan}}, \bibinfo {author}
  {\bibfnamefont {D.~N.}\ \bibnamefont {Christodoulides}}, \ and\ \bibinfo
  {author} {\bibfnamefont {M.}~\bibnamefont {Segev}},\ }\href {\doibase
  10.1126/science.aar4003} {\bibfield  {journal} {\bibinfo  {journal}
  {Science}\ }\textbf {\bibinfo {volume} {359}},\ \bibinfo {pages} {eaar4003}
  (\bibinfo {year} {2018})}\BibitemShut {NoStop}%
\bibitem [{\citenamefont {Bandres}\ \emph {et~al.}(2018)\citenamefont
  {Bandres}, \citenamefont {Wittek}, \citenamefont {Harari}, \citenamefont
  {Parto}, \citenamefont {Ren}, \citenamefont {Segev}, \citenamefont
  {Christodoulides},\ and\ \citenamefont {Khajavikhan}}]{Miguel_2019}%
  \BibitemOpen
  \bibfield  {author} {\bibinfo {author} {\bibfnamefont {M.~A.}\ \bibnamefont
  {Bandres}}, \bibinfo {author} {\bibfnamefont {S.}~\bibnamefont {Wittek}},
  \bibinfo {author} {\bibfnamefont {G.}~\bibnamefont {Harari}}, \bibinfo
  {author} {\bibfnamefont {M.}~\bibnamefont {Parto}}, \bibinfo {author}
  {\bibfnamefont {J.}~\bibnamefont {Ren}}, \bibinfo {author} {\bibfnamefont
  {M.}~\bibnamefont {Segev}}, \bibinfo {author} {\bibfnamefont {D.~N.}\
  \bibnamefont {Christodoulides}}, \ and\ \bibinfo {author} {\bibfnamefont
  {M.}~\bibnamefont {Khajavikhan}},\ }\href {\doibase 10.1126/science.aar4005}
  {\bibfield  {journal} {\bibinfo  {journal} {Science}\ }\textbf {\bibinfo
  {volume} {359}},\ \bibinfo {pages} {eaar4005} (\bibinfo {year}
  {2018})}\BibitemShut {NoStop}%
\bibitem [{\citenamefont {Behr}\ \emph {et~al.}(2019)\citenamefont {Behr},
  \citenamefont {Benner},\ and\ \citenamefont {Heiland}}]{Behr_2019}%
  \BibitemOpen
  \bibfield  {author} {\bibinfo {author} {\bibfnamefont {M.}~\bibnamefont
  {Behr}}, \bibinfo {author} {\bibfnamefont {P.}~\bibnamefont {Benner}}, \ and\
  \bibinfo {author} {\bibfnamefont {J.}~\bibnamefont {Heiland}},\ }\href
  {\doibase 10.1007/s10092-019-0348-x} {\bibfield  {journal} {\bibinfo
  {journal} {Calcolo}\ }\textbf {\bibinfo {volume} {56}},\ \bibinfo {pages}
  {51} (\bibinfo {year} {2019})}\BibitemShut {NoStop}%
\bibitem [{\citenamefont {Walls}\ and\ \citenamefont
  {Milburn}(2008)}]{Milburn_book1}%
  \BibitemOpen
  \bibfield  {author} {\bibinfo {author} {\bibfnamefont {D.}~\bibnamefont
  {Walls}}\ and\ \bibinfo {author} {\bibfnamefont {G.~J.}\ \bibnamefont
  {Milburn}},\ }\href@noop {} {\emph {\bibinfo {title} {Quantum pptics}}}\
  (\bibinfo  {publisher} {Springer-Verlag Berlin Heidelberg},\ \bibinfo {year}
  {2008})\BibitemShut {NoStop}%
\bibitem [{\citenamefont {Wiseman}\ and\ \citenamefont
  {Milburn}(2010)}]{Milburn_book2}%
  \BibitemOpen
  \bibfield  {author} {\bibinfo {author} {\bibfnamefont {H.~M.}\ \bibnamefont
  {Wiseman}}\ and\ \bibinfo {author} {\bibfnamefont {G.~J.}\ \bibnamefont
  {Milburn}},\ }\href@noop {} {\emph {\bibinfo {title} {Quantum measurement and
  control}}}\ (\bibinfo  {publisher} {Cambridge Universtiy Press, New York},\
  \bibinfo {year} {2010})\BibitemShut {NoStop}%
\bibitem [{\citenamefont {Lindblad}(1976)}]{lindblad1976}%
  \BibitemOpen
  \bibfield  {author} {\bibinfo {author} {\bibfnamefont {G.}~\bibnamefont
  {Lindblad}},\ }\href {\doibase 10.1007/bf01608499} {\bibfield  {journal}
  {\bibinfo  {journal} {Communications in Mathematical Physics}\ }\textbf
  {\bibinfo {volume} {48}},\ \bibinfo {pages} {119} (\bibinfo {year}
  {1976})}\BibitemShut {NoStop}%
\bibitem [{\citenamefont {Gorini}\ \emph {et~al.}(1976)\citenamefont {Gorini},
  \citenamefont {Kossakowski},\ and\ \citenamefont {Sudarshan}}]{GKS1976}%
  \BibitemOpen
  \bibfield  {author} {\bibinfo {author} {\bibfnamefont {V.}~\bibnamefont
  {Gorini}}, \bibinfo {author} {\bibfnamefont {A.}~\bibnamefont {Kossakowski}},
  \ and\ \bibinfo {author} {\bibfnamefont {E.~C.~G.}\ \bibnamefont
  {Sudarshan}},\ }\href {\doibase 10.1063/1.522979} {\bibfield  {journal}
  {\bibinfo  {journal} {Journal of Mathematical Physics}\ }\textbf {\bibinfo
  {volume} {17}},\ \bibinfo {pages} {821} (\bibinfo {year} {1976})}\BibitemShut
  {NoStop}%
\bibitem [{\citenamefont {Gorini}\ \emph {et~al.}(1978)\citenamefont {Gorini},
  \citenamefont {Frigerio}, \citenamefont {Verri}, \citenamefont
  {Kossakowski},\ and\ \citenamefont {Sudarshan}}]{Gorini_1978}%
  \BibitemOpen
  \bibfield  {author} {\bibinfo {author} {\bibfnamefont {V.}~\bibnamefont
  {Gorini}}, \bibinfo {author} {\bibfnamefont {A.}~\bibnamefont {Frigerio}},
  \bibinfo {author} {\bibfnamefont {M.}~\bibnamefont {Verri}}, \bibinfo
  {author} {\bibfnamefont {A.}~\bibnamefont {Kossakowski}}, \ and\ \bibinfo
  {author} {\bibfnamefont {E.}~\bibnamefont {Sudarshan}},\ }\href {\doibase
  https://doi.org/10.1016/0034-4877(78)90050-2} {\bibfield  {journal} {\bibinfo
   {journal} {Reports on Mathematical Physics}\ }\textbf {\bibinfo {volume}
  {13}},\ \bibinfo {pages} {149} (\bibinfo {year} {1978})}\BibitemShut
  {NoStop}%
\bibitem [{\citenamefont {Chung}\ and\ \citenamefont
  {Peschel}(2001)}]{Peschel_2001}%
  \BibitemOpen
  \bibfield  {author} {\bibinfo {author} {\bibfnamefont {M.-C.}\ \bibnamefont
  {Chung}}\ and\ \bibinfo {author} {\bibfnamefont {I.}~\bibnamefont
  {Peschel}},\ }\href {\doibase 10.1103/PhysRevB.64.064412} {\bibfield
  {journal} {\bibinfo  {journal} {Phys. Rev. B}\ }\textbf {\bibinfo {volume}
  {64}},\ \bibinfo {pages} {064412} (\bibinfo {year} {2001})}\BibitemShut
  {NoStop}%
\bibitem [{\citenamefont {Dhar}\ \emph {et~al.}(2012)\citenamefont {Dhar},
  \citenamefont {Saito},\ and\ \citenamefont {H\"anggi}}]{Dhar_2012}%
  \BibitemOpen
  \bibfield  {author} {\bibinfo {author} {\bibfnamefont {A.}~\bibnamefont
  {Dhar}}, \bibinfo {author} {\bibfnamefont {K.}~\bibnamefont {Saito}}, \ and\
  \bibinfo {author} {\bibfnamefont {P.}~\bibnamefont {H\"anggi}},\ }\href
  {\doibase 10.1103/PhysRevE.85.011126} {\bibfield  {journal} {\bibinfo
  {journal} {Phys. Rev. E}\ }\textbf {\bibinfo {volume} {85}},\ \bibinfo
  {pages} {011126} (\bibinfo {year} {2012})}\BibitemShut {NoStop}%
\bibitem [{\citenamefont {Eisler}\ and\ \citenamefont
  {Peschel}(2017)}]{Peschel_2017}%
  \BibitemOpen
  \bibfield  {author} {\bibinfo {author} {\bibfnamefont {V.}~\bibnamefont
  {Eisler}}\ and\ \bibinfo {author} {\bibfnamefont {I.}~\bibnamefont
  {Peschel}},\ }\href {\doibase 10.1088/1751-8121/aa76b5} {\bibfield  {journal}
  {\bibinfo  {journal} {Journal of Physics A: Mathematical and Theoretical}\
  }\textbf {\bibinfo {volume} {50}},\ \bibinfo {pages} {284003} (\bibinfo
  {year} {2017})}\BibitemShut {NoStop}%
\bibitem [{\citenamefont {Thingna}\ \emph {et~al.}(2012)\citenamefont
  {Thingna}, \citenamefont {Wang},\ and\ \citenamefont
  {H{\"a}nggi}}]{Juzar_2012}%
  \BibitemOpen
  \bibfield  {author} {\bibinfo {author} {\bibfnamefont {J.}~\bibnamefont
  {Thingna}}, \bibinfo {author} {\bibfnamefont {J.-S.}\ \bibnamefont {Wang}}, \
  and\ \bibinfo {author} {\bibfnamefont {P.}~\bibnamefont {H{\"a}nggi}},\
  }\href {\doibase 10.1063/1.4718706} {\bibfield  {journal} {\bibinfo
  {journal} {The Journal of Chemical Physics}\ }\textbf {\bibinfo {volume}
  {136}},\ \bibinfo {pages} {194110} (\bibinfo {year} {2012})}\BibitemShut
  {NoStop}%
\bibitem [{\citenamefont {Thingna}\ \emph {et~al.}(2013)\citenamefont
  {Thingna}, \citenamefont {Wang},\ and\ \citenamefont
  {H\"anggi}}]{Juzar_2013}%
  \BibitemOpen
  \bibfield  {author} {\bibinfo {author} {\bibfnamefont {J.}~\bibnamefont
  {Thingna}}, \bibinfo {author} {\bibfnamefont {J.-S.}\ \bibnamefont {Wang}}, \
  and\ \bibinfo {author} {\bibfnamefont {P.}~\bibnamefont {H\"anggi}},\ }\href
  {\doibase 10.1103/PhysRevE.88.052127} {\bibfield  {journal} {\bibinfo
  {journal} {Phys. Rev. E}\ }\textbf {\bibinfo {volume} {88}},\ \bibinfo
  {pages} {052127} (\bibinfo {year} {2013})}\BibitemShut {NoStop}%
\bibitem [{\citenamefont {Xu}\ \emph {et~al.}(2017)\citenamefont {Xu},
  \citenamefont {Thingna},\ and\ \citenamefont {Wang}}]{Juzar_2017}%
  \BibitemOpen
  \bibfield  {author} {\bibinfo {author} {\bibfnamefont {X.}~\bibnamefont
  {Xu}}, \bibinfo {author} {\bibfnamefont {J.}~\bibnamefont {Thingna}}, \ and\
  \bibinfo {author} {\bibfnamefont {J.-S.}\ \bibnamefont {Wang}},\ }\href
  {\doibase 10.1103/PhysRevB.95.035428} {\bibfield  {journal} {\bibinfo
  {journal} {Phys. Rev. B}\ }\textbf {\bibinfo {volume} {95}},\ \bibinfo
  {pages} {035428} (\bibinfo {year} {2017})}\BibitemShut {NoStop}%
\bibitem [{\citenamefont {Chan}\ \emph {et~al.}(2014)\citenamefont {Chan},
  \citenamefont {Lin}, \citenamefont {Yelin},\ and\ \citenamefont
  {Lukin}}]{Chan_2014}%
  \BibitemOpen
  \bibfield  {author} {\bibinfo {author} {\bibfnamefont {C.-K.}\ \bibnamefont
  {Chan}}, \bibinfo {author} {\bibfnamefont {G.-D.}\ \bibnamefont {Lin}},
  \bibinfo {author} {\bibfnamefont {S.~F.}\ \bibnamefont {Yelin}}, \ and\
  \bibinfo {author} {\bibfnamefont {M.~D.}\ \bibnamefont {Lukin}},\ }\href
  {\doibase 10.1103/PhysRevA.89.042117} {\bibfield  {journal} {\bibinfo
  {journal} {Phys. Rev. A}\ }\textbf {\bibinfo {volume} {89}},\ \bibinfo
  {pages} {042117} (\bibinfo {year} {2014})}\BibitemShut {NoStop}%
\bibitem [{\citenamefont {Giusteri}\ \emph {et~al.}(2017)\citenamefont
  {Giusteri}, \citenamefont {Recrosi}, \citenamefont {Schaller},\ and\
  \citenamefont {Celardo}}]{Giusteri_2017}%
  \BibitemOpen
  \bibfield  {author} {\bibinfo {author} {\bibfnamefont {G.~G.}\ \bibnamefont
  {Giusteri}}, \bibinfo {author} {\bibfnamefont {F.}~\bibnamefont {Recrosi}},
  \bibinfo {author} {\bibfnamefont {G.}~\bibnamefont {Schaller}}, \ and\
  \bibinfo {author} {\bibfnamefont {G.~L.}\ \bibnamefont {Celardo}},\ }\href
  {\doibase 10.1103/PhysRevE.96.012113} {\bibfield  {journal} {\bibinfo
  {journal} {Phys. Rev. E}\ }\textbf {\bibinfo {volume} {96}},\ \bibinfo
  {pages} {012113} (\bibinfo {year} {2017})}\BibitemShut {NoStop}%
\bibitem [{\citenamefont {Ko\l{}ody\ifmmode~\acute{n}\else \'{n}\fi{}ski}\
  \emph {et~al.}(2018)\citenamefont {Ko\l{}ody\ifmmode~\acute{n}\else
  \'{n}\fi{}ski}, \citenamefont {Brask}, \citenamefont {Perarnau-Llobet},\ and\
  \citenamefont {Bylicka}}]{Bogna_2018}%
  \BibitemOpen
  \bibfield  {author} {\bibinfo {author} {\bibfnamefont {J.}~\bibnamefont
  {Ko\l{}ody\ifmmode~\acute{n}\else \'{n}\fi{}ski}}, \bibinfo {author}
  {\bibfnamefont {J.~B.}\ \bibnamefont {Brask}}, \bibinfo {author}
  {\bibfnamefont {M.}~\bibnamefont {Perarnau-Llobet}}, \ and\ \bibinfo {author}
  {\bibfnamefont {B.}~\bibnamefont {Bylicka}},\ }\href {\doibase
  10.1103/PhysRevA.97.062124} {\bibfield  {journal} {\bibinfo  {journal} {Phys.
  Rev. A}\ }\textbf {\bibinfo {volume} {97}},\ \bibinfo {pages} {062124}
  (\bibinfo {year} {2018})}\BibitemShut {NoStop}%
\bibitem [{\citenamefont {Maguire}\ \emph {et~al.}(2019)\citenamefont
  {Maguire}, \citenamefont {Iles-Smith},\ and\ \citenamefont
  {Nazir}}]{Maguire_2019}%
  \BibitemOpen
  \bibfield  {author} {\bibinfo {author} {\bibfnamefont {H.}~\bibnamefont
  {Maguire}}, \bibinfo {author} {\bibfnamefont {J.}~\bibnamefont {Iles-Smith}},
  \ and\ \bibinfo {author} {\bibfnamefont {A.}~\bibnamefont {Nazir}},\ }\href
  {\doibase 10.1103/PhysRevLett.123.093601} {\bibfield  {journal} {\bibinfo
  {journal} {Phys. Rev. Lett.}\ }\textbf {\bibinfo {volume} {123}},\ \bibinfo
  {pages} {093601} (\bibinfo {year} {2019})}\BibitemShut {NoStop}%
\bibitem [{\citenamefont {Emary}(2007)}]{Emary_2007}%
  \BibitemOpen
  \bibfield  {author} {\bibinfo {author} {\bibfnamefont {C.}~\bibnamefont
  {Emary}},\ }\href {\doibase 10.1103/PhysRevB.76.245319} {\bibfield  {journal}
  {\bibinfo  {journal} {Phys. Rev. B}\ }\textbf {\bibinfo {volume} {76}},\
  \bibinfo {pages} {245319} (\bibinfo {year} {2007})}\BibitemShut {NoStop}%
\bibitem [{\citenamefont {Bu{\v{c}}a}\ and\ \citenamefont
  {Prosen}(2012)}]{Buca_2012}%
  \BibitemOpen
  \bibfield  {author} {\bibinfo {author} {\bibfnamefont {B.}~\bibnamefont
  {Bu{\v{c}}a}}\ and\ \bibinfo {author} {\bibfnamefont {T.}~\bibnamefont
  {Prosen}},\ }\href {\doibase 10.1088/1367-2630/14/7/073007} {\bibfield
  {journal} {\bibinfo  {journal} {New Journal of Physics}\ }\textbf {\bibinfo
  {volume} {14}},\ \bibinfo {pages} {073007} (\bibinfo {year}
  {2012})}\BibitemShut {NoStop}%
\bibitem [{\citenamefont {Finkelstein-Shapiro}\ \emph
  {et~al.}(2019)\citenamefont {Finkelstein-Shapiro}, \citenamefont {Felicetti},
  \citenamefont {Hansen}, \citenamefont {Pullerits},\ and\ \citenamefont
  {Keller}}]{Keller_2019}%
  \BibitemOpen
  \bibfield  {author} {\bibinfo {author} {\bibfnamefont {D.}~\bibnamefont
  {Finkelstein-Shapiro}}, \bibinfo {author} {\bibfnamefont {S.}~\bibnamefont
  {Felicetti}}, \bibinfo {author} {\bibfnamefont {T.}~\bibnamefont {Hansen}},
  \bibinfo {author} {\bibfnamefont {T.~o.}\ \bibnamefont {Pullerits}}, \ and\
  \bibinfo {author} {\bibfnamefont {A.}~\bibnamefont {Keller}},\ }\href
  {\doibase 10.1103/PhysRevA.99.053829} {\bibfield  {journal} {\bibinfo
  {journal} {Phys. Rev. A}\ }\textbf {\bibinfo {volume} {99}},\ \bibinfo
  {pages} {053829} (\bibinfo {year} {2019})}\BibitemShut {NoStop}%
\bibitem [{\citenamefont {Quach}\ and\ \citenamefont
  {Munro}(2020)}]{Quach_2020}%
  \BibitemOpen
  \bibfield  {author} {\bibinfo {author} {\bibfnamefont {J.~Q.}\ \bibnamefont
  {Quach}}\ and\ \bibinfo {author} {\bibfnamefont {W.~J.}\ \bibnamefont
  {Munro}},\ }\href {\doibase 10.1103/PhysRevApplied.14.024092} {\bibfield
  {journal} {\bibinfo  {journal} {Phys. Rev. Applied}\ }\textbf {\bibinfo
  {volume} {14}},\ \bibinfo {pages} {024092} (\bibinfo {year}
  {2020})}\BibitemShut {NoStop}%
\bibitem [{\citenamefont {Talkner}(1986)}]{Talkner_1986}%
  \BibitemOpen
  \bibfield  {author} {\bibinfo {author} {\bibfnamefont {P.}~\bibnamefont
  {Talkner}},\ }\href {\doibase https://doi.org/10.1016/0003-4916(86)90207-1}
  {\bibfield  {journal} {\bibinfo  {journal} {Annals of Physics}\ }\textbf
  {\bibinfo {volume} {167}},\ \bibinfo {pages} {390} (\bibinfo {year}
  {1986})}\BibitemShut {NoStop}%
\bibitem [{\citenamefont {Ford}\ and\ \citenamefont
  {O'Connell}(1996)}]{Ford_1996}%
  \BibitemOpen
  \bibfield  {author} {\bibinfo {author} {\bibfnamefont {G.~W.}\ \bibnamefont
  {Ford}}\ and\ \bibinfo {author} {\bibfnamefont {R.~F.}\ \bibnamefont
  {O'Connell}},\ }\href {\doibase 10.1103/PhysRevLett.77.798} {\bibfield
  {journal} {\bibinfo  {journal} {Phys. Rev. Lett.}\ }\textbf {\bibinfo
  {volume} {77}},\ \bibinfo {pages} {798} (\bibinfo {year} {1996})}\BibitemShut
  {NoStop}%
\bibitem [{\citenamefont {Guarnieri}\ \emph {et~al.}(2014)\citenamefont
  {Guarnieri}, \citenamefont {Smirne},\ and\ \citenamefont
  {Vacchini}}]{Giac_2014}%
  \BibitemOpen
  \bibfield  {author} {\bibinfo {author} {\bibfnamefont {G.}~\bibnamefont
  {Guarnieri}}, \bibinfo {author} {\bibfnamefont {A.}~\bibnamefont {Smirne}}, \
  and\ \bibinfo {author} {\bibfnamefont {B.}~\bibnamefont {Vacchini}},\ }\href
  {\doibase 10.1103/PhysRevA.90.022110} {\bibfield  {journal} {\bibinfo
  {journal} {Phys. Rev. A}\ }\textbf {\bibinfo {volume} {90}},\ \bibinfo
  {pages} {022110} (\bibinfo {year} {2014})}\BibitemShut {NoStop}%
\bibitem [{\citenamefont {Khan}\ \emph {et~al.}(2021)\citenamefont {Khan},
  \citenamefont {Agarwalla},\ and\ \citenamefont {Jain}}]{Bijay_2021}%
  \BibitemOpen
  \bibfield  {author} {\bibinfo {author} {\bibfnamefont {S.}~\bibnamefont
  {Khan}}, \bibinfo {author} {\bibfnamefont {B.~K.}\ \bibnamefont {Agarwalla}},
  \ and\ \bibinfo {author} {\bibfnamefont {S.}~\bibnamefont {Jain}},\
  }\href@noop {} {\  (\bibinfo {year} {2021})},\ \Eprint
  {http://arxiv.org/abs/2111.14879} {arXiv:2111.14879 [quant-ph]} \BibitemShut
  {NoStop}%
\bibitem [{\citenamefont {Purkayastha}\ \emph {et~al.}(2018)\citenamefont
  {Purkayastha}, \citenamefont {Sanyal}, \citenamefont {Dhar},\ and\
  \citenamefont {Kulkarni}}]{Archak_AAH_2018}%
  \BibitemOpen
  \bibfield  {author} {\bibinfo {author} {\bibfnamefont {A.}~\bibnamefont
  {Purkayastha}}, \bibinfo {author} {\bibfnamefont {S.}~\bibnamefont {Sanyal}},
  \bibinfo {author} {\bibfnamefont {A.}~\bibnamefont {Dhar}}, \ and\ \bibinfo
  {author} {\bibfnamefont {M.}~\bibnamefont {Kulkarni}},\ }\href {\doibase
  10.1103/PhysRevB.97.174206} {\bibfield  {journal} {\bibinfo  {journal} {Phys.
  Rev. B}\ }\textbf {\bibinfo {volume} {97}},\ \bibinfo {pages} {174206}
  (\bibinfo {year} {2018})}\BibitemShut {NoStop}%
\bibitem [{\citenamefont {Salazar}\ and\ \citenamefont
  {Landi}(2020)}]{Landi_2020}%
  \BibitemOpen
  \bibfield  {author} {\bibinfo {author} {\bibfnamefont {D.~S.~P.}\
  \bibnamefont {Salazar}}\ and\ \bibinfo {author} {\bibfnamefont {G.~T.}\
  \bibnamefont {Landi}},\ }\href {\doibase 10.1103/PhysRevResearch.2.033090}
  {\bibfield  {journal} {\bibinfo  {journal} {Phys. Rev. Research}\ }\textbf
  {\bibinfo {volume} {2}},\ \bibinfo {pages} {033090} (\bibinfo {year}
  {2020})}\BibitemShut {NoStop}%
\bibitem [{\citenamefont {Purkayastha}\ and\ \citenamefont
  {Dubi}(2017)}]{Archak_2017}%
  \BibitemOpen
  \bibfield  {author} {\bibinfo {author} {\bibfnamefont {A.}~\bibnamefont
  {Purkayastha}}\ and\ \bibinfo {author} {\bibfnamefont {Y.}~\bibnamefont
  {Dubi}},\ }\href {\doibase 10.1103/PhysRevB.96.085425} {\bibfield  {journal}
  {\bibinfo  {journal} {Phys. Rev. B}\ }\textbf {\bibinfo {volume} {96}},\
  \bibinfo {pages} {085425} (\bibinfo {year} {2017})}\BibitemShut {NoStop}%
\end{thebibliography}%
\end{document}